\documentclass[longauth]{aa}
\usepackage{graphicx}
\usepackage{txfonts}
\usepackage[colorlinks=true, linkcolor=blue, citecolor=cyan, urlcolor=blue]{hyperref}
\usepackage[nameinlink]{cleveref}

\providecommand{\linktodm}{https://gea.esac.esa.int/archive/documentation/GDR3/Gaia_archive/chap_datamodel}
\providecommand{\linktoparam}[2]{\href{\linktodm/sec_dm_main_source_catalogue/ssec_dm_#1.html\##1-#2}{\ \texttt{\small#2}\xspace}}
\providecommand{\linktotable}[1]{\href{\linktodm/sec_dm_main_source_catalogue/ssec_dm_#1.html}{\ \texttt{\small#1}\xspace}}
\providecommand{\linktodmfpr}{https://gea.esac.esa.int/archive/documentation/FPR/chap_datamodel}
\providecommand{\linktoparamfpr}[2]{\href{\linktodmfpr/sec_dm_focused_product_release/ssec_dm_#1.html\##1-#2}{\ \texttt{\small#2}\xspace}}
\providecommand{\linktotablefpr}[1]{\href{\linktodmfpr/sec_dm_focused_product_release/ssec_dm_#1.html}{\texttt{\ \small#1}\xspace}}
\providecommand{\orcit}[1]{\protect\href{https://orcid.org/#1}{\protect\includegraphics[width=8pt]{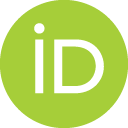}}}
\providecommand{\gaia}{{\it Gaia}}
\providecommand{\hip}{\textsc{Hipparcos}}
\providecommand{\tyc}{{\it Tycho}}
\providecommand{\tyctwo}{{\it Tycho-2}}

\makeatletter
\renewcommand*\maketitle{%
  \thispagestyle{firstpage}
\begingroup
    \if@wideboxfn
    \setlength\bibindent{1.4\parindent}
    \else
    \setlength\bibindent{\parindent}
    \fi
    \renewcommand*\thefootnote{\@fnsymbol\c@footnote}%
    \renewcommand\@makefntext[1]{%
    \ifaa@longfn\hsize\textwidth\fi
    \noindent
    \hb@xt@\bibindent{\hss\@makefnmark\enspace}##1}
  \ifaa@twocolumn
  \begin{aa@strip}
    \aa@maketitle
    \@thanks
  \end{aa@strip}
  \else
    \begingroup
      \let\thanks\footnote
      \aa@maketitle
    \endgroup
  \fi
\endgroup
  \setcounter{footnote}{0}%
}
\makeatother

\begin{document} 
    \title{{\it Gaia} Focused Product Release:\\Sources from Service Interface Function image analysis}
    \subtitle{Half a million new sources in omega Centauri
    \thanks{A table listing remaining duplicates between the SIF CF and {\it Gaia} DR3 catalogue (see Sect. \ref{sec:issues}) is available in electronic form at the CDS via anonymous ftp to cdsarc.cds.unistra.fr or via https://cdsarc.cds.unistra.fr/cgi-bin/qcat?J/A+A/}
    }
    \titlerunning{{\it Gaia} FPR: SIF CF sources in $\omega$ Cen}
    \authorrunning{Gaia Collaboration, K.~Weingrill, A.~Mints et al.}
\author{
{\it Gaia} Collaboration
\and K.        ~Weingrill                     \orcit{0000-0002-8163-2493}\inst{\ref{inst:0001}}
\and A.        ~Mints                         \orcit{0000-0002-8440-1455}\inst{\ref{inst:0001}}
\and J.        ~Casta\~{n}eda                 \orcit{0000-0001-7820-946X}\inst{\ref{inst:0003},\ref{inst:0004},\ref{inst:0005}}
\and Z.        ~Kostrzewa-Rutkowska           \inst{\ref{inst:0006}}
\and M.        ~Davidson                      \orcit{0000-0001-9271-4411}\inst{\ref{inst:0007}}
\and F.        ~De Angeli                     \orcit{0000-0003-1879-0488}\inst{\ref{inst:0008}}
\and J.        ~Hern\'{a}ndez                 \orcit{0000-0002-0361-4994}\inst{\ref{inst:0009}}
\and F.        ~Torra                         \orcit{0000-0002-8429-299X}\inst{\ref{inst:0003},\ref{inst:0004},\ref{inst:0005}}
\and M.        ~Ramos-Lerate                  \orcit{0009-0005-4677-8031}\inst{\ref{inst:0013}}
\and C.        ~Babusiaux                     \orcit{0000-0002-7631-348X}\inst{\ref{inst:0014}}
\and M.        ~Biermann                      \orcit{0000-0002-5791-9056}\inst{\ref{inst:0015}}
\and C.        ~Crowley                       \orcit{0000-0002-9391-9360}\inst{\ref{inst:0016}}
\and D.W.      ~Evans                         \orcit{0000-0002-6685-5998}\inst{\ref{inst:0008}}
\and L.        ~Lindegren                     \orcit{0000-0002-5443-3026}\inst{\ref{inst:0018}}
\and J.M.      ~Mart\'{i}n-Fleitas            \orcit{0000-0002-8594-569X}\inst{\ref{inst:0019}}
\and L.        ~Palaversa                     \orcit{0000-0003-3710-0331}\inst{\ref{inst:0020}}
\and D.        ~Ruz Mieres                    \orcit{0000-0002-9455-157X}\inst{\ref{inst:0008}}
\and K.        ~Tisani\'{c}                   \orcit{0000-0001-6382-4937}\inst{\ref{inst:0020}}
\and A.G.A.    ~Brown                         \orcit{0000-0002-7419-9679}\inst{\ref{inst:0006}}
\and A.        ~Vallenari                     \orcit{0000-0003-0014-519X}\inst{\ref{inst:0024}}
\and T.        ~Prusti                        \orcit{0000-0003-3120-7867}\inst{\ref{inst:0025}}
\and J.H.J.    ~de Bruijne                    \orcit{0000-0001-6459-8599}\inst{\ref{inst:0025}}
\and F.        ~Arenou                        \orcit{0000-0003-2837-3899}\inst{\ref{inst:0027}}
\and A.        ~Barbier                       \orcit{0009-0004-0983-931X}\inst{\ref{inst:0028}}
\and O.L.      ~Creevey                       \orcit{0000-0003-1853-6631}\inst{\ref{inst:0029}}
\and C.        ~Ducourant                     \orcit{0000-0003-4843-8979}\inst{\ref{inst:0030}}
\and L.        ~Eyer                          \orcit{0000-0002-0182-8040}\inst{\ref{inst:0031}}
\and R.        ~Guerra                        \orcit{0000-0002-9850-8982}\inst{\ref{inst:0009}}
\and A.        ~Hutton                        \inst{\ref{inst:0019}}
\and C.        ~Jordi                         \orcit{0000-0001-5495-9602}\inst{\ref{inst:0004},\ref{inst:0035},\ref{inst:0005}}
\and S.A.      ~Klioner                       \orcit{0000-0003-4682-7831}\inst{\ref{inst:0037}}
\and U.        ~Lammers                       \orcit{0000-0001-8309-3801}\inst{\ref{inst:0009}}
\and X.        ~Luri                          \orcit{0000-0001-5428-9397}\inst{\ref{inst:0004},\ref{inst:0035},\ref{inst:0005}}
\and F.        ~Mignard                       \inst{\ref{inst:0029}}
\and S.        ~Randich                       \orcit{0000-0003-2438-0899}\inst{\ref{inst:0043}}
\and P.        ~Sartoretti                    \orcit{0000-0002-6574-7565}\inst{\ref{inst:0027}}
\and R.        ~Smiljanic                     \orcit{0000-0003-0942-7855}\inst{\ref{inst:0045}}
\and P.        ~Tanga                         \orcit{0000-0002-2718-997X}\inst{\ref{inst:0029}}
\and N.A.      ~Walton                        \orcit{0000-0003-3983-8778}\inst{\ref{inst:0008}}
\and C.A.L.    ~Bailer-Jones                  \inst{\ref{inst:0048}}
\and U.        ~Bastian                       \orcit{0000-0002-8667-1715}\inst{\ref{inst:0015}}
\and M.        ~Cropper                       \orcit{0000-0003-4571-9468}\inst{\ref{inst:0050}}
\and R.        ~Drimmel                       \orcit{0000-0002-1777-5502}\inst{\ref{inst:0051}}
\and D.        ~Katz                          \orcit{0000-0001-7986-3164}\inst{\ref{inst:0027}}
\and C.        ~Soubiran                      \orcit{0000-0003-3304-8134}\inst{\ref{inst:0030}}
\and F.        ~van Leeuwen                   \orcit{0000-0003-1781-4441}\inst{\ref{inst:0008}}
\and M.        ~Audard                        \orcit{0000-0003-4721-034X}\inst{\ref{inst:0031},\ref{inst:0056}}
\and J.        ~Bakker                        \inst{\ref{inst:0009}}
\and R.        ~Blomme                        \orcit{0000-0002-2526-346X}\inst{\ref{inst:0058}}
\and C.        ~Fabricius                     \orcit{0000-0003-2639-1372}\inst{\ref{inst:0005},\ref{inst:0004},\ref{inst:0035}}
\and M.        ~Fouesneau                     \orcit{0000-0001-9256-5516}\inst{\ref{inst:0048}}
\and Y.        ~Fr\'{e}mat                    \orcit{0000-0002-4645-6017}\inst{\ref{inst:0058}}
\and L.        ~Galluccio                     \orcit{0000-0002-8541-0476}\inst{\ref{inst:0029}}
\and A.        ~Guerrier                      \inst{\ref{inst:0028}}
\and E.        ~Masana                        \orcit{0000-0002-4819-329X}\inst{\ref{inst:0005},\ref{inst:0004},\ref{inst:0035}}
\and R.        ~Messineo                      \inst{\ref{inst:0069}}
\and C.        ~Nicolas                       \inst{\ref{inst:0028}}
\and K.        ~Nienartowicz                  \orcit{0000-0001-5415-0547}\inst{\ref{inst:0071},\ref{inst:0056}}
\and F.        ~Pailler                       \orcit{0000-0002-4834-481X}\inst{\ref{inst:0028}}
\and P.        ~Panuzzo                       \orcit{0000-0002-0016-8271}\inst{\ref{inst:0027}}
\and F.        ~Riclet                        \inst{\ref{inst:0028}}
\and W.        ~Roux                          \orcit{0000-0002-7816-1950}\inst{\ref{inst:0028}}
\and G.M.      ~Seabroke                      \orcit{0000-0003-4072-9536}\inst{\ref{inst:0050}}
\and R.        ~Sordo                         \orcit{0000-0003-4979-0659}\inst{\ref{inst:0024}}
\and F.        ~Th\'{e}venin                  \orcit{0000-0002-5032-2476}\inst{\ref{inst:0029}}
\and G.        ~Gracia-Abril                  \inst{\ref{inst:0080},\ref{inst:0015}}
\and J.        ~Portell                       \orcit{0000-0002-8886-8925}\inst{\ref{inst:0004},\ref{inst:0035},\ref{inst:0005}}
\and D.        ~Teyssier                      \orcit{0000-0002-6261-5292}\inst{\ref{inst:0013}}
\and M.        ~Altmann                       \orcit{0000-0002-0530-0913}\inst{\ref{inst:0015},\ref{inst:0087}}
\and K.        ~Benson                        \inst{\ref{inst:0050}}
\and J.        ~Berthier                      \orcit{0000-0003-1846-6485}\inst{\ref{inst:0089}}
\and P.W.      ~Burgess                       \orcit{0009-0002-6668-4559}\inst{\ref{inst:0008}}
\and D.        ~Busonero                      \orcit{0000-0002-3903-7076}\inst{\ref{inst:0051}}
\and G.        ~Busso                         \orcit{0000-0003-0937-9849}\inst{\ref{inst:0008}}
\and H.        ~C\'{a}novas                   \orcit{0000-0001-7668-8022}\inst{\ref{inst:0013}}
\and B.        ~Carry                         \orcit{0000-0001-5242-3089}\inst{\ref{inst:0029}}
\and N.        ~Cheek                         \inst{\ref{inst:0095}}
\and G.        ~Clementini                    \orcit{0000-0001-9206-9723}\inst{\ref{inst:0096}}
\and Y.        ~Damerdji                      \orcit{0000-0002-3107-4024}\inst{\ref{inst:0097},\ref{inst:0098}}
\and P.        ~de Teodoro                    \inst{\ref{inst:0009}}
\and L.        ~Delchambre                    \orcit{0000-0003-2559-408X}\inst{\ref{inst:0097}}
\and A.        ~Dell'Oro                      \orcit{0000-0003-1561-9685}\inst{\ref{inst:0043}}
\and E.        ~Fraile Garcia                 \orcit{0000-0001-7742-9663}\inst{\ref{inst:0102}}
\and D.        ~Garabato                      \orcit{0000-0002-7133-6623}\inst{\ref{inst:0103}}
\and P.        ~Garc\'{i}a-Lario              \orcit{0000-0003-4039-8212}\inst{\ref{inst:0009}}
\and N.        ~Garralda Torres               \inst{\ref{inst:0105}}
\and P.        ~Gavras                        \orcit{0000-0002-4383-4836}\inst{\ref{inst:0102}}
\and R.        ~Haigron                       \inst{\ref{inst:0027}}
\and N.C.      ~Hambly                        \orcit{0000-0002-9901-9064}\inst{\ref{inst:0007}}
\and D.L.      ~Harrison                      \orcit{0000-0001-8687-6588}\inst{\ref{inst:0008},\ref{inst:0110}}
\and D.        ~Hatzidimitriou                \orcit{0000-0002-5415-0464}\inst{\ref{inst:0111}}
\and S.T.      ~Hodgkin                       \orcit{0000-0002-5470-3962}\inst{\ref{inst:0008}}
\and B.        ~Holl                          \orcit{0000-0001-6220-3266}\inst{\ref{inst:0031},\ref{inst:0056}}
\and S.        ~Jamal                         \orcit{0000-0002-3929-6668}\inst{\ref{inst:0048}}
\and S.        ~Jordan                        \orcit{0000-0001-6316-6831}\inst{\ref{inst:0015}}
\and A.        ~Krone-Martins                 \orcit{0000-0002-2308-6623}\inst{\ref{inst:0117},\ref{inst:0118}}
\and A.C.      ~Lanzafame                     \orcit{0000-0002-2697-3607}\inst{\ref{inst:0119},\ref{inst:0120}}
\and W.        ~L\"{ o}ffler                  \inst{\ref{inst:0015}}
\and A.        ~Lorca                         \orcit{0000-0002-7985-250X}\inst{\ref{inst:0019}}
\and O.        ~Marchal                       \orcit{ 0000-0001-7461-892}\inst{\ref{inst:0123}}
\and P.M.      ~Marrese                       \orcit{0000-0002-8162-3810}\inst{\ref{inst:0124},\ref{inst:0125}}
\and A.        ~Moitinho                      \orcit{0000-0003-0822-5995}\inst{\ref{inst:0118}}
\and K.        ~Muinonen                      \orcit{0000-0001-8058-2642}\inst{\ref{inst:0127},\ref{inst:0128}}
\and M.        ~Nu\~{n}ez Campos              \inst{\ref{inst:0019}}
\and I.        ~Oreshina-Slezak               \inst{\ref{inst:0029}}
\and P.        ~Osborne                       \orcit{0000-0003-4482-3538}\inst{\ref{inst:0008}}
\and E.        ~Pancino                       \orcit{0000-0003-0788-5879}\inst{\ref{inst:0043},\ref{inst:0125}}
\and T.        ~Pauwels                       \inst{\ref{inst:0058}}
\and A.        ~Recio-Blanco                  \orcit{0000-0002-6550-7377}\inst{\ref{inst:0029}}
\and M.        ~Riello                        \orcit{0000-0002-3134-0935}\inst{\ref{inst:0008}}
\and L.        ~Rimoldini                     \orcit{0000-0002-0306-585X}\inst{\ref{inst:0056}}
\and A.C.      ~Robin                         \orcit{0000-0001-8654-9499}\inst{\ref{inst:0138}}
\and T.        ~Roegiers                      \orcit{0000-0002-1231-4440}\inst{\ref{inst:0139}}
\and L.M.      ~Sarro                         \orcit{0000-0002-5622-5191}\inst{\ref{inst:0140}}
\and M.        ~Schultheis                    \orcit{0000-0002-6590-1657}\inst{\ref{inst:0029}}
\and C.        ~Siopis                        \orcit{0000-0002-6267-2924}\inst{\ref{inst:0142}}
\and M.        ~Smith                         \inst{\ref{inst:0050}}
\and A.        ~Sozzetti                      \orcit{0000-0002-7504-365X}\inst{\ref{inst:0051}}
\and E.        ~Utrilla                       \inst{\ref{inst:0019}}
\and M.        ~van Leeuwen                   \orcit{0000-0001-9698-2392}\inst{\ref{inst:0008}}
\and U.        ~Abbas                         \orcit{0000-0002-5076-766X}\inst{\ref{inst:0051}}
\and P.        ~\'{A}brah\'{a}m               \orcit{0000-0001-6015-646X}\inst{\ref{inst:0148},\ref{inst:0149}}
\and A.        ~Abreu Aramburu                \orcit{0000-0003-3959-0856}\inst{\ref{inst:0105}}
\and C.        ~Aerts                         \orcit{0000-0003-1822-7126}\inst{\ref{inst:0151},\ref{inst:0152},\ref{inst:0048}}
\and G.        ~Altavilla                     \orcit{0000-0002-9934-1352}\inst{\ref{inst:0124},\ref{inst:0125}}
\and M.A.      ~\'{A}lvarez                   \orcit{0000-0002-6786-2620}\inst{\ref{inst:0103}}
\and J.        ~Alves                         \orcit{0000-0002-4355-0921}\inst{\ref{inst:0157}}
\and F.        ~Anders                        \inst{\ref{inst:0004},\ref{inst:0035},\ref{inst:0005}}
\and R.I.      ~Anderson                      \orcit{0000-0001-8089-4419}\inst{\ref{inst:0161}}
\and T.        ~Antoja                        \orcit{0000-0003-2595-5148}\inst{\ref{inst:0004},\ref{inst:0035},\ref{inst:0005}}
\and D.        ~Baines                        \orcit{0000-0002-6923-3756}\inst{\ref{inst:0165}}
\and S.G.      ~Baker                         \orcit{0000-0002-6436-1257}\inst{\ref{inst:0050}}
\and Z.        ~Balog                         \orcit{0000-0003-1748-2926}\inst{\ref{inst:0015},\ref{inst:0048}}
\and C.        ~Barache                       \inst{\ref{inst:0087}}
\and D.        ~Barbato                       \inst{\ref{inst:0031},\ref{inst:0051}}
\and M.        ~Barros                        \orcit{0000-0002-9728-9618}\inst{\ref{inst:0172}}
\and M.A.      ~Barstow                       \orcit{0000-0002-7116-3259}\inst{\ref{inst:0173}}
\and S.        ~Bartolom\'{e}                 \orcit{0000-0002-6290-6030}\inst{\ref{inst:0005},\ref{inst:0004},\ref{inst:0035}}
\and D.        ~Bashi                         \orcit{0000-0002-9035-2645}\inst{\ref{inst:0177},\ref{inst:0178}}
\and N.        ~Bauchet                       \orcit{0000-0002-2307-8973}\inst{\ref{inst:0027}}
\and N.        ~Baudeau                       \inst{\ref{inst:0180}}
\and U.        ~Becciani                      \orcit{0000-0002-4389-8688}\inst{\ref{inst:0119}}
\and L.R.      ~Bedin                         \inst{\ref{inst:0024}}
\and I.        ~Bellas-Velidis                \inst{\ref{inst:0183}}
\and M.        ~Bellazzini                    \orcit{0000-0001-8200-810X}\inst{\ref{inst:0096}}
\and W.        ~Beordo                        \orcit{0000-0002-5094-1306}\inst{\ref{inst:0051},\ref{inst:0186}}
\and A.        ~Berihuete                     \orcit{0000-0002-8589-4423}\inst{\ref{inst:0187}}
\and M.        ~Bernet                        \orcit{0000-0001-7503-1010}\inst{\ref{inst:0004},\ref{inst:0035},\ref{inst:0005}}
\and C.        ~Bertolotto                    \inst{\ref{inst:0069}}
\and S.        ~Bertone                       \orcit{0000-0001-9885-8440}\inst{\ref{inst:0051}}
\and L.        ~Bianchi                       \orcit{0000-0002-7999-4372}\inst{\ref{inst:0193}}
\and A.        ~Binnenfeld                    \orcit{0000-0002-9319-3838}\inst{\ref{inst:0194}}
\and A.        ~Blazere                       \inst{\ref{inst:0195}}
\and T.        ~Boch                          \orcit{0000-0001-5818-2781}\inst{\ref{inst:0123}}
\and A.        ~Bombrun                       \inst{\ref{inst:0016}}
\and S.        ~Bouquillon                    \inst{\ref{inst:0087},\ref{inst:0199}}
\and A.        ~Bragaglia                     \orcit{0000-0002-0338-7883}\inst{\ref{inst:0096}}
\and J.        ~Braine                        \orcit{0000-0003-1740-1284}\inst{\ref{inst:0030}}
\and L.        ~Bramante                      \inst{\ref{inst:0069}}
\and E.        ~Breedt                        \orcit{0000-0001-6180-3438}\inst{\ref{inst:0008}}
\and A.        ~Bressan                       \orcit{0000-0002-7922-8440}\inst{\ref{inst:0204}}
\and N.        ~Brouillet                     \orcit{0000-0002-3274-7024}\inst{\ref{inst:0030}}
\and E.        ~Brugaletta                    \orcit{0000-0003-2598-6737}\inst{\ref{inst:0119}}
\and B.        ~Bucciarelli                   \orcit{0000-0002-5303-0268}\inst{\ref{inst:0051},\ref{inst:0186}}
\and A.G.      ~Butkevich                     \orcit{0000-0002-4098-3588}\inst{\ref{inst:0051}}
\and R.        ~Buzzi                         \orcit{0000-0001-9389-5701}\inst{\ref{inst:0051}}
\and E.        ~Caffau                        \orcit{0000-0001-6011-6134}\inst{\ref{inst:0027}}
\and R.        ~Cancelliere                   \orcit{0000-0002-9120-3799}\inst{\ref{inst:0212}}
\and S.        ~Cannizzo                      \inst{\ref{inst:0213}}
\and T.        ~Cantat-Gaudin                 \orcit{0000-0001-8726-2588}\inst{\ref{inst:0048}}
\and R.        ~Carballo                      \orcit{0000-0001-7412-2498}\inst{\ref{inst:0215}}
\and T.        ~Carlucci                      \inst{\ref{inst:0087}}
\and M.I.      ~Carnerero                     \orcit{0000-0001-5843-5515}\inst{\ref{inst:0051}}
\and J.M.      ~Carrasco                      \orcit{0000-0002-3029-5853}\inst{\ref{inst:0005},\ref{inst:0004},\ref{inst:0035}}
\and J.        ~Carretero                     \orcit{0000-0002-3130-0204}\inst{\ref{inst:0221},\ref{inst:0222}}
\and S.        ~Carton                        \inst{\ref{inst:0213}}
\and L.        ~Casamiquela                   \orcit{0000-0001-5238-8674}\inst{\ref{inst:0030},\ref{inst:0027}}
\and M.        ~Castellani                    \orcit{0000-0002-7650-7428}\inst{\ref{inst:0124}}
\and A.        ~Castro-Ginard                 \orcit{0000-0002-9419-3725}\inst{\ref{inst:0006}}
\and V.        ~Cesare                        \orcit{0000-0003-1119-4237}\inst{\ref{inst:0119}}
\and P.        ~Charlot                       \orcit{0000-0002-9142-716X}\inst{\ref{inst:0030}}
\and L.        ~Chemin                        \orcit{0000-0002-3834-7937}\inst{\ref{inst:0230}}
\and V.        ~Chiaramida                    \inst{\ref{inst:0069}}
\and A.        ~Chiavassa                     \orcit{0000-0003-3891-7554}\inst{\ref{inst:0029}}
\and N.        ~Chornay                       \orcit{0000-0002-8767-3907}\inst{\ref{inst:0008},\ref{inst:0056}}
\and R.        ~Collins                       \orcit{0000-0001-8437-1703}\inst{\ref{inst:0007}}
\and G.        ~Contursi                      \orcit{0000-0001-5370-1511}\inst{\ref{inst:0029}}
\and W.J.      ~Cooper                        \orcit{0000-0003-3501-8967}\inst{\ref{inst:0237},\ref{inst:0051}}
\and T.        ~Cornez                        \inst{\ref{inst:0213}}
\and M.        ~Crosta                        \orcit{0000-0003-4369-3786}\inst{\ref{inst:0051},\ref{inst:0241}}
\and C.        ~Dafonte                       \orcit{0000-0003-4693-7555}\inst{\ref{inst:0103}}
\and P.        ~de Laverny                    \orcit{0000-0002-2817-4104}\inst{\ref{inst:0029}}
\and F.        ~De Luise                      \orcit{0000-0002-6570-8208}\inst{\ref{inst:0245}}
\and R.        ~De March                      \orcit{0000-0003-0567-842X}\inst{\ref{inst:0069}}
\and R.        ~de Souza                      \orcit{0009-0007-7669-0254}\inst{\ref{inst:0247}}
\and A.        ~de Torres                     \inst{\ref{inst:0016}}
\and E.F.      ~del Peloso                    \inst{\ref{inst:0015}}
\and M.        ~Delbo                         \orcit{0000-0002-8963-2404}\inst{\ref{inst:0029}}
\and A.        ~Delgado                       \inst{\ref{inst:0102}}
\and T.E.      ~Dharmawardena                 \orcit{0000-0002-9583-5216}\inst{\ref{inst:0048}}
\and S.        ~Diakite                       \inst{\ref{inst:0253}}
\and C.        ~Diener                        \inst{\ref{inst:0008}}
\and E.        ~Distefano                     \orcit{0000-0002-2448-2513}\inst{\ref{inst:0119}}
\and C.        ~Dolding                       \inst{\ref{inst:0050}}
\and K.        ~Dsilva                        \orcit{0000-0002-1476-9772}\inst{\ref{inst:0142}}
\and J.        ~Dur\'{a}n                     \inst{\ref{inst:0102}}
\and H.        ~Enke                          \orcit{0000-0002-2366-8316}\inst{\ref{inst:0001}}
\and P.        ~Esquej                        \orcit{0000-0001-8195-628X}\inst{\ref{inst:0102}}
\and C.        ~Fabre                         \inst{\ref{inst:0195}}
\and M.        ~Fabrizio                      \orcit{0000-0001-5829-111X}\inst{\ref{inst:0124},\ref{inst:0125}}
\and S.        ~Faigler                       \orcit{0000-0002-8368-5724}\inst{\ref{inst:0177}}
\and M.        ~Fatovi\'{c}                   \orcit{0000-0003-1911-4326}\inst{\ref{inst:0020}}
\and G.        ~Fedorets                      \orcit{0000-0002-8418-4809}\inst{\ref{inst:0127},\ref{inst:0267}}
\and J.        ~Fern\'{a}ndez-Hern\'{a}ndez   \inst{\ref{inst:0102}}
\and P.        ~Fernique                      \orcit{0000-0002-3304-2923}\inst{\ref{inst:0123}}
\and F.        ~Figueras                      \orcit{0000-0002-3393-0007}\inst{\ref{inst:0004},\ref{inst:0035},\ref{inst:0005}}
\and Y.        ~Fournier                      \orcit{0000-0002-6633-9088}\inst{\ref{inst:0001}}
\and C.        ~Fouron                        \inst{\ref{inst:0180}}
\and M.        ~Gai                           \orcit{0000-0001-9008-134X}\inst{\ref{inst:0051}}
\and M.        ~Galinier                      \orcit{0000-0001-7920-0133}\inst{\ref{inst:0029}}
\and A.        ~Garcia-Gutierrez              \inst{\ref{inst:0005},\ref{inst:0004},\ref{inst:0035}}
\and M.        ~Garc\'{i}a-Torres             \orcit{0000-0002-6867-7080}\inst{\ref{inst:0280}}
\and A.        ~Garofalo                      \orcit{0000-0002-5907-0375}\inst{\ref{inst:0096}}
\and E.        ~Gerlach                       \orcit{0000-0002-9533-2168}\inst{\ref{inst:0037}}
\and R.        ~Geyer                         \orcit{0000-0001-6967-8707}\inst{\ref{inst:0037}}
\and P.        ~Giacobbe                      \orcit{0000-0001-7034-7024}\inst{\ref{inst:0051}}
\and G.        ~Gilmore                       \orcit{0000-0003-4632-0213}\inst{\ref{inst:0008},\ref{inst:0286}}
\and S.        ~Girona                        \orcit{0000-0002-1975-1918}\inst{\ref{inst:0287}}
\and G.        ~Giuffrida                     \orcit{0000-0002-8979-4614}\inst{\ref{inst:0124}}
\and R.        ~Gomel                         \inst{\ref{inst:0177}}
\and A.        ~Gomez                         \orcit{0000-0002-3796-3690}\inst{\ref{inst:0103}}
\and J.        ~Gonz\'{a}lez-N\'{u}\~{n}ez    \orcit{0000-0001-5311-5555}\inst{\ref{inst:0291}}
\and I.        ~Gonz\'{a}lez-Santamar\'{i}a   \orcit{0000-0002-8537-9384}\inst{\ref{inst:0103}}
\and E.        ~Gosset                        \inst{\ref{inst:0097},\ref{inst:0294}}
\and M.        ~Granvik                       \orcit{0000-0002-5624-1888}\inst{\ref{inst:0127},\ref{inst:0296}}
\and V.        ~Gregori Barrera               \inst{\ref{inst:0005},\ref{inst:0004},\ref{inst:0035}}
\and R.        ~Guti\'{e}rrez-S\'{a}nchez     \orcit{0009-0003-1500-4733}\inst{\ref{inst:0013}}
\and M.        ~Haywood                       \orcit{0000-0003-0434-0400}\inst{\ref{inst:0027}}
\and A.        ~Helmer                        \inst{\ref{inst:0213}}
\and A.        ~Helmi                         \orcit{0000-0003-3937-7641}\inst{\ref{inst:0303}}
\and K.        ~Henares                       \inst{\ref{inst:0165}}
\and S.L.      ~Hidalgo                       \orcit{0000-0002-0002-9298}\inst{\ref{inst:0305},\ref{inst:0306}}
\and T.        ~Hilger                        \orcit{0000-0003-1646-0063}\inst{\ref{inst:0037}}
\and D.        ~Hobbs                         \orcit{0000-0002-2696-1366}\inst{\ref{inst:0018}}
\and C.        ~Hottier                       \orcit{0000-0002-3498-3944}\inst{\ref{inst:0027}}
\and H.E.      ~Huckle                        \inst{\ref{inst:0050}}
\and M.        ~Jab\l{}o\'{n}ska              \orcit{0000-0001-6962-4979}\inst{\ref{inst:0311},\ref{inst:0312}}
\and F.        ~Jansen                        \inst{\ref{inst:0313}}
\and \'{O}.    ~Jim\'{e}nez-Arranz            \orcit{0000-0001-7434-5165}\inst{\ref{inst:0004},\ref{inst:0035},\ref{inst:0005}}
\and J.        ~Juaristi Campillo             \inst{\ref{inst:0015}}
\and S.        ~Khanna                        \orcit{0000-0002-2604-4277}\inst{\ref{inst:0051},\ref{inst:0303}}
\and G.        ~Kordopatis                    \orcit{0000-0002-9035-3920}\inst{\ref{inst:0029}}
\and \'{A}     ~K\'{o}sp\'{a}l                \orcit{0000-0001-7157-6275}\inst{\ref{inst:0148},\ref{inst:0048},\ref{inst:0149}}
\and M.        ~Kun                           \orcit{0000-0002-7538-5166}\inst{\ref{inst:0148}}
\and S.        ~Lambert                       \orcit{0000-0001-6759-5502}\inst{\ref{inst:0087}}
\and A.F.      ~Lanza                         \orcit{0000-0001-5928-7251}\inst{\ref{inst:0119}}
\and J.-F.     ~Le Campion                    \inst{\ref{inst:0030}}
\and Y.        ~Lebreton                      \orcit{0000-0002-4834-2144}\inst{\ref{inst:0328},\ref{inst:0329}}
\and T.        ~Lebzelter                     \orcit{0000-0002-0702-7551}\inst{\ref{inst:0157}}
\and S.        ~Leccia                        \orcit{0000-0001-5685-6930}\inst{\ref{inst:0331}}
\and I.        ~Lecoeur-Taibi                 \orcit{0000-0003-0029-8575}\inst{\ref{inst:0056}}
\and G.        ~Lecoutre                      \inst{\ref{inst:0138}}
\and S.        ~Liao                          \orcit{0000-0002-9346-0211}\inst{\ref{inst:0334},\ref{inst:0051},\ref{inst:0336}}
\and L.        ~Liberato                      \orcit{0000-0003-3433-6269}\inst{\ref{inst:0029},\ref{inst:0338}}
\and E.        ~Licata                        \orcit{0000-0002-5203-0135}\inst{\ref{inst:0051}}
\and H.E.P.    ~Lindstr{\o}m                  \orcit{0009-0004-8864-5459}\inst{\ref{inst:0051},\ref{inst:0341},\ref{inst:0342}}
\and T.A.      ~Lister                        \orcit{0000-0002-3818-7769}\inst{\ref{inst:0343}}
\and E.        ~Livanou                       \orcit{0000-0003-0628-2347}\inst{\ref{inst:0111}}
\and A.        ~Lobel                         \orcit{0000-0001-5030-019X}\inst{\ref{inst:0058}}
\and C.        ~Loup                          \inst{\ref{inst:0123}}
\and L.        ~Mahy                          \orcit{0000-0003-0688-7987}\inst{\ref{inst:0058}}
\and R.G.      ~Mann                          \orcit{0000-0002-0194-325X}\inst{\ref{inst:0007}}
\and M.        ~Manteiga                      \orcit{0000-0002-7711-5581}\inst{\ref{inst:0349}}
\and J.M.      ~Marchant                      \orcit{0000-0002-3678-3145}\inst{\ref{inst:0350}}
\and M.        ~Marconi                       \orcit{0000-0002-1330-2927}\inst{\ref{inst:0331}}
\and D.        ~Mar\'{i}n Pina                \orcit{0000-0001-6482-1842}\inst{\ref{inst:0004},\ref{inst:0035},\ref{inst:0005}}
\and S.        ~Marinoni                      \orcit{0000-0001-7990-6849}\inst{\ref{inst:0124},\ref{inst:0125}}
\and D.J.      ~Marshall                      \orcit{0000-0003-3956-3524}\inst{\ref{inst:0357}}
\and J.        ~Mart\'{i}n Lozano             \orcit{0009-0001-2435-6680}\inst{\ref{inst:0095}}
\and G.        ~Marton                        \orcit{0000-0002-1326-1686}\inst{\ref{inst:0148}}
\and N.        ~Mary                          \inst{\ref{inst:0213}}
\and A.        ~Masip                         \orcit{0000-0003-1419-0020}\inst{\ref{inst:0005},\ref{inst:0004},\ref{inst:0035}}
\and D.        ~Massari                       \orcit{0000-0001-8892-4301}\inst{\ref{inst:0096}}
\and A.        ~Mastrobuono-Battisti          \orcit{0000-0002-2386-9142}\inst{\ref{inst:0027}}
\and T.        ~Mazeh                         \orcit{0000-0002-3569-3391}\inst{\ref{inst:0177}}
\and P.J.      ~McMillan                      \orcit{0000-0002-8861-2620}\inst{\ref{inst:0018}}
\and J.        ~Meichsner                     \orcit{0000-0002-9900-7864}\inst{\ref{inst:0037}}
\and S.        ~Messina                       \orcit{0000-0002-2851-2468}\inst{\ref{inst:0119}}
\and D.        ~Michalik                      \orcit{0000-0002-7618-6556}\inst{\ref{inst:0025}}
\and N.R.      ~Millar                        \inst{\ref{inst:0008}}
\and D.        ~Molina                        \orcit{0000-0003-4814-0275}\inst{\ref{inst:0035},\ref{inst:0004},\ref{inst:0005}}
\and R.        ~Molinaro                      \orcit{0000-0003-3055-6002}\inst{\ref{inst:0331}}
\and L.        ~Moln\'{a}r                    \orcit{0000-0002-8159-1599}\inst{\ref{inst:0148},\ref{inst:0377},\ref{inst:0149}}
\and G.        ~Monari                        \orcit{0000-0002-6863-0661}\inst{\ref{inst:0123}}
\and M.        ~Mongui\'{o}                   \orcit{0000-0002-4519-6700}\inst{\ref{inst:0004},\ref{inst:0035},\ref{inst:0005}}
\and P.        ~Montegriffo                   \orcit{0000-0001-5013-5948}\inst{\ref{inst:0096}}
\and A.        ~Montero                       \inst{\ref{inst:0095}}
\and R.        ~Mor                           \orcit{0000-0002-8179-6527}\inst{\ref{inst:0385},\ref{inst:0035},\ref{inst:0005}}
\and A.        ~Mora                          \inst{\ref{inst:0019}}
\and R.        ~Morbidelli                    \orcit{0000-0001-7627-4946}\inst{\ref{inst:0051}}
\and T.        ~Morel                         \orcit{0000-0002-8176-4816}\inst{\ref{inst:0097}}
\and D.        ~Morris                        \orcit{0000-0002-1952-6251}\inst{\ref{inst:0007}}
\and N.        ~Mowlavi                       \orcit{0000-0003-1578-6993}\inst{\ref{inst:0031}}
\and D.        ~Munoz                         \inst{\ref{inst:0213}}
\and T.        ~Muraveva                      \orcit{0000-0002-0969-1915}\inst{\ref{inst:0096}}
\and C.P.      ~Murphy                        \inst{\ref{inst:0009}}
\and I.        ~Musella                       \orcit{0000-0001-5909-6615}\inst{\ref{inst:0331}}
\and Z.        ~Nagy                          \orcit{0000-0002-3632-1194}\inst{\ref{inst:0148}}
\and S.        ~Nieto                         \inst{\ref{inst:0102}}
\and L.        ~Noval                         \inst{\ref{inst:0213}}
\and A.        ~Ogden                         \inst{\ref{inst:0008}}
\and C.        ~Ordenovic                     \inst{\ref{inst:0029}}
\and C.        ~Pagani                        \orcit{0000-0001-5477-4720}\inst{\ref{inst:0402}}
\and I.        ~Pagano                        \orcit{0000-0001-9573-4928}\inst{\ref{inst:0119}}
\and P.A.      ~Palicio                       \orcit{0000-0002-7432-8709}\inst{\ref{inst:0029}}
\and L.        ~Pallas-Quintela               \orcit{0000-0001-9296-3100}\inst{\ref{inst:0103}}
\and A.        ~Panahi                        \orcit{0000-0001-5850-4373}\inst{\ref{inst:0177}}
\and C.        ~Panem                         \inst{\ref{inst:0028}}
\and S.        ~Payne-Wardenaar               \inst{\ref{inst:0015}}
\and L.        ~Pegoraro                      \inst{\ref{inst:0028}}
\and A.        ~Penttil\"{ a}                 \orcit{0000-0001-7403-1721}\inst{\ref{inst:0127}}
\and P.        ~Pesciullesi                   \inst{\ref{inst:0102}}
\and A.M.      ~Piersimoni                    \orcit{0000-0002-8019-3708}\inst{\ref{inst:0245}}
\and M.        ~Pinamonti                     \orcit{0000-0002-4445-1845}\inst{\ref{inst:0051}}
\and F.-X.     ~Pineau                        \orcit{0000-0002-2335-4499}\inst{\ref{inst:0123}}
\and E.        ~Plachy                        \orcit{0000-0002-5481-3352}\inst{\ref{inst:0148},\ref{inst:0377},\ref{inst:0149}}
\and G.        ~Plum                          \inst{\ref{inst:0027}}
\and E.        ~Poggio                        \orcit{0000-0003-3793-8505}\inst{\ref{inst:0029},\ref{inst:0051}}
\and D.        ~Pourbaix$^\dagger$            \orcit{0000-0002-3020-1837}\inst{\ref{inst:0142},\ref{inst:0294}}
\and A.        ~Pr\v{s}a                      \orcit{0000-0002-1913-0281}\inst{\ref{inst:0423}}
\and L.        ~Pulone                        \orcit{0000-0002-5285-998X}\inst{\ref{inst:0124}}
\and E.        ~Racero                        \orcit{0000-0002-6101-9050}\inst{\ref{inst:0095},\ref{inst:0426}}
\and M.        ~Rainer                        \orcit{0000-0002-8786-2572}\inst{\ref{inst:0043},\ref{inst:0428}}
\and C.M.      ~Raiteri                       \orcit{0000-0003-1784-2784}\inst{\ref{inst:0051}}
\and P.        ~Ramos                         \orcit{0000-0002-5080-7027}\inst{\ref{inst:0430},\ref{inst:0004},\ref{inst:0005}}
\and M.        ~Ratajczak                     \orcit{0000-0002-3218-2684}\inst{\ref{inst:0311}}
\and P.        ~Re Fiorentin                  \orcit{0000-0002-4995-0475}\inst{\ref{inst:0051}}
\and S.        ~Regibo                        \orcit{0000-0001-7227-9563}\inst{\ref{inst:0151}}
\and C.        ~Reyl\'{e}                     \orcit{0000-0003-2258-2403}\inst{\ref{inst:0138}}
\and V.        ~Ripepi                        \orcit{0000-0003-1801-426X}\inst{\ref{inst:0331}}
\and A.        ~Riva                          \orcit{0000-0002-6928-8589}\inst{\ref{inst:0051}}
\and H.-W.     ~Rix                           \orcit{0000-0003-4996-9069}\inst{\ref{inst:0048}}
\and G.        ~Rixon                         \orcit{0000-0003-4399-6568}\inst{\ref{inst:0008}}
\and N.        ~Robichon                      \orcit{0000-0003-4545-7517}\inst{\ref{inst:0027}}
\and C.        ~Robin                         \inst{\ref{inst:0213}}
\and M.        ~Romero-G\'{o}mez              \orcit{0000-0003-3936-1025}\inst{\ref{inst:0004},\ref{inst:0035},\ref{inst:0005}}
\and N.        ~Rowell                        \orcit{0000-0003-3809-1895}\inst{\ref{inst:0007}}
\and F.        ~Royer                         \orcit{0000-0002-9374-8645}\inst{\ref{inst:0027}}
\and K.A.      ~Rybicki                       \orcit{0000-0002-9326-9329}\inst{\ref{inst:0448}}
\and G.        ~Sadowski                      \orcit{0000-0002-3411-1003}\inst{\ref{inst:0142}}
\and A.        ~S\'{a}ez N\'{u}\~{n}ez        \orcit{0009-0001-6078-0868}\inst{\ref{inst:0005},\ref{inst:0004},\ref{inst:0035}}
\and A.        ~Sagrist\`{a} Sell\'{e}s       \orcit{0000-0001-6191-2028}\inst{\ref{inst:0015}}
\and J.        ~Sahlmann                      \orcit{0000-0001-9525-3673}\inst{\ref{inst:0102}}
\and V.        ~Sanchez Gimenez               \orcit{0000-0003-1797-3557}\inst{\ref{inst:0005},\ref{inst:0004},\ref{inst:0035}}
\and N.        ~Sanna                         \orcit{0000-0001-9275-9492}\inst{\ref{inst:0043}}
\and R.        ~Santove\~{n}a                 \orcit{0000-0002-9257-2131}\inst{\ref{inst:0103}}
\and M.        ~Sarasso                       \orcit{0000-0001-5121-0727}\inst{\ref{inst:0051}}
\and C.        ~Sarrate Riera                 \inst{\ref{inst:0003},\ref{inst:0004},\ref{inst:0005}}
\and E.        ~Sciacca                       \orcit{0000-0002-5574-2787}\inst{\ref{inst:0119}}
\and J.C.      ~Segovia                       \inst{\ref{inst:0095}}
\and D.        ~S\'{e}gransan                 \orcit{0000-0003-2355-8034}\inst{\ref{inst:0031}}
\and S.        ~Shahaf                        \orcit{0000-0001-9298-8068}\inst{\ref{inst:0448}}
\and A.        ~Siebert                       \orcit{0000-0001-8059-2840}\inst{\ref{inst:0123},\ref{inst:0469}}
\and L.        ~Siltala                       \orcit{0000-0002-6938-794X}\inst{\ref{inst:0127}}
\and E.        ~Slezak                        \inst{\ref{inst:0029}}
\and R.L.      ~Smart                         \orcit{0000-0002-4424-4766}\inst{\ref{inst:0051},\ref{inst:0237}}
\and O.N.      ~Snaith                        \orcit{0000-0003-1414-1296}\inst{\ref{inst:0027},\ref{inst:0475}}
\and E.        ~Solano                        \orcit{0000-0003-1885-5130}\inst{\ref{inst:0476}}
\and F.        ~Solitro                       \inst{\ref{inst:0069}}
\and D.        ~Souami                        \orcit{0000-0003-4058-0815}\inst{\ref{inst:0328},\ref{inst:0479}}
\and J.        ~Souchay                       \inst{\ref{inst:0087}}
\and L.        ~Spina                         \orcit{0000-0002-9760-6249}\inst{\ref{inst:0024}}
\and E.        ~Spitoni                       \orcit{0000-0001-9715-5727}\inst{\ref{inst:0029},\ref{inst:0483}}
\and F.        ~Spoto                         \orcit{0000-0001-7319-5847}\inst{\ref{inst:0484}}
\and L.A.      ~Squillante                    \inst{\ref{inst:0069}}
\and I.A.      ~Steele                        \orcit{0000-0001-8397-5759}\inst{\ref{inst:0350}}
\and H.        ~Steidelm\"{ u}ller            \inst{\ref{inst:0037}}
\and J.        ~Surdej                        \orcit{0000-0002-7005-1976}\inst{\ref{inst:0097}}
\and L.        ~Szabados                      \orcit{0000-0002-2046-4131}\inst{\ref{inst:0148}}
\and F.        ~Taris                         \inst{\ref{inst:0087}}
\and M.B.      ~Taylor                        \orcit{0000-0002-4209-1479}\inst{\ref{inst:0491}}
\and R.        ~Teixeira                      \orcit{0000-0002-6806-6626}\inst{\ref{inst:0247}}
\and L.        ~Tolomei                       \orcit{0000-0002-3541-3230}\inst{\ref{inst:0069}}
\and G.        ~Torralba Elipe                \orcit{0000-0001-8738-194X}\inst{\ref{inst:0103},\ref{inst:0495},\ref{inst:0496}}
\and M.        ~Trabucchi                     \orcit{0000-0002-1429-2388}\inst{\ref{inst:0497},\ref{inst:0031}}
\and M.        ~Tsantaki                      \orcit{0000-0002-0552-2313}\inst{\ref{inst:0043}}
\and A.        ~Ulla                          \orcit{0000-0001-6424-5005}\inst{\ref{inst:0500},\ref{inst:0501}}
\and N.        ~Unger                         \orcit{0000-0003-3993-7127}\inst{\ref{inst:0031}}
\and O.        ~Vanel                         \orcit{0000-0002-7898-0454}\inst{\ref{inst:0027}}
\and A.        ~Vecchiato                     \orcit{0000-0003-1399-5556}\inst{\ref{inst:0051}}
\and D.        ~Vicente                       \orcit{0000-0002-1584-1182}\inst{\ref{inst:0287}}
\and S.        ~Voutsinas                     \inst{\ref{inst:0007}}
\and M.        ~Weiler                        \inst{\ref{inst:0005},\ref{inst:0004},\ref{inst:0035}}
\and \L{}.     ~Wyrzykowski                   \orcit{0000-0002-9658-6151}\inst{\ref{inst:0311}}
\and H.        ~Zhao                          \orcit{0000-0003-2645-6869}\inst{\ref{inst:0029},\ref{inst:0512}}
\and J.        ~Zorec                         \orcit{0000-0003-1257-6915}\inst{\ref{inst:0513}}
\and T.        ~Zwitter                       \orcit{0000-0002-2325-8763}\inst{\ref{inst:0514}}
\and L.        ~Balaguer-N\'{u}\~{n}ez      \orcit{0000-0001-9789-7069}\inst{\ref{inst:0005},\ref{inst:0004},\ref{inst:0035}}
\and N.        ~Leclerc                      \orcit{0009-0001-5569-6098}\inst{\ref{inst:0027}}
\and S.        ~Morgenthaler               \orcit{0009-0005-6349-3716}\inst{\ref{inst:0555}}
\and G.        ~Robert                         \inst{\ref{inst:0213}}
\and S.        ~Zucker                    \orcit{0000-0003-3173-3138}\inst{\ref{inst:0194}}
}
\institute{
     Leibniz Institute for Astrophysics Potsdam (AIP), An der Sternwarte 16, 14482 Potsdam, Germany\relax                                                                                                                                                                                                                                                                                                            \label{inst:0001}
\and DAPCOM Data Services, c. dels Vilabella, 5-7, 08500 Vic, Barcelona, Spain\relax                                                                                                                                                                                                                                                                                                                                 \label{inst:0003}
\and Institut de Ci\`{e}ncies del Cosmos (ICCUB), Universitat  de  Barcelona  (UB), Mart\'{i} i  Franqu\`{e}s  1, 08028 Barcelona, Spain\relax                                                                                                                                                                                                                                                                       \label{inst:0004}
\and Institut d'Estudis Espacials de Catalunya (IEEC), c. Gran Capit\`{a}, 2-4, 08034 Barcelona, Spain\relax                                                                                                                                                                                                                                                                                                         \label{inst:0005}
\and Leiden Observatory, Leiden University, Niels Bohrweg 2, 2333 CA Leiden, The Netherlands\relax                                                                                                                                                                                                                                                                                                                   \label{inst:0006}
\and Institute for Astronomy, University of Edinburgh, Royal Observatory, Blackford Hill, Edinburgh EH9 3HJ, United Kingdom\relax                                                                                                                                                                                                                                                                                    \label{inst:0007}
\and Institute of Astronomy, University of Cambridge, Madingley Road, Cambridge CB3 0HA, United Kingdom\relax                                                                                                                                                                                                                                                                                                        \label{inst:0008}
\and European Space Agency (ESA), European Space Astronomy Centre (ESAC), Camino bajo del Castillo, s/n, Urbanizaci\'{o}n Villafranca del Castillo, Villanueva de la Ca\~{n}ada, 28692 Madrid, Spain\relax                                                                                                                                                                                                           \label{inst:0009}
\and Telespazio UK S.L. for European Space Agency (ESA), Camino bajo del Castillo, s/n, Urbanizaci\'{o}n Villafranca del Castillo, Villanueva de la Ca\~{n}ada, 28692 Madrid, Spain\relax                                                                                                                                                                                                                            \label{inst:0013}
\and Univ. Grenoble Alpes, CNRS, IPAG, 38000 Grenoble, France\relax                                                                                                                                                                                                                                                                                                                                                  \label{inst:0014}
\and Astronomisches Rechen-Institut, Zentrum f\"{ u}r Astronomie der Universit\"{ a}t Heidelberg, M\"{ o}nchhofstr. 12-14, 69120 Heidelberg, Germany\relax                                                                                                                                                                                                                                                           \label{inst:0015}
\and HE Space Operations BV for European Space Agency (ESA), Camino bajo del Castillo, s/n, Urbanizaci\'{o}n Villafranca del Castillo, Villanueva de la Ca\~{n}ada, 28692 Madrid, Spain\relax                                                                                                                                                                                                                        \label{inst:0016}
\and Lund Observatory, Division of Astrophysics, Department of Physics, Lund University, Box 43, 22100 Lund, Sweden\relax                                                                                                                                                                                                                                                                                            \label{inst:0018}
\and Aurora Technology for European Space Agency (ESA), Camino bajo del Castillo, s/n, Urbanizaci\'{o}n Villafranca del Castillo, Villanueva de la Ca\~{n}ada, 28692 Madrid, Spain\relax                                                                                                                                                                                                                             \label{inst:0019}
\and Ru{\dj}er Bo\v{s}kovi\'{c} Institute, Bijeni\v{c}ka cesta 54, 10000 Zagreb, Croatia\relax                                                                                                                                                                                                                                                                                                                       \label{inst:0020}
\and INAF - Osservatorio astronomico di Padova, Vicolo Osservatorio 5, 35122 Padova, Italy\relax                                                                                                                                                                                                                                                                                                                     \label{inst:0024}
\and European Space Agency (ESA), European Space Research and Technology Centre (ESTEC), Keplerlaan 1, 2201AZ, Noordwijk, The Netherlands\relax                                                                                                                                                                                                                                                                      \label{inst:0025}
\and GEPI, Observatoire de Paris, Universit\'{e} PSL, CNRS, 5 Place Jules Janssen, 92190 Meudon, France\relax                                                                                                                                                                                                                                                                                                        \label{inst:0027}
\and CNES Centre Spatial de Toulouse, 18 avenue Edouard Belin, 31401 Toulouse Cedex 9, France\relax                                                                                                                                                                                                                                                                                                                  \label{inst:0028}
\and Universit\'{e} C\^{o}te d'Azur, Observatoire de la C\^{o}te d'Azur, CNRS, Laboratoire Lagrange, Bd de l'Observatoire, CS 34229, 06304 Nice Cedex 4, France\relax                                                                                                                                                                                                                                                \label{inst:0029}
\and Laboratoire d'astrophysique de Bordeaux, Univ. Bordeaux, CNRS, B18N, all{\'e}e Geoffroy Saint-Hilaire, 33615 Pessac, France\relax                                                                                                                                                                                                                                                                               \label{inst:0030}
\and Department of Astronomy, University of Geneva, Chemin Pegasi 51, 1290 Versoix, Switzerland\relax                                                                                                                                                                                                                                                                                                                \label{inst:0031}
\and Departament de F\'{i}sica Qu\`{a}ntica i Astrof\'{i}sica (FQA), Universitat de Barcelona (UB), c. Mart\'{i} i Franqu\`{e}s 1, 08028 Barcelona, Spain\relax                                                                                                                                                                                                                                                      \label{inst:0035}
\and Lohrmann Observatory, Technische Universit\"{ a}t Dresden, Mommsenstra{\ss}e 13, 01062 Dresden, Germany\relax                                                                                                                                                                                                                                                                                                   \label{inst:0037}
\and INAF - Osservatorio Astrofisico di Arcetri, Largo Enrico Fermi 5, 50125 Firenze, Italy\relax                                                                                                                                                                                                                                                                                                                    \label{inst:0043}
\and Nicolaus Copernicus Astronomical Center, Polish Academy of Sciences, ul. Bartycka 18, 00-716 Warsaw, Poland\relax                                                                                                                                                                                                                                                                                               \label{inst:0045}
\and Max Planck Institute for Astronomy, K\"{ o}nigstuhl 17, 69117 Heidelberg, Germany\relax                                                                                                                                                                                                                                                                                                                         \label{inst:0048}
\and Mullard Space Science Laboratory, University College London, Holmbury St Mary, Dorking, Surrey RH5 6NT, United Kingdom\relax                                                                                                                                                                                                                                                                                    \label{inst:0050}
\and INAF - Osservatorio Astrofisico di Torino, via Osservatorio 20, 10025 Pino Torinese (TO), Italy\relax                                                                                                                                                                                                                                                                                                           \label{inst:0051}
\and Department of Astronomy, University of Geneva, Chemin d'Ecogia 16, 1290 Versoix, Switzerland\relax                                                                                                                                                                                                                                                                                                              \label{inst:0056}
\and Royal Observatory of Belgium, Ringlaan 3, 1180 Brussels, Belgium\relax                                                                                                                                                                                                                                                                                                                                          \label{inst:0058}
\and ALTEC S.p.a, Corso Marche, 79,10146 Torino, Italy\relax                                                                                                                                                                                                                                                                                                                                                         \label{inst:0069}
\and Sednai S\`{a}rl, Geneva, Switzerland\relax                                                                                                                                                                                                                                                                                                                                                                      \label{inst:0071}
\and Gaia DPAC Project Office, ESAC, Camino bajo del Castillo, s/n, Urbanizaci\'{o}n Villafranca del Castillo, Villanueva de la Ca\~{n}ada, 28692 Madrid, Spain\relax                                                                                                                                                                                                                                                \label{inst:0080}
\and SYRTE, Observatoire de Paris, Universit\'{e} PSL, CNRS, Sorbonne Universit\'{e}, LNE, 61 avenue de l'Observatoire 75014 Paris, France\relax                                                                                                                                                                                                                                                                     \label{inst:0087}
\and IMCCE, Observatoire de Paris, Universit\'{e} PSL, CNRS, Sorbonne Universit{\'e}, Univ. Lille, 77 av. Denfert-Rochereau, 75014 Paris, France\relax                                                                                                                                                                                                                                                               \label{inst:0089}
\and Serco Gesti\'{o}n de Negocios for European Space Agency (ESA), Camino bajo del Castillo, s/n, Urbanizaci\'{o}n Villafranca del Castillo, Villanueva de la Ca\~{n}ada, 28692 Madrid, Spain\relax                                                                                                                                                                                                                 \label{inst:0095}
\and INAF - Osservatorio di Astrofisica e Scienza dello Spazio di Bologna, via Piero Gobetti 93/3, 40129 Bologna, Italy\relax                                                                                                                                                                                                                                                                                        \label{inst:0096}
\and Institut d'Astrophysique et de G\'{e}ophysique, Universit\'{e} de Li\`{e}ge, 19c, All\'{e}e du 6 Ao\^{u}t, B-4000 Li\`{e}ge, Belgium\relax                                                                                                                                                                                                                                                                      \label{inst:0097}
\and CRAAG - Centre de Recherche en Astronomie, Astrophysique et G\'{e}ophysique, Route de l'Observatoire Bp 63 Bouzareah 16340 Algiers, Algeria\relax                                                                                                                                                                                                                                                               \label{inst:0098}
\and RHEA for European Space Agency (ESA), Camino bajo del Castillo, s/n, Urbanizaci\'{o}n Villafranca del Castillo, Villanueva de la Ca\~{n}ada, 28692 Madrid, Spain\relax                                                                                                                                                                                                                                          \label{inst:0102}
\and CIGUS CITIC - Department of Computer Science and Information Technologies, University of A Coru\~{n}a, Campus de Elvi\~{n}a s/n, A Coru\~{n}a, 15071, Spain\relax                                                                                                                                                                                                                                               \label{inst:0103}
\and ATG Europe for European Space Agency (ESA), Camino bajo del Castillo, s/n, Urbanizaci\'{o}n Villafranca del Castillo, Villanueva de la Ca\~{n}ada, 28692 Madrid, Spain\relax                                                                                                                                                                                                                                    \label{inst:0105}
\and Kavli Institute for Cosmology Cambridge, Institute of Astronomy, Madingley Road, Cambridge, CB3 0HA\relax                                                                                                                                                                                                                                                                                                       \label{inst:0110}
\and Department of Astrophysics, Astronomy and Mechanics, National and Kapodistrian University of Athens, Panepistimiopolis, Zografos, 15783 Athens, Greece\relax                                                                                                                                                                                                                                                    \label{inst:0111}
\and Donald Bren School of Information and Computer Sciences, University of California, Irvine, CA 92697, USA\relax                                                                                                                                                                                                                                                                                                  \label{inst:0117}
\and CENTRA, Faculdade de Ci\^{e}ncias, Universidade de Lisboa, Edif. C8, Campo Grande, 1749-016 Lisboa, Portugal\relax                                                                                                                                                                                                                                                                                              \label{inst:0118}
\and INAF - Osservatorio Astrofisico di Catania, via S. Sofia 78, 95123 Catania, Italy\relax                                                                                                                                                                                                                                                                                                                         \label{inst:0119}
\and Dipartimento di Fisica e Astronomia ""Ettore Majorana"", Universit\`{a} di Catania, Via S. Sofia 64, 95123 Catania, Italy\relax                                                                                                                                                                                                                                                                                 \label{inst:0120}
\and Universit\'{e} de Strasbourg, CNRS, Observatoire astronomique de Strasbourg, UMR 7550,  11 rue de l'Universit\'{e}, 67000 Strasbourg, France\relax                                                                                                                                                                                                                                                              \label{inst:0123}
\and INAF - Osservatorio Astronomico di Roma, Via Frascati 33, 00078 Monte Porzio Catone (Roma), Italy\relax                                                                                                                                                                                                                                                                                                         \label{inst:0124}
\and Space Science Data Center - ASI, Via del Politecnico SNC, 00133 Roma, Italy\relax                                                                                                                                                                                                                                                                                                                               \label{inst:0125}
\and Department of Physics, University of Helsinki, P.O. Box 64, 00014 Helsinki, Finland\relax                                                                                                                                                                                                                                                                                                                       \label{inst:0127}
\and Finnish Geospatial Research Institute FGI, Vuorimiehentie 5, 02150 Espoo, Finland\relax                                                                                                                                                                                                                                                                                                                         \label{inst:0128}
\and Institut UTINAM CNRS UMR6213, Universit\'{e} de Franche-Comt\'{e}, OSU THETA Franche-Comt\'{e} Bourgogne, Observatoire de Besan\c{c}on, BP1615, 25010 Besan\c{c}on Cedex, France\relax                                                                                                                                                                                                                          \label{inst:0138}
\and HE Space Operations BV for European Space Agency (ESA), Keplerlaan 1, 2201AZ, Noordwijk, The Netherlands\relax                                                                                                                                                                                                                                                                                                  \label{inst:0139}
\and Dpto. de Inteligencia Artificial, UNED, c/ Juan del Rosal 16, 28040 Madrid, Spain\relax                                                                                                                                                                                                                                                                                                                         \label{inst:0140}
\and Institut d'Astronomie et d'Astrophysique, Universit\'{e} Libre de Bruxelles CP 226, Boulevard du Triomphe, 1050 Brussels, Belgium\relax                                                                                                                                                                                                                                                                         \label{inst:0142}
\and Konkoly Observatory, Research Centre for Astronomy and Earth Sciences, E\"{ o}tv\"{ o}s Lor{\'a}nd Research Network (ELKH), MTA Centre of Excellence, Konkoly Thege Mikl\'{o}s \'{u}t 15-17, 1121 Budapest, Hungary\relax                                                                                                                                                                                       \label{inst:0148}
\and ELTE E\"{ o}tv\"{ o}s Lor\'{a}nd University, Institute of Physics, 1117, P\'{a}zm\'{a}ny P\'{e}ter s\'{e}t\'{a}ny 1A, Budapest, Hungary\relax                                                                                                                                                                                                                                                                   \label{inst:0149}
\and Instituut voor Sterrenkunde, KU Leuven, Celestijnenlaan 200D, 3001 Leuven, Belgium\relax                                                                                                                                                                                                                                                                                                                        \label{inst:0151}
\and Department of Astrophysics/IMAPP, Radboud University, P.O.Box 9010, 6500 GL Nijmegen, The Netherlands\relax                                                                                                                                                                                                                                                                                                     \label{inst:0152}
\and University of Vienna, Department of Astrophysics, T\"{ u}rkenschanzstra{\ss}e 17, A1180 Vienna, Austria\relax                                                                                                                                                                                                                                                                                                   \label{inst:0157}
\and Institute of Physics, Ecole Polytechnique F\'ed\'erale de Lausanne (EPFL), Observatoire de Sauverny, 1290 Versoix, Switzerland\relax                                                                                                                                                                                                                                                                            \label{inst:0161}
\and Quasar Science Resources for European Space Agency (ESA), Camino bajo del Castillo, s/n, Urbanizaci\'{o}n Villafranca del Castillo, Villanueva de la Ca\~{n}ada, 28692 Madrid, Spain\relax                                                                                                                                                                                                                      \label{inst:0165}
\and LASIGE, Faculdade de Ci\^{e}ncias, Universidade de Lisboa, Edif. C6, Campo Grande, 1749-016 Lisboa, Portugal\relax                                                                                                                                                                                                                                                                                              \label{inst:0172}
\and School of Physics and Astronomy , University of Leicester, University Road, Leicester LE1 7RH, United Kingdom\relax                                                                                                                                                                                                                                                                                             \label{inst:0173}
\and School of Physics and Astronomy, Tel Aviv University, Tel Aviv 6997801, Israel\relax                                                                                                                                                                                                                                                                                                                            \label{inst:0177}
\and Cavendish Laboratory, JJ Thomson Avenue, Cambridge CB3 0HE, United Kingdom\relax                                                                                                                                                                                                                                                                                                                                \label{inst:0178}
\and Telespazio for CNES Centre Spatial de Toulouse, 18 avenue Edouard Belin, 31401 Toulouse Cedex 9, France\relax                                                                                                                                                                                                                                                                                                   \label{inst:0180}
\and National Observatory of Athens, I. Metaxa and Vas. Pavlou, Palaia Penteli, 15236 Athens, Greece\relax                                                                                                                                                                                                                                                                                                           \label{inst:0183}
\and University of Turin, Department of Physics, Via Pietro Giuria 1, 10125 Torino, Italy\relax                                                                                                                                                                                                                                                                                                                      \label{inst:0186}
\and Depto. Estad\'istica e Investigaci\'on Operativa. Universidad de C\'adiz, Avda. Rep\'ublica Saharaui s/n, 11510 Puerto Real, C\'adiz, Spain\relax                                                                                                                                                                                                                                                               \label{inst:0187}
\and EURIX S.r.l., Corso Vittorio Emanuele II 61, 10128, Torino, Italy\relax                                                                                                                                                                                                                                                                                                                                         \label{inst:0193}
\and Porter School of the Environment and Earth Sciences, Tel Aviv University, Tel Aviv 6997801, Israel\relax                                                                                                                                                                                                                                                                                                        \label{inst:0194}
\and ATOS for CNES Centre Spatial de Toulouse, 18 avenue Edouard Belin, 31401 Toulouse Cedex 9, France\relax                                                                                                                                                                                                                                                                                                         \label{inst:0195}
\and LFCA/DAS,Universidad de Chile,CNRS,Casilla 36-D, Santiago, Chile\relax                                                                                                                                                                                                                                                                                                                                          \label{inst:0199}
\and SISSA - Scuola Internazionale Superiore di Studi Avanzati, via Bonomea 265, 34136 Trieste, Italy\relax                                                                                                                                                                                                                                                                                                          \label{inst:0204}
\and University of Turin, Department of Computer Sciences, Corso Svizzera 185, 10149 Torino, Italy\relax                                                                                                                                                                                                                                                                                                             \label{inst:0212}
\and Thales Services for CNES Centre Spatial de Toulouse, 18 avenue Edouard Belin, 31401 Toulouse Cedex 9, France\relax                                                                                                                                                                                                                                                                                              \label{inst:0213}
\and Dpto. de Matem\'{a}tica Aplicada y Ciencias de la Computaci\'{o}n, Univ. de Cantabria, ETS Ingenieros de Caminos, Canales y Puertos, Avda. de los Castros s/n, 39005 Santander, Spain\relax                                                                                                                                                                                                                     \label{inst:0215}
\and Institut de F\'{i}sica d'Altes Energies (IFAE), The Barcelona Institute of Science and Technology, Campus UAB, 08193 Bellaterra (Barcelona), Spain\relax                                                                                                                                                                                                                                                        \label{inst:0221}
\and Port d'Informaci\'{o} Cient\'{i}fica (PIC), Campus UAB, C. Albareda s/n, 08193 Bellaterra (Barcelona), Spain\relax                                                                                                                                                                                                                                                                                              \label{inst:0222}
\and Instituto de Astrof\'{i}sica, Universidad Andres Bello, Fernandez Concha 700, Las Condes, Santiago RM, Chile\relax                                                                                                                                                                                                                                                                                              \label{inst:0230}
\and Centre for Astrophysics Research, University of Hertfordshire, College Lane, AL10 9AB, Hatfield, United Kingdom\relax                                                                                                                                                                                                                                                                                           \label{inst:0237}
\and University of Turin, Mathematical Department ""G.Peano"", Via Carlo Alberto 10, 10123 Torino, Italy\relax                                                                                                                                                                                                                                                                                                       \label{inst:0241}
\and INAF - Osservatorio Astronomico d'Abruzzo, Via Mentore Maggini, 64100 Teramo, Italy\relax                                                                                                                                                                                                                                                                                                                       \label{inst:0245}
\and Instituto de Astronomia, Geof\`{i}sica e Ci\^{e}ncias Atmosf\'{e}ricas, Universidade de S\~{a}o Paulo, Rua do Mat\~{a}o, 1226, Cidade Universitaria, 05508-900 S\~{a}o Paulo, SP, Brazil\relax                                                                                                                                                                                                                  \label{inst:0247}
\and M\'{e}socentre de calcul de Franche-Comt\'{e}, Universit\'{e} de Franche-Comt\'{e}, 16 route de Gray, 25030 Besan\c{c}on Cedex, France\relax                                                                                                                                                                                                                                                                    \label{inst:0253}
\and Astrophysics Research Centre, School of Mathematics and Physics, Queen's University Belfast, Belfast BT7 1NN, UK\relax                                                                                                                                                                                                                                                                                          \label{inst:0267}
\and Data Science and Big Data Lab, Pablo de Olavide University, 41013, Seville, Spain\relax                                                                                                                                                                                                                                                                                                                         \label{inst:0280}
\and Institute of Astrophysics, FORTH, Crete, Greece\relax                                                                                                                                                                                                                                                                                                                                                           \label{inst:0286}
\and Barcelona Supercomputing Center (BSC), Pla\c{c}a Eusebi G\"{ u}ell 1-3, 08034-Barcelona, Spain\relax                                                                                                                                                                                                                                                                                                            \label{inst:0287}
\and ETSE Telecomunicaci\'{o}n, Universidade de Vigo, Campus Lagoas-Marcosende, 36310 Vigo, Galicia, Spain\relax                                                                                                                                                                                                                                                                                                     \label{inst:0291}
\and F.R.S.-FNRS, Rue d'Egmont 5, 1000 Brussels, Belgium\relax                                                                                                                                                                                                                                                                                                                                                       \label{inst:0294}
\and Asteroid Engineering Laboratory, Lule\aa{} University of Technology, Box 848, S-981 28 Kiruna, Sweden\relax                                                                                                                                                                                                                                                                                                     \label{inst:0296}
\and Kapteyn Astronomical Institute, University of Groningen, Landleven 12, 9747 AD Groningen, The Netherlands\relax                                                                                                                                                                                                                                                                                                 \label{inst:0303}
\and IAC - Instituto de Astrofisica de Canarias, Via L\'{a}ctea s/n, 38200 La Laguna S.C., Tenerife, Spain\relax                                                                                                                                                                                                                                                                                                     \label{inst:0305}
\and Department of Astrophysics, University of La Laguna, Via L\'{a}ctea s/n, 38200 La Laguna S.C., Tenerife, Spain\relax                                                                                                                                                                                                                                                                                            \label{inst:0306}
\and Astronomical Observatory, University of Warsaw,  Al. Ujazdowskie 4, 00-478 Warszawa, Poland\relax                                                                                                                                                                                                                                                                                                               \label{inst:0311}
\and Research School of Astronomy \& Astrophysics, Australian National University, Cotter Road, Weston, ACT 2611, Australia\relax                                                                                                                                                                                                                                                                                     \label{inst:0312}
\and European Space Agency (ESA, retired), European Space Research and Technology Centre (ESTEC), Keplerlaan 1, 2201AZ, Noordwijk, The Netherlands\relax                                                                                                                                                                                                                                                             \label{inst:0313}
\and LESIA, Observatoire de Paris, Universit\'{e} PSL, CNRS, Sorbonne Universit\'{e}, Universit\'{e} de Paris, 5 Place Jules Janssen, 92190 Meudon, France\relax                                                                                                                                                                                                                                                     \label{inst:0328}
\and Universit\'{e} Rennes, CNRS, IPR (Institut de Physique de Rennes) - UMR 6251, 35000 Rennes, France\relax                                                                                                                                                                                                                                                                                                        \label{inst:0329}
\and INAF - Osservatorio Astronomico di Capodimonte, Via Moiariello 16, 80131, Napoli, Italy\relax                                                                                                                                                                                                                                                                                                                   \label{inst:0331}
\and Shanghai Astronomical Observatory, Chinese Academy of Sciences, 80 Nandan Road, Shanghai 200030, People's Republic of China\relax                                                                                                                                                                                                                                                                               \label{inst:0334}
\and University of Chinese Academy of Sciences, No.19(A) Yuquan Road, Shijingshan District, Beijing 100049, People's Republic of China\relax                                                                                                                                                                                                                                                                         \label{inst:0336}
\and S\~{a}o Paulo State University, Grupo de Din\^{a}mica Orbital e Planetologia, CEP 12516-410, Guaratinguet\'{a}, SP, Brazil\relax                                                                                                                                                                                                                                                                                \label{inst:0338}
\and Niels Bohr Institute, University of Copenhagen, Juliane Maries Vej 30, 2100 Copenhagen {\O}, Denmark\relax                                                                                                                                                                                                                                                                                                      \label{inst:0341}
\and DXC Technology, Retortvej 8, 2500 Valby, Denmark\relax                                                                                                                                                                                                                                                                                                                                                          \label{inst:0342}
\and Las Cumbres Observatory, 6740 Cortona Drive Suite 102, Goleta, CA 93117, USA\relax                                                                                                                                                                                                                                                                                                                              \label{inst:0343}
\and CIGUS CITIC, Department of Nautical Sciences and Marine Engineering, University of A Coru\~{n}a, Paseo de Ronda 51, 15071, A Coru\~{n}a, Spain\relax                                                                                                                                                                                                                                                            \label{inst:0349}
\and Astrophysics Research Institute, Liverpool John Moores University, 146 Brownlow Hill, Liverpool L3 5RF, United Kingdom\relax                                                                                                                                                                                                                                                                                    \label{inst:0350}
\and IRAP, Universit\'{e} de Toulouse, CNRS, UPS, CNES, 9 Av. colonel Roche, BP 44346, 31028 Toulouse Cedex 4, France\relax                                                                                                                                                                                                                                                                                          \label{inst:0357}
\and MTA CSFK Lend\"{ u}let Near-Field Cosmology Research Group, Konkoly Observatory, MTA Research Centre for Astronomy and Earth Sciences, Konkoly Thege Mikl\'{o}s \'{u}t 15-17, 1121 Budapest, Hungary\relax                                                                                                                                                                                                      \label{inst:0377}
\and Pervasive Technologies s.l., c. Saragossa 118, 08006 Barcelona, Spain\relax                                                                                                                                                                                                                                                                                                                                     \label{inst:0385}
\and School of Physics and Astronomy, University of Leicester, University Road, Leicester LE1 7RH, United Kingdom\relax                                                                                                                                                                                                                                                                                              \label{inst:0402}
\and Villanova University, Department of Astrophysics and Planetary Science, 800 E Lancaster Avenue, Villanova PA 19085, USA\relax                                                                                                                                                                                                                                                                                   \label{inst:0423}
\and Departmento de F\'{i}sica de la Tierra y Astrof\'{i}sica, Universidad Complutense de Madrid, 28040 Madrid, Spain\relax                                                                                                                                                                                                                                                                                          \label{inst:0426}
\and INAF - Osservatorio Astronomico di Brera, via E. Bianchi, 46, 23807 Merate (LC), Italy\relax                                                                                                                                                                                                                                                                                                                    \label{inst:0428}
\and National Astronomical Observatory of Japan, 2-21-1 Osawa, Mitaka, Tokyo 181-8588, Japan\relax                                                                                                                                                                                                                                                                                                                   \label{inst:0430}
\and Department of Particle Physics and Astrophysics, Weizmann Institute of Science, Rehovot 7610001, Israel\relax                                                                                                                                                                                                                                                                                                   \label{inst:0448}
\and Centre de Donn\'{e}es Astronomique de Strasbourg, Strasbourg, France\relax                                                                                                                                                                                                                                                                                                                                      \label{inst:0469}
\and University of Exeter, School of Physics and Astronomy, Stocker Road, Exeter, EX2 7SJ, United Kingdom\relax                                                                                                                                                                                                                                                                                                      \label{inst:0475}
\and Departamento de Astrof\'{i}sica, Centro de Astrobiolog\'{i}a (CSIC-INTA), ESA-ESAC. Camino Bajo del Castillo s/n. 28692 Villanueva de la Ca\~{n}ada, Madrid, Spain\relax                                                                                                                                                                                                                                        \label{inst:0476}
\and naXys, Department of Mathematics, University of Namur, Rue de Bruxelles 61, 5000 Namur, Belgium\relax                                                                                                                                                                                                                                                                                                           \label{inst:0479}
\and INAF. Osservatorio Astronomico di Trieste, via G.B. Tiepolo 11, 34131, Trieste, Italy\relax                                                                                                                                                                                                                                                                                                                     \label{inst:0483}
\and Harvard-Smithsonian Center for Astrophysics, 60 Garden St., MS 15, Cambridge, MA 02138, USA\relax                                                                                                                                                                                                                                                                                                               \label{inst:0484}
\and H H Wills Physics Laboratory, University of Bristol, Tyndall Avenue, Bristol BS8 1TL, United Kingdom\relax                                                                                                                                                                                                                                                                                                      \label{inst:0491}
\and Escuela de Arquitectura y Polit\'{e}cnica - Universidad Europea de Valencia, Spain\relax                                                                                                                                                                                                                                                                                                                        \label{inst:0495}
\and Escuela Superior de Ingenier\'{i}a y Tecnolog\'{i}a - Universidad Internacional de la Rioja, Spain\relax                                                                                                                                                                                                                                                                                                        \label{inst:0496}
\and Department of Physics and Astronomy G. Galilei, University of Padova, Vicolo dell'Osservatorio 3, 35122, Padova, Italy\relax                                                                                                                                                                                                                                                                                    \label{inst:0497}
\and Applied Physics Department, Universidade de Vigo, 36310 Vigo, Spain\relax                                                                                                                                                                                                                                                                                                                                       \label{inst:0500}
\and Instituto de F{'i}sica e Ciencias Aeroespaciais (IFCAE), Universidade de Vigo‚ \'{A} Campus de As Lagoas, 32004 Ourense, Spain\relax                                                                                                                                                                                                                                                                          \label{inst:0501}
\and Purple Mountain Observatory, Chinese Academy of Sciences, Nanjing 210023, China\relax                                                                                                                                                                                                                                                                                                                           \label{inst:0512}
\and Sorbonne Universit\'{e}, CNRS, UMR7095, Institut d'Astrophysique de Paris, 98bis bd. Arago, 75014 Paris, France\relax                                                                                                                                                                                                                                                                                           \label{inst:0513}
\and Faculty of Mathematics and Physics, University of Ljubljana, Jadranska ulica 19, 1000 Ljubljana, Slovenia\relax                                                                                                                                                                                                                                                                                                 \label{inst:0514}
\and Institute of Mathematics, Ecole Polytechnique F\'ed\'erale de Lausanne (EPFL), Switzerland\relax
\label{inst:0555}
}    
    
\date{Received ; accepted }

\abstract
    {{\it Gaia}'s readout window strategy is challenged by very dense fields in the sky. Therefore, in addition to standard {\it Gaia} observations, full Sky Mapper (SM) images were recorded for nine selected regions in the sky. A new software pipeline exploits these Service Interface Function (SIF) images of crowded fields (CFs),  making use of the availability of the full two-dimensional (2D) information. This new pipeline produced half a million additional {\it Gaia} sources in the region of the omega Centauri ($\omega$ Cen) cluster, which are published with this Focused Product Release. We discuss the dedicated SIF CF data reduction pipeline, validate its data products, and introduce their {\it Gaia} archive table.}
    {Our aim is to improve the completeness of the {\it Gaia} source inventory in a very dense region in the sky, $\omega$ Cen.}
    {An adapted version of {\it Gaia}'s Source Detection and Image Parameter Determination software located sources in the 2D SIF CF images. These source detections were clustered and assigned to new SIF CF or existing {\it Gaia} sources by {\it Gaia}'s cross-match software. For the new sources, astrometry was calculated using the Astrometric Global Iterative Solution software, and photometry was obtained in the {\it Gaia} DR3 reference system. We validated the results by comparing them to the public {\it Gaia} DR3 catalogue and external Hubble Space Telescope data.}
    {With this Focused Product Release, 526\,587 new sources have been added to the {\it Gaia} catalogue in $\omega$ Cen. Apart from positions and brightnesses, the additional catalogue contains parallaxes and proper motions, but no meaningful colour information. While SIF CF source parameters generally have a lower precision than nominal {\it Gaia} sources, in the cluster centre they increase the depth of the combined catalogue by three magnitudes and improve the source density by a factor of ten.}
    {This first SIF CF data publication already adds great value to the {\it Gaia} catalogue. It demonstrates what to expect for the fourth {\it Gaia} catalogue, which will contain additional sources for all nine SIF CF regions.}
\maketitle

\section{Introduction}

\label{sec:introduction}On June 13, 2022, the Data Processing and Analysis Consortium (DPAC) of the European Space Agency (ESA) cornerstone mission {\it Gaia} published Data Release 3 (DR3), containing astrometry and photometry for more than 1.8 billion sources \citep{2021A&A...650C...3G,2023A&A...674A...1G}. This third {\it Gaia} data release provides incredible insights into our home galaxy and beyond. 

The {\it Gaia} catalogue has been used for several studies in the omega Centauri ($\omega$ Cen) cluster. Examples are the 
the measurement of the cluster parallax and calibration of the luminosity of the tip of the red giant branch \citep{2021ApJ...908L...5S}; 
investigations on stellar kinematics to describe the distribution of both the dark and luminous mass components in the cluster \citep{2022MNRAS.511.4251E}; 
analysis of the kinematics of stellar sub-populations in the cluster \citep{2020ApJ...898..147C,2020A&A...637A..46S} and
identification of a stellar stream originating in $\omega$ Cen \citep{2019NatAs...3..667I,2019ApJ...872..152I}; 
as well as combinations of {\it Gaia} data with other survey catalogues \citep{2021MNRAS.505.3549S,2022ApJ...939L..20P}, just to mention a few. 

Generally, the completeness of the {\it Gaia} catalogue is very high. A small number of densely populated regions, one of which is $\omega$ Cen, are an exception though. {\it Gaia}'s astrometric crowding limit is around 1\,050\,000 objects deg$^{-2}$ \citep{2016A&A...595A...1G}, which is about one object every 12 square arcseconds. For denser areas, the readout window strategy reaches its limitations. Therefore, for the most crowded regions in the sky, clear holes are visible in the catalogue data. With this Focused Product Release (FPR), the hole in the core of the $\omega$ Cen cluster is now filled up with half a million additional sources. These were created with a new software pipeline, written and operated by {\it Gaia} DPAC, exploiting so far unused two-dimensional (2D) images of {\it Gaia}'s Sky Mapper (SM) instrument. These so-called Service Interface Function (SIF) images were recorded for selected crowded fields (CFs) in the sky; for more details, readers can refer to Section 6.6 in \citet{2016A&A...595A...1G} and Section 1.1.3 in \citet{2022gdr3.reptE...1D}.

In standard operation, referred to as 'nominal', {\it Gaia} uses dedicated readout windows obtained with the Astrometric Field (AF) instrument, the Blue and Red Photometer (BP/RP), and the Radial Velocity Spectrograph (RVS). Compared to AF readout windows, which are collapsed in the across-scan (AC) direction for sources fainter than 13 mag, the new pipeline profits immensely from the availability of the full 2D information, even though SIF images are acquired in samples binned from $2\times2$ pixels and thus of a lower quality in the along-scan direction compared to the nominal {\it Gaia} data. Not being limited by a readout window, SIF CF processing uses the full 2D information about neighbouring sources to obtain accurate astrometry and photometry. The aim of this paper is to describe the implementation of the SIF CF pipeline and validate its performance. It starts with information on the SIF CF images, the main input of the pipeline, in Sect. \ref{sec:images}, followed by a description of the SIF CF data reduction in Sect. \ref{sec:pipeline}.

Most of the SIF CF pipeline is based on the nominal {\it Gaia} software systems that have been used to produce the {\it Gaia} catalogues released so far. For SIF CF processing, these software systems were adapted to process extended 2D images: The SIF CF image parameter determination (SIF CF IPD) registers all source detections in an image (Sect. \ref{sec:ipd}). The cross-match between these detections was done by SIF CF XM (Sect. \ref{sec:xm}) and a coverage map was calculated from the image boundaries to identify reliable sources (Sect. \ref{sec:coverage}). For these, {\it Gaia's} Astrometric Global Iterative Solution (AGIS) produced valid astrometry (Sect. \ref{sec:agis}), and robust photometry was obtained with a dedicated ad hoc procedure using the {\it Gaia} DR3 photometry as a reference (Sect. \ref{sec:phot}). In Sect. \ref{sec:filters}, we list all filters that were applied to the SIF CF data. While a full description of the nominal {\it Gaia} software systems can be found in the corresponding, linked papers, here we concentrate on the adaptations implemented for the SIF CF processing.

A validation of the SIF CF data products is provided in Sect. \ref{sec:validation}. 
It was done by comparing SIF CF data to nominal {\it Gaia} DR3 data and to a dedicated Hubble Space Telescope (HST) dataset that covers the core of $\omega$ Cen. After basic information regarding the number of detections and sources created by the SIF CF pipeline (Sect. \ref{sec:numbers}), photometry was validated  comparing SIF CF statistics to nominal {\it Gaia} (Sect. \ref{sec:photometry}). Astrometry was then examined in terms of positional (Sect. \ref{sec:position}), proper motion (Sect. \ref{sec:propermotion}), and parallax (Sect. \ref{sec:parallax}) uncertainties. In the following section, we describe the HST dataset and its usage (Sect. \ref{sec:bellini}), address the small-scale completeness of SIF CF data compared to {\it Gaia} DR3 and HST data (Sect. \ref{sec:smallscale}), the source density and depth (Sect. \ref{sec:depth}), and the reliability and completeness of the SIF CF sample (Sect. \ref{sec:completeness}). 
Known issues are discussed in Sect. \ref{sec:issues}.

In Sect. \ref{sec:output} the SIF CF data product, available in the {\it Gaia} archive table\linktotablefpr{crowded\_field\_source}, is described. Finally, a summary is provided in Sect. \ref{sec:conclusion}.

\section{SIF CF images}

\label{sec:images}During commissioning of the {\it Gaia} satellite in 2014, data gaps were visible
in the very dense centres of dense regions in the sky. While in the outskirts of $\omega$ Cen the star detections are usually well separated, multiple sources overlap in the core of the cluster, leading to conflicting readout windows, windows containing multiple sources or no readout at all for the nominal processing, see Section 3.3.8 in \citet{2016A&A...595A...1G}.

A short feasibility study was conducted and showed that the analysis of two-dimensional SM images could enhance the completeness of the {\it Gaia} catalogue in these regions by roughly two magnitudes. Indeed, this turned out to be a conservative estimate. Originally, full SM images were scheduled for transmission to ground for calibration purposes only, but the improvement was such that the DPAC Executive Board together with the {\it Gaia} Science Team decided to record SIF images in addition to the nominal readout windows for a few selected regions on the sky.

These regions include the surroundings of 230 very bright stars (VBS) and nine crowded fields (CFs). 

\begin{table}[tbh]
    \caption{Regions selected for SIF CF observation}
    \label{tab:regions}
    \centering
    \begin{tabular}{l l l}
    \hline\hline 
    \noalign{\smallskip}
    {\textbf{Region}}               & {\textbf{Acq. Start}} & \textbf{Area}\\
                                    &                       & [deg$^2$]\\    
    \hline 
    \noalign{\smallskip}
      Omega Centauri (NGC\,5139)    &  2015-01-01 & 1.83\\
      Baade’s Window (Bulge)        &  2015-01-01 & 1.0\\
      Large Magellanic Cloud        &  2016-07-01 & 2.22\\
      Small Magellanic Cloud        &  2016-07-01 & 1.6\\
      Sagittarius I (Bulge)         &  2016-07-01 & 1.1\\
      47\,Tuc (NGC\,104)            &  2016-07-01 & 1.72\\
      M\,22 (NGC\,6656)             &  2016-07-01 & 0.75\\
      NGC\,4372                     &  2016-07-01 & 1.13\\
      M\,4 (NGC\,6121)              &  2016-07-01 & 0.95\\
    \hline
    \end{tabular}
\end{table}

\Cref{tab:regions} lists the dense regions in the sky selected for {\it Gaia} SIF CF data acquisition together with the starting date of their observation, see Section 1.1.3 in \citet{2022gdr3.reptE...1D}. From that date onward during most scans of one of the satellite fields of view (FOVs) containing one of the listed regions, SIF CF images were recorded. 
For $\omega$ Cen and Baade's Window data record started 18 months earlier than for the other regions, because they were selected as test regions.
Additionally, Baade's window and Sgr I were both observed once on 15 October 2014, see again Section 1.1.3 in \citet{2022gdr3.reptE...1D}.

While acquisition of SIF CF images is still ongoing, the $\omega$ Cen sources presented here were observed in images obtained between 1 January 2015 and 8 January 2020.

\Cref{tab:regions} also provides the approximate sizes of the observed areas in square degrees. The shapes of these areas vary. The total data volume for all SIF images, including CF, VBS, and regular calibration images amounts to roughly 60\,GB\,/\,year.

SIF images are read as samples of $2\times2$ pixels, which decreases the possible precision of SIF CF observations compared to fully resolved AF observations. Regarding the size of a SIF  CF image, in the AC direction it is always the same: the full readout of the illuminated part of the CCD detector of 983 samples. In the along-scan (AL) direction the image size depends on the time span of the crossing of the satellite on the selected region. It can vary from a few seconds to more than one minute, leading to images between 860 and 32\,768 samples wide. 
This is possible, because {\it Gaia} works in Time Delayed Integration (TDI) mode, which means that charge is transferred to the adjacent pixel with the same speed with which the image of a source moves to the next pixel due to the satellite rotation until it reaches the readout register. The Sky Mapper is the first instrument to receive the light of a star entering {\it Gaia}'s FOV, and the SM is the only {\it Gaia} instrument for which CCDs are illuminated by a single FOV only.

The limitation of the SIF CF processing is the absence of colour and spectral information for the sources. All information is based on the SM images only. The advantages of 2D SIF images compared to nominal readout windows are obvious: SIF CF data suffer neither from being collapsed to one dimension before download nor from truncation of data, as can happen for overlapping windows, and each SIF CF image contains sources from one FOV only.

\section{SIF CF pipeline}

\label{sec:pipeline} In March 2019, a small team was assembled to analyse the SIF CF data, and between September 2021 and August 2022 the SIF CF pipeline had its first operational run, the results of which are published in this FPR. It used a total run time of 22 days distributed across the various software systems involved.

\paragraph{Initial investigation phase}  The first step towards a SIF CF data pipeline consisted of choosing a software system to be used for source detection in the future pipeline. Well known software packages like DAOPHOT \citep{1987PASP...99..191S} and SExtractor \citep{1996A&AS..117..393B}, as well as an adapted version of {\it Gaia}'s nominal 'Detection and Image Parameter Determination' (IPD) software \citep{2015hsa8.conf..792C}
 were tested on a set of selected SIF CF images. Five independent research groups from Bologna, Groningen, Padova, Edinburgh, and Potsdam participated in the tests.

In terms of source detection, all three software tools delivered comparable results. Roughly 80\,000 sources were found in a very crowded SIF CF area. As none of the methods outperformed the others, it was decided to use the adapted IPD code as the basis for the SIF CF pipeline. With comparable detection numbers, the reason for this choice was the most realistic description of the Point Spread Function (PSF) that {\it Gaia}'s IPD provides. A good description of the PSF is crucial for accurate astrometry and photometry, and for IPD the PSF is based on all SM data available and not limited to the SIF CF images themselves.
The SIF CF source detection and Image Parameter Determination software (SIF CF IPD) was then revised and improved regarding run time and performance for SIF CF images (see Sect. \ref{sec:ipd}).

\paragraph{Basic SIF pipeline}  Before the SIF CF data reduction can start, the original SIF data packets transmitted from the {\it Gaia} satellite need to be restructured. This is done with the 'BasicSifPipeline' provided by and run at the European Space Astronomy Centre (ESAC) in Villafranca. The BasicSifPipeline performs an initial data transformation and rewrites the SIF image data as DPAC datamodel, that can be read by the first SIF CF software system, SIF CF IPD.

\subsection{SIF CF IPD} 

\label{sec:ipd} The source-finding software, SIF CF IPD, is based on {\it Gaia}'s nominal Intermediate Data Update (IDU) Detection and Image Parameter Determination (IPD), see Section 3.3.6 in \citet{2022gdr3.reptE...3C} and \citet{2015hsa8.conf..792C}. In the following, we describe the modifications implemented in the SIF CF IPD code to profit from the two dimensional image information.

\paragraph{ Background}  After subtracting the bias and correcting for the dark signal, the background is estimated on a grid of $200\times60$ samples, as an average of all samples below a certain threshold calculated from Poisson noise and total detection noise. Between the grid points, the background is spline-interpolated. 

\paragraph{ Detection}  A first source detection run locates new sources via cross-correlation of a PSF kernel with the SIF image. For a full description of {\it Gaia}'s PSF model see \citet{2021A&A...649A..11R} and Section 3.3.5 in \citet{2022gdr3.reptE...3C}. Gaia's CCD pixels have rectangular shapes, covering an area of 59 mas in AL and 177 mas in AC. This leads to different angular resolutions in AL and AC direction.

Before each detection iteration, an artificial image containing the background and the PSF fits of already detected sources, the so-called scene, is subtracted from the original image. At the beginning of source detection, the scene contains only the background.  

\paragraph{ PSF fit}  Once a source is found, a simple initial estimate of its position and flux is determined by fitting the SM PSF model to the detection. As the SIF CF images fully cover a large sky area, they do not contain the usual {\it Gaia} readout windows. In order to re-use as much DPAC software and calibrations as possible though, we used artificial readout windows centred on the detection for the PSF fitting. The PSF fit of the source is then added to the scene. Sources are processed in order of apparent brightness, starting with the brightest source. This is done for all source detections one after the other, taking the already known sources into account via the updated scene.

\paragraph{ Adapted window size}  As the SM PSF is reliably defined only for an SM window of $9\times6$ samples (equivalent to $18\times12$ CCD pixels), these dimensions determine the maximum window size that can be used for the PSF fit. For sources with an estimated flux brighter than 18 mag, the full $9\times6$ sample SM PSF model is used. For fainter sources the PSF model is restricted more and more in the AL direction, until for detections of 18.8 mag or fainter the minimum PSF model of $6\times6$ samples is used. This is done, because fainter sources get progressively more affected by their neighbours and imperfect scene estimates, thus for them it is better to focus on the central part of the window, where the majority of the source flux is located. 

In order to place the detected source into the scene, its location and flux estimate are used together with the full window size PSF model for both bright and faint sources. The PSF spikes of the brighter stars extend far beyond the area for which a reliable PSF calibration is available. Hence, those spikes cannot be properly described with the available PSF model, and this leads to steps in the scene at the borders of the PSF model window. In the next source detection run, such a step could be interpreted as a new source, which is itself fitted with the PSF model then. In consequence, a set of spurious source detections are located along the spikes of the PSF. Those spurious detections are filtered out by our detection fraction filter with high efficiency (see Sect. \ref{sec:filters}), since in different scans the spikes are at different orientations relative to the scene.

\paragraph{ Iterations}  Unlike the nominal IPD, SIF CF IPD has the ability to review the source detections and PSF fits in the context of the source surroundings. Once all sources are located and fit for the first time, the PSF fits are repeated, starting again with the brightest source. In this second fit iteration, the information about the source surroundings is improved, since now even for the brightest source there is a full scene available, consisting of modelled sources and background. 

When the second PSF fit is completed for all sources, the source detection is run again. It now takes into account the knowledge about all already detected and fitted sources. This allows the pipeline to detect fainter sources and sources closer to bright ones. The second source detection run is followed by another two PSF fit iterations for all sources, old and newly detected ones.

Finally, a third block of one source detection and two PSF fit runs is performed. The combination of three source detection and six PSF fit runs turned out to be the best trade-off between performance improvement and run time increase. In every SIF CF image, between 100 to 700\,000 source detections are found; this extreme range is a result of the drastic differences in image size rather than simply the source density within the images.

When source detection and PSF fit runs are completed, some basic filters are applied to remove internal duplicates and detections located outside the CCD detectors. The full list of filters is given in Sect. \ref{sec:filters}.

\paragraph{}In short, SIF CF IPD iterates source finding and PSF fitting for all sources in the SIF CF image ordered by source brightness. In doing so, it takes into account an artificial image constructed from PSF models of already detected sources and background to create well defined new source detections.

\subsection{SIF CF XM} 

\label{sec:xm} The SIF CF Crossmatch (SIF CF XM) module assigns the SIF CF detections to already known and new sources. During this process, clusters of SIF CF detections belonging to the same physical light source on the sky are formed, and a unique source identifier is assigned to each such cluster of detections.

The SIF CF XM code is a slightly altered version of IDU XM, see \citet{2021A&A...649A..10T} and Section 3.4.13 in \citet{2022gdr3.reptE...3C}. The SIF CF version does not include the first XM stage (the detection classifier), which distinguishes real from spurious detections. As explained in the previous section, SIF CF detections are found on-ground by SIF CF IPD from SIF CF images and thus are not affected by limitations of the {\it Gaia} on-board detection software and processing resources. The on-ground detection performance significantly reduces the appearance of spurious detections and their patterns do not match those from nominal data. Subsequent XM stages use the same algorithms as the IDU XM but with different parameters and source catalogue as described below.

\paragraph{ Parameters}  The main difference between nominal IDU XM and SIF CF XM processing is the choice of input parameters regarding detection clustering. These parameter differences are caused by the fact that the default XM parameters are optimised for typical source densities of the {\it Gaia} catalogue, and SIF CF XM operates in a much denser environment for which a different set of parameters is needed. To find the best parameter set, SIF CF XM was run with various parameter settings, improving the clustering performance on the dense regions of SIF CF data. The results were compared to an HST reference catalogue (\citealt{Bellini2017}, see Sect. \ref{sec:bellini}). The optimal set of parameters increased the number of matches to the reference catalogue, and hence decreased the number of unmatched sources from both, SIF CF and HST. 

\paragraph{ Nominal source input}  Instead of providing just the SIF CF detections as input to the SIF CF XM software, we additionally fed a subset of nominal {\it Gaia} sources to it, that is, exactly those nominal sources that were already published in {\it Gaia} DR3 for the $\omega$ Cen region. SIF CF detections could then be assigned to either a newly created SIF CF source or to an existing nominal source, and SIF CF XM returned two easily distinguishable source sets:
\begin{list}{$\bullet$}{}  
    \item Known {\it Gaia} DR3 sources together with their matched SIF CF detections. As these sources are published in {\it Gaia} DR3 already, they are excluded from further processing in the SIF CF pipeline, and not published again in this FPR. They are used as input for the photometric overlap calibration though, and also for internal validation.
    \item Newly created SIF CF sources with their matched SIF CF detections. These are new sources with respect to {\it Gaia} DR3. Astrometry and photometry are computed for them, and these sources are published in this FPR.
\end{list}

Contrary to what is done in the nominal XM processing, the nominal input sources are not altered by SIF CF XM. Therefore the SIF CF XM configuration is changed to prevent merging or splitting these sources.

The reason to discard sources that have a nominal match in {\it Gaia} DR3 from the output list and not to publish them is that they are already present in DR3. These sources are created, used internally for the photometric overlap calibration, and then discarded. Doing so ensures the FPR data to be a truly disjunct extension to {\it Gaia} DR3. 

\paragraph{ No detection mixing}  All SIF CF sources are based on SIF CF detections only, and thus there is no mixing of SIF CF and nominal {\it Gaia} data in any of the sources. {\it Gaia} DR3 sources are based on nominal {\it Gaia} SM and AF observations only, and SIF CF sources are derived from SIF CF observations only.

\paragraph{ Source identifiers}  The SIF CF XM module assigns a unique source identifier, the sourceId, see Section 20.1.1 in \citet{2022gdr3.reptE..20H}, to all newly created SIF CF sources. This sourceId follows the same scheme as for nominal {\it Gaia} sources and codes a spatial index, the source provenance (code of the data processing centre or system where it was created), a running number, and a component number.\footnote{The SIF CF sourceIds use an as of yet unused provenance code (code = 6), which allows one to easily distinguish SIF CF sourceIds from nominal {\it Gaia} sourceIds using the following expression: \\ \texttt{(sourceId >> 32) \& 7 == 6}.}

\paragraph{}For a more detailed description of the different stages and algorithms used in XM, we refer to \cite{2021A&A...649A..10T}. The next step in the chain addresses the calculation of coverage maps for the SIF CF images.

\subsection{SIF CF Coverage} 

\begin{figure}[tbhp]
        \begin{center}
                \includegraphics[angle=0, width=\linewidth]{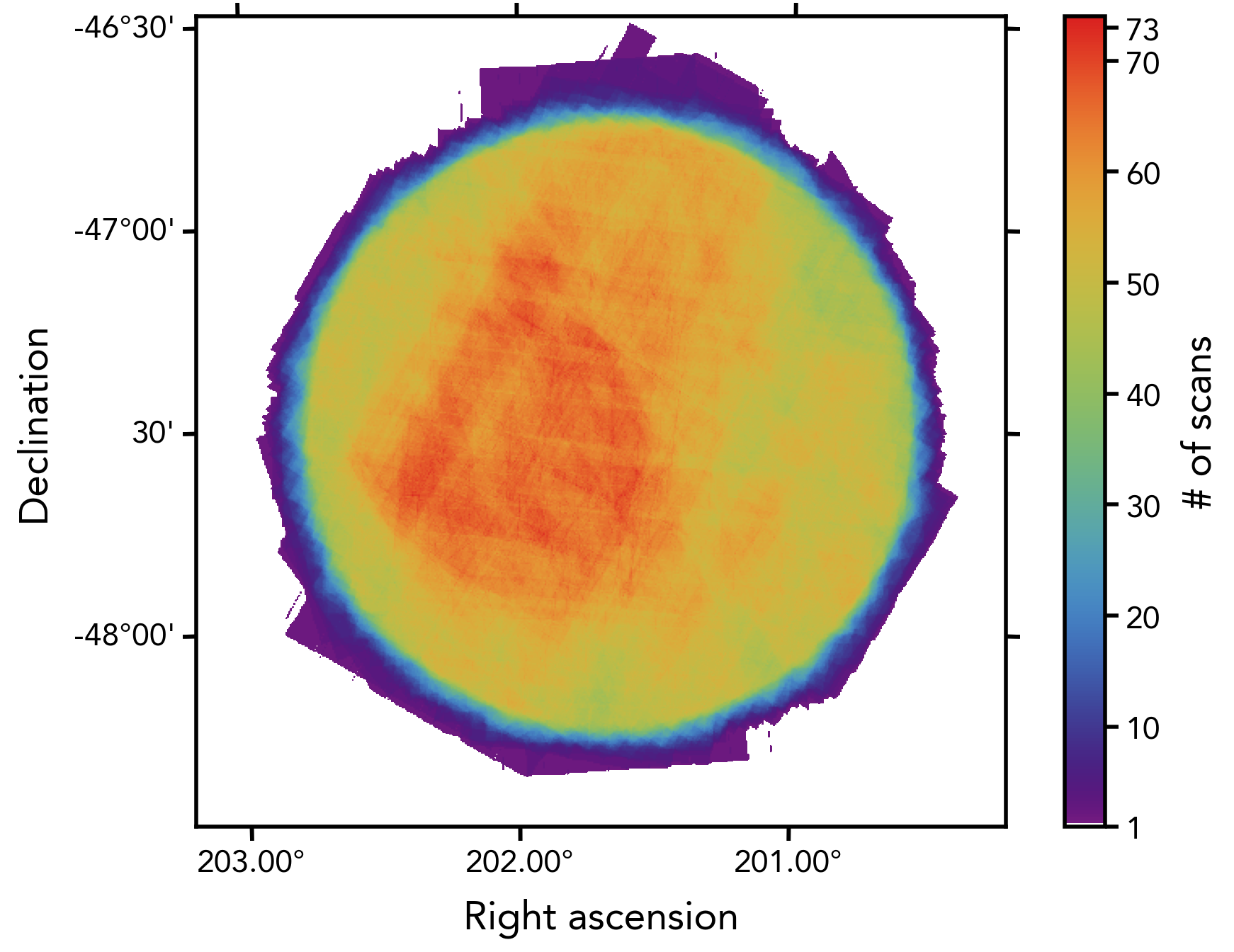}
                \caption{$\omega$ Cen area coloured by the number of SIF CF scans.}\label{fig:coverage}
        \end{center}
\end{figure}

\label{sec:coverage} The SIF CF Coverage task filters unreliable sources by calculating the number of possible detections for each source location\linktoparamfpr{crowded_field_source}{n\_scans}, and comparing it to the number of actual detections assigned to the source\linktoparamfpr{crowded_field_source}{matched\_transits}. In order to assess the number of possible detections, which is the number of SIF CF images covering the precise location within a SIF CF region, the image boundaries are transferred to sky coordinates. As SIF CF images can be very long, the image is described as a polygon rather than a rectangle. The combination of all image descriptions provides the number of scans for each source location (see Fig. \ref{fig:coverage}).

We define the image boundary as a polygon in the Cartesian space of right ascension and declination. This is not equivalent to a polygon in spherical coordinates.
Minor attitude inconsistencies, rounding errors etc. add to this effect. As a consequence, for some sources located at the image borders,\linktoparamfpr{crowded_field_source}{n\_scans} might be incorrect. In a test dataset, this was true for just one out of $\sim$\,200\,000 sources. Even if errors add up for several images, numbers are so small that it seems tolerable, and it was decided not to implement proper spherical polygons in this task.

\subsection{SIF CF AGIS} 

\label{sec:agis} The Astrometric Global Iterative Solution (AGIS) software provides precise astrometry for all nominal {\it Gaia} sources. A detailed description of AGIS can be found in \cite{lindegren2021} and Section 4.1.1 in \citet{2022gdr3.reptE...4H}. The nominal AGIS software ran unaltered for the SIF CF sources.

The operational SIF CF AGIS run for this FPR took place in March 2022 and created AGIS solutions for all SIF CF sources published in this FPR. For SIF CF, in the AGIS run only the astrometry of the sources is updated. This is known as the AGIS secondary solution. When running AGIS in secondary mode, it needs a calibrated initial attitude and a geometrical calibration model as input. These are not updated by the SIF CF AGIS pipeline but supplied from a previous nominal AGIS run in its primary mode. Not a single source failed to converge. 

The SIF CF AGIS run used a six parameter solution (position, proper motion, parallax, and pseudo colour) with a pseudo-colour prior\footnote{As for faint sources the colour is less crucial than for bright ones, and SIF CF sources are generally rather faint, the default nominal pseudo colour was used as colour prior to optimise software run time.} of $1.43\pm 0.1 \,\,{\mu}m^{-1}$. The quality of the pseudo colour measurements obtained without any colour information, thus purely from astrometry, is rather low. It is not advised to use these as a proxy for the colour.

Apart from the astrometry for the newly created SIF CF sources, SIF CF AGIS also produced AGIS solutions for those SIF CF detections assigned to an already known {\it Gaia} source. These SIF CF solutions of already published {\it Gaia} DR3 sources, were used for the photometric calibration of SIF CF sources (see Sect. \ref{sec:phot}) and for internal validation. They were discarded after use in order to avoid a duplication of existing {\it Gaia} DR3 sources in the FPR catalogue.

\subsection{SIF CF photometry} 

\label{sec:phot} In order to ensure consistency between the SIF CF photometry and the most-recently released {\it Gaia} catalogue, the photometric calibration of the SIF CF data follows a more traditional approach compared to the one for {\it Gaia} DR3. Rather than using the data themselves to define the photometric system, a set of calibrators was selected among those unpublished SIF CF sources that had counterparts in {\it Gaia} DR3. The {\it Gaia} DR3 photometry was then used as a reference to calibrate the SIF CF observations.\footnote{An alternative approach would involve applying the photometric calibrations derived in the nominal photometric processing \citep{2021A&A...649A...3R} for the SM CCDs. However, for this first run of the SIF CF processing, several inconsistencies in the IPD, most importantly the use of a different PSF library with respect to the nominal run, made this option not applicable. This choice will be re-evaluated for future releases.}

The time evolution of the photometric calibration was taken into account by splitting the input data into groups of observations separated by more than 30 satellite revolutions and not longer than six revolutions. For each group, one calibration per CCD (and thus per FOV) was solved for independently. The calibration model is described by a quadratic function in AC coordinate and provides the ratio between observed SIF CF and {\it Gaia} DR3 reference flux. No magnitude term was introduced in this step.

Applying these calibrations to the observed fluxes effectively translates them to the photometric system of {\it Gaia} DR3, removing all differential effects due to the evolution in time of the instrument, differences between the CCDs, and other elements in the optical paths. This allows the procedure to accumulate all single observations of a source to produce a mean source photometry.

\begin{figure}[thbp]
        \begin{center}
                \includegraphics[angle=0, width=\linewidth]{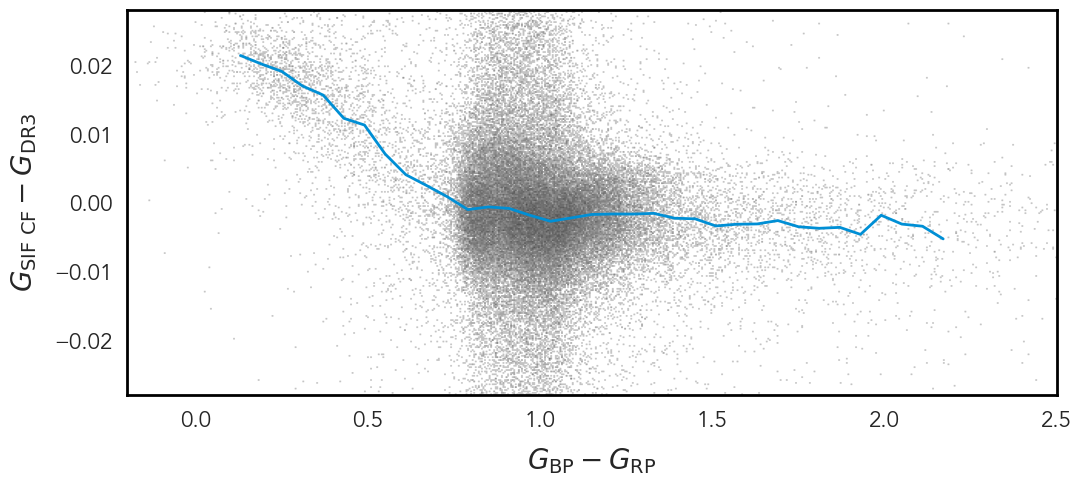}
                
                \includegraphics[angle=0, width=\linewidth]{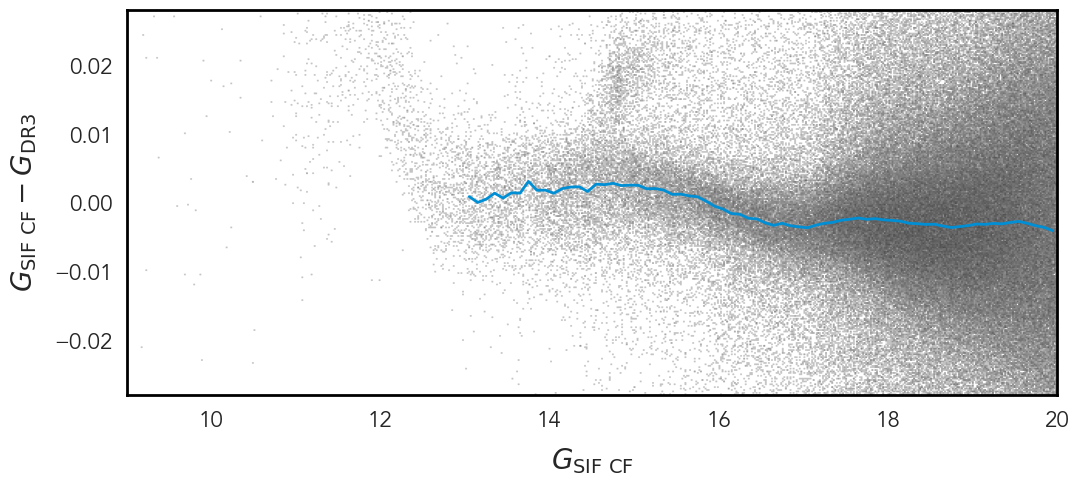}
                \caption{Colour and magnitude terms between the SIF CF calibrated magnitudes and the reference magnitudes from {\it Gaia} DR3. The magnitude terms have been corrected for in this FPR.}\label{fig:phot_colmagterms}
        \end{center}
\end{figure}

Among all sources from {\it Gaia} DR3 with observations in SIF CF mode, only those with\linktotable{gaia\_source}.\linktoparam{gaia\_source}{phot\_variable\_flag} \texttt{\small ='NOT\_AVAILABLE'}\xspace, $15<G<20$ and $0.7 < G_{\rm BP} - G_{\rm RP} < 4.0$ were used as calibrators. The magnitude and colour ranges were defined to avoid strong magnitude and colour terms that might affect the observed fluxes with respect to the reference photometry. 
These are visible in Fig. \ref{fig:phot_colmagterms}, where the difference between the SIF CF calibrated magnitude and the reference one are shown versus colour for a range of magnitudes in the top panel, and versus magnitude for a range of colours in the bottom panel. 

Using the colour information available for the calibrators, an ad hoc calibration could be derived. However, as no colour information is available for SIF CF sources, it is not possible to apply such calibration. 
As a result of this, the calibrated photometry for new SIF CF sources bluer than 0.7 in $G_{\rm BP} - G_{\rm RP}$ will suffer from a systematic colour term of magnitude similar to that shown by the blue line in the top panel of Fig. \ref{fig:phot_colmagterms}. This includes sources in the Blue Horizontal Branch.
These are visible in the bottom panel of Fig. \ref{fig:phot_colmagterms} as a population of sources that, starting at around G magnitude $15$, show large (of the order of a few percent) differences between the SIF CF calibrated magnitudes and the reference {\it Gaia} DR3 magnitudes.

No attempt was made to calibrate the strong magnitude term for $G<13$ as only a handful of SIF CF sources exist in that magnitude range. The magnitude term affecting fainter sources instead was calibrated out and removed from the mean source photometry. 

Once the initial mean photometry was obtained applying the calibrations to all observations, we evaluated the remaining effects not described by our model and estimated the remaining magnitude term. It is calibrated empirically as a simple offset to be applied to the SIF CF mean photometry in order to bring it into agreement with the corresponding {\it Gaia} DR3 values using sources for which mean photometry is available in both {\it Gaia} DR3 and SIF CF in the magnitude range $[13, 20]$ in $G$ and colour $0.7 < G_{\rm BP} - G_{\rm RP} < 4.0$. When applying this additional correction to the mean photometry, sources outside the magnitude range were corrected using the value for the closest magnitude included in the range. When applying the correction to sources in the covered magnitude range, interpolation is used between the bin values. 

\begin{figure}[t!]
        \begin{center}
                \includegraphics[angle=0, width=\linewidth]{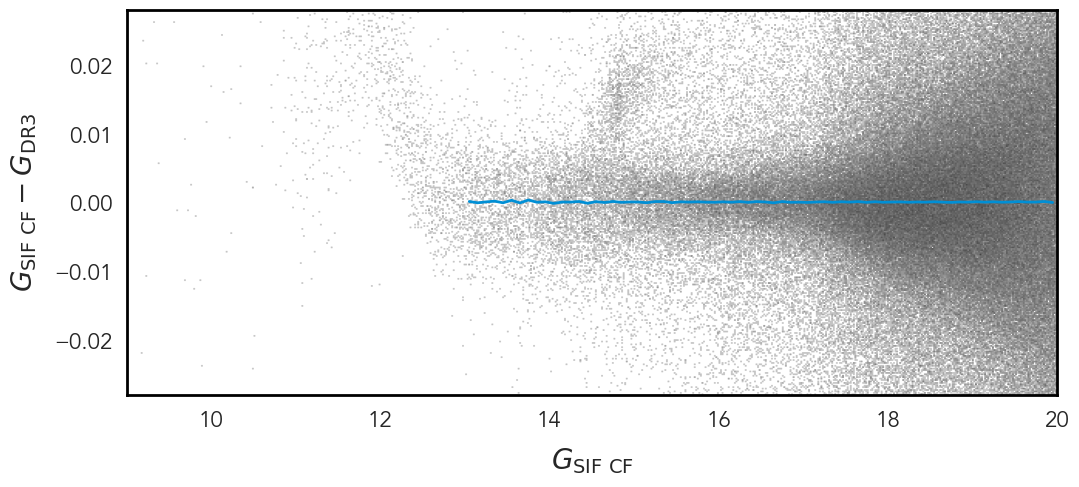}
                \caption{Comparison between SIF CF calibrated magnitudes and reference magnitudes from {\it Gaia} DR3 for the sources in common after the removal of the magnitude term.}\label{fig:phot_magtermsafter}
        \end{center}
\end{figure}

\Cref{fig:phot_magtermsafter} shows the same comparison as the bottom panel of Fig. \ref{fig:phot_colmagterms} after application of the magnitude term calibration. The correction was applied to the mean source photometry and to all relevant quantities, which are to median, minimum, maximum, and quartiles. 
It has no effect on photometric uncertainties as the error on the magnitude term is insignificant with respect to the uncertainties of the photometric measurements.

It is probably useful to stress again that the calibrations described in this section are not published in this FPR, but ensure consistency between the SIF CF and {\it Gaia} DR3 photometry. Systematics affecting the reference photometry, that is {\it Gaia} DR3, will also be present in the SIF CF FPR catalogue. Interested users are therefore encouraged to refer to the {\it Gaia} Early Data Release 3 \citep{2021A&A...649A...3R, 2021A&A...649A...5F} and DR3 \citep{2023A&A...674A...1G} papers where the known systematics are discussed.

\subsection{Filters} 

\label{sec:filters} Unlike in nominal {\it Gaia} data releases, no specific filtering is applied as a last step of the data processing. Some detections and sources did get discarded though during operation of the SIF CF pipeline:

\paragraph{ Filters at detection level}  In SIF CF IPD and SIF CF XM we removed the following detections:
\begin{itemize}
    \item Detections for which the PSF fit did not converge or found a source position outside the hypothetical observation window. Most likely these sources are actually just noise.
    \item Detections with internal duplicates. For the very rare cases, in which after the status cleaning two sources are located within the same subsample region, only the brightest source is kept. Most likely the fitting procedure here used two PSFs to fit one source.
    \item Detections that are located outside the SM CCD detector.
    \item Detections that would result outside the other instrument CCD detectors, as these cannot be processed by the downstream systems.
    \item Detections with less than 50 counts, which corresponds to a magnitude of 22.5, because they are most likely spurious.
    \item Detections with a positional uncertainty of more than 1 pixel and/or AL offset from PSF fit window centre of more than 1.2 pixels.
\end{itemize}

\paragraph{ Filters at source level}  In SIF CF XM, SIF CF Coverage, and SIF CF AGIS post processing we removed the following sources:
\begin{itemize}
    \item Sources with less than 11 observations in total. This we regarded as the minimum number of detections required for a reliable source. 
    \item Sources with a detection fraction of less than 50\,\% (see Sect. \ref{sec:coverage}). Sources that were observed in less than half of the cases that they could have been observed are most likely spurious. 
    \item Sources with positional error bars larger than 100 mas (30 sources removed).
    \item Sources within $0\farcs16$ distance from a brighter star. This was done irrespective of whether the brighter star was a {\it Gaia} DR3 source or a SIF CF source. We do not trust sources that are too close to brighter ones, as these can easily result from spike remnants left over when subtracting the scene containing the PSF model of a bright source from its observed image before source detection (673 sources removed).
    \item Non-converged sources (this did not actually happen).
\end{itemize}

The specification of SIF CF filters follows the nominal processing rules as closely as possible. The last three filters, applied during AGIS post-processing, are in fact identical except for the precise distance to a brighter star. This distance was increased from the nominal value of $0\farcs12$ to $0\farcs16$ for SIF CF.

After filtering, 526\,587 SIF CF sources persisted in $\omega$ Cen. These are published with this FPR.

\section{Validation}

\label{sec:validation} In order to assure the quality of the SIF CF data product, a scientific validation has been performed. For this, we defined a fixed colour for each dataset that we use. Except for the last figure, throughout the whole document we use yellow for {\it Gaia} DR3 data, violet for SIF CF data published in this FPR, and dark blue for the combination of these two. Additionally, there are unpublished SIF CF data, which were used for photometric calibration only, in blue and HST reference data in teal.

\subsection{Numbers}\label{sec:numbers}

 The number of detections in a single image ranges from 208 to 471\,039. In total, SIF CF IPD produced 66\,660\,921 detections from 2\,329 SIF CF images of $\omega$ Cen. From these, 45\,747\,947 detections were matched to a source, and 27\,500\,167 of these were matched to a new source not present in {\it Gaia} DR3. 

 Originally, SIF CF XM created 848\,988 sources, of which 527\,290 sources persisted after filtering for already known {\it Gaia} DR3 sources, the detection fraction, and minimum number of observations (see Sect. \ref{sec:filters}). Filtering for sources too close to brighter stars and sources with high uncertainties removed another 703 sources. The remaining 526\,587 sources were consolidated from 27\,476\,327 detections.

\begin{figure}[bhtp]
        \includegraphics[angle=0, width=\linewidth]{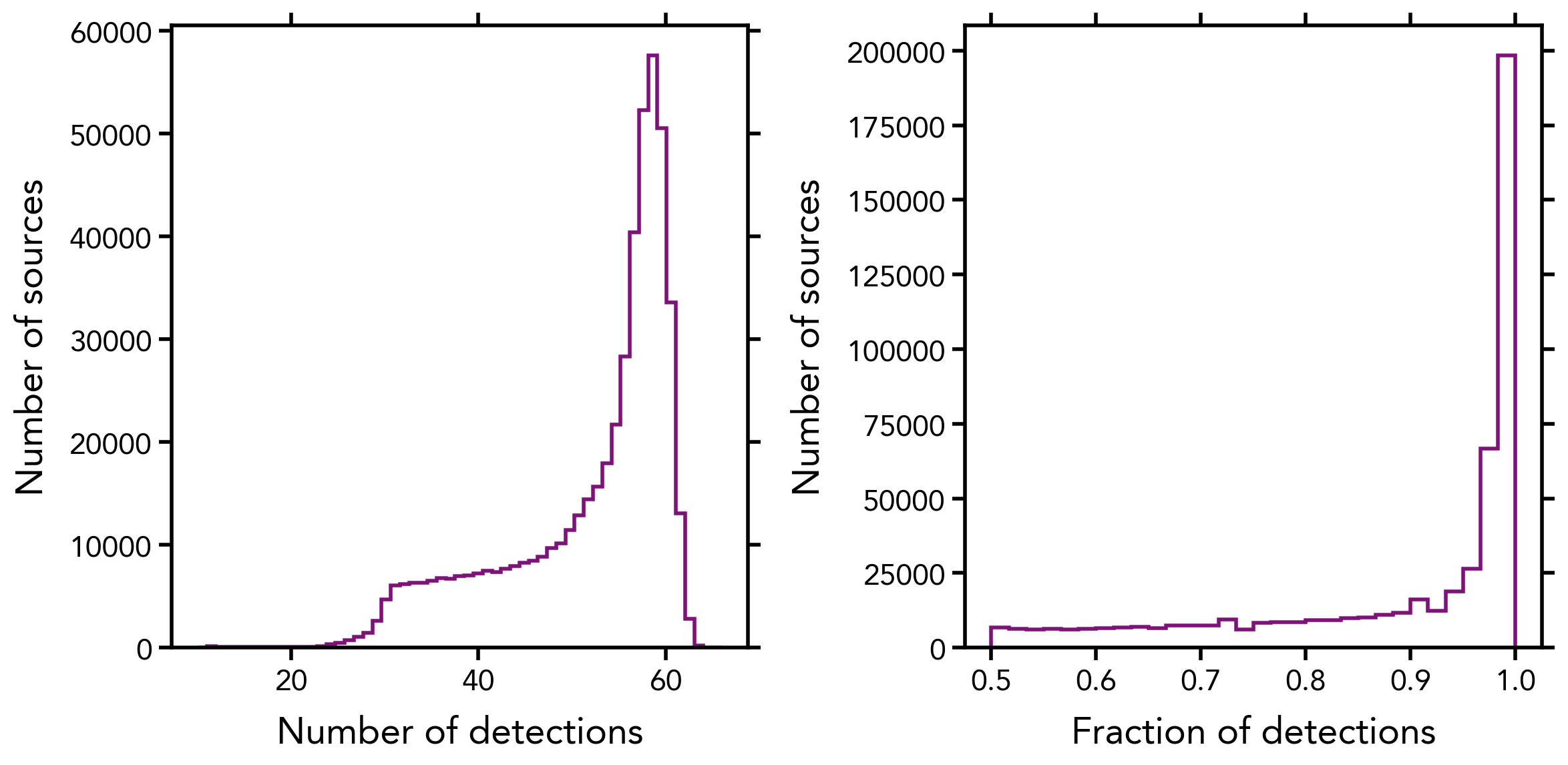}
        \caption{Number of detections and detection fraction. Distribution of sources with a certain amount of detections (left) and detection fraction (right) for SIF CF sources.} \label{fig:nobs}
\end{figure}

\Cref{fig:nobs} shows the distribution of the absolute number of detections for the SIF CF sources (left) and the fraction of the maximum possible observations (right). Most SIF CF sources have a very good detection fraction; they are observed in almost every image covering their location. A smaller amount of sources is observed in 50 to 90 percent of the scans. The absolute number of detections (left) shows that SIF CF sources are generally well probed, in most cases with 50 - 60 detections.

\subsection{Photometry}\label{sec:photometry}

To evaluate the effect of the photometric calibration on the uncertainties, it is interesting to see how these change with the application of the calibration, and how they compare with expectations from {\it Gaia} DR3. In the first instance, this can be done for the set of calibrators, for which the expected and actual uncertainties from {\it Gaia} DR3 are available. Both are valuable to quantify the effects of crowding on the photometric uncertainties.

\begin{figure}[htb]
        \begin{center}
                \includegraphics[angle=0, width=\linewidth]{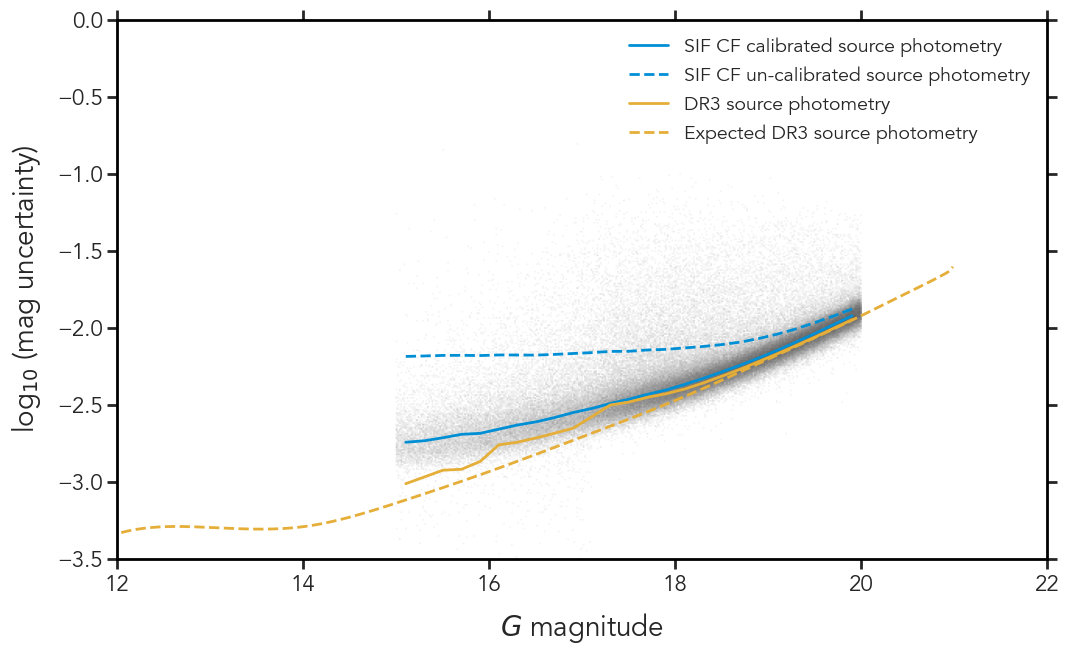}
                \caption{Median G magnitude uncertainty versus G magnitude. G magnitude uncertainty for all calibrators with data density in grey, see the legend and text for a description of the various overlaid curves.} \label{fig:phot_calibrators}
        \end{center}
\end{figure}
\begin{figure}[htb]
        \begin{center}
                \includegraphics[angle=0, width=\linewidth]{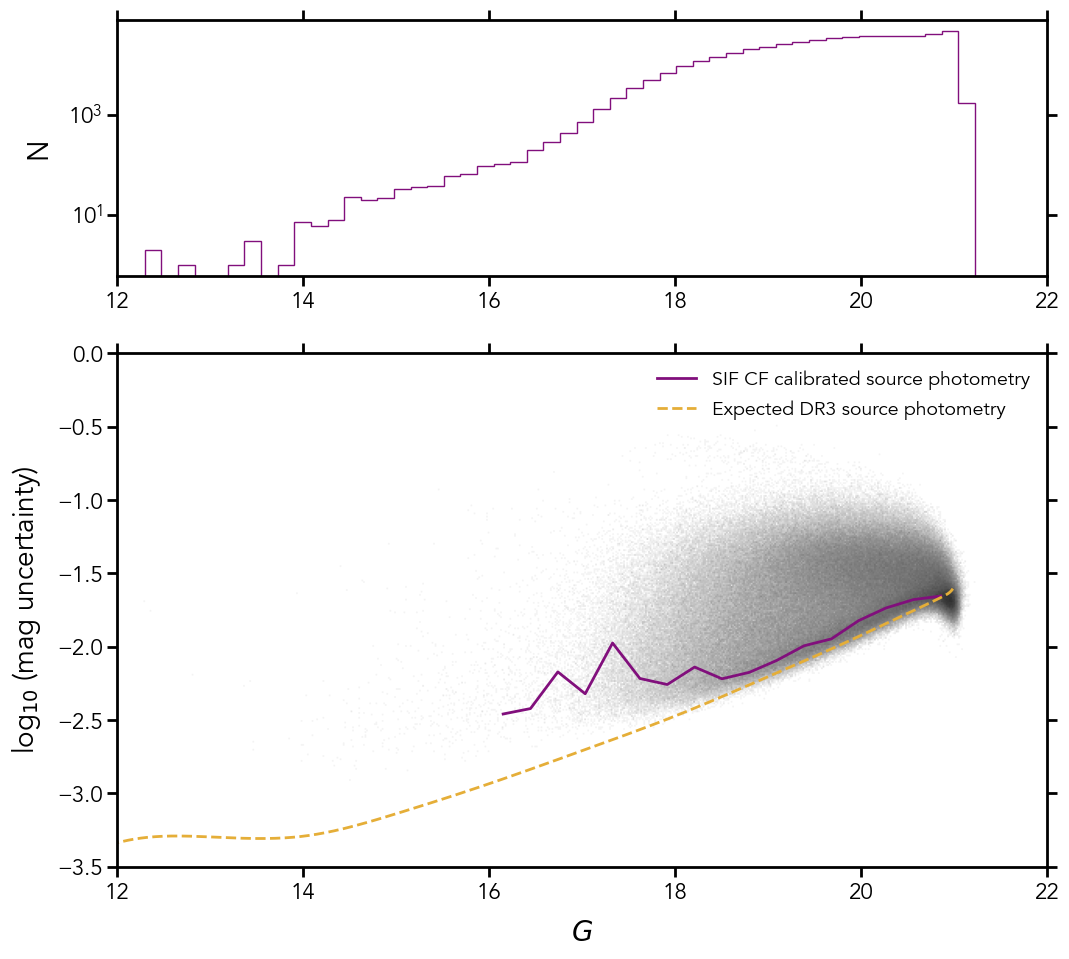}
                \caption{Distribution of magnitude and magnitude uncertainty for SIF CF sources in $\omega$ Cen. Top: Magnitude distribution. Bottom: Distribution of the mean G magnitude uncertainty versus G magnitude with data density in grey, see the legend and text for a description of the various overlaid curves.} \label{fig:phot_newsifsources}
        \end{center}
\end{figure}

 The density plot in Fig. \ref{fig:phot_calibrators} shows the distribution of photometric uncertainties versus {\it Gaia} DR3 magnitude for all the sources selected as calibrators. The dashed blue line illustrates the median photometric error for the uncalibrated SIF CF data. In this case, the source photometry is generated by simply taking a weighted average of the fluxes as measured by SIF CF IDU. The continuous blue line presents the median uncertainty of the calibrated SIF CF photometry. The density distribution of these data is provided in grey. The continuous yellow line provides the median photometric error measured from the {\it Gaia} DR3 photometry for the same set of sources. Finally, the dashed yellow line shows the expected {\it Gaia} DR3 photometry. 
 
 We provide this last curve to offer an idea of the quality of the photometry for the sources used as calibrators with respect to the {\it Gaia} DR3 photometric catalogue. It is expected that the {\it Gaia} DR3 photometry for sources in a region centred on $\omega$ Cen will be affected by crowding. All photometric errors have been scaled to correspond to a number of 50 observations, which is close to the average number of observations for the SIF CF FPR catalogue.

The G magnitude distribution of all SIF CF sources is presented in the top panel of Fig. \ref{fig:phot_newsifsources}.
The bottom panel of the same figure shows the distribution of the mode of the photometric uncertainties for all SIF CF sources. The data density is given in grey. As for these sources no photometry is available from {\it Gaia} DR3, only the expected nominal {\it Gaia} DR3 photometric uncertainties are provided for comparison (dashed yellow line). The continuous violet line in this plot depicts the mode rather than the median (as in Fig. \ref{fig:phot_calibrators}) given the rather skewed distribution at the faint end. 

The comparison of the blue dashed and continuous lines in Fig. \ref{fig:phot_calibrators} demonstrates that the applied calibration is efficient in reducing the scatter (as inferred from the measured uncertainties) between observed fluxes for a given source. The remaining scatter is larger than for the {\it Gaia} DR3 photometry. This is expected due to the different observing conditions, as only measurements from SM CCDs contribute to the SIF CF photometry. A larger scatter can also be due to the fact that all of these sources are observed in dense regions, where a high fraction of sources is blended and different scanning directions may give rise to different contamination from nearby objects. This is also apparent from the larger {\it Gaia} DR3 photometric uncertainties characterising the calibrators in this region with respect to the level expected from the entire catalogue.

\subsection{Position}\label{sec:position}

\begin{figure}[b]
        \begin{center}
                \includegraphics[angle=0, width=1.0\linewidth]{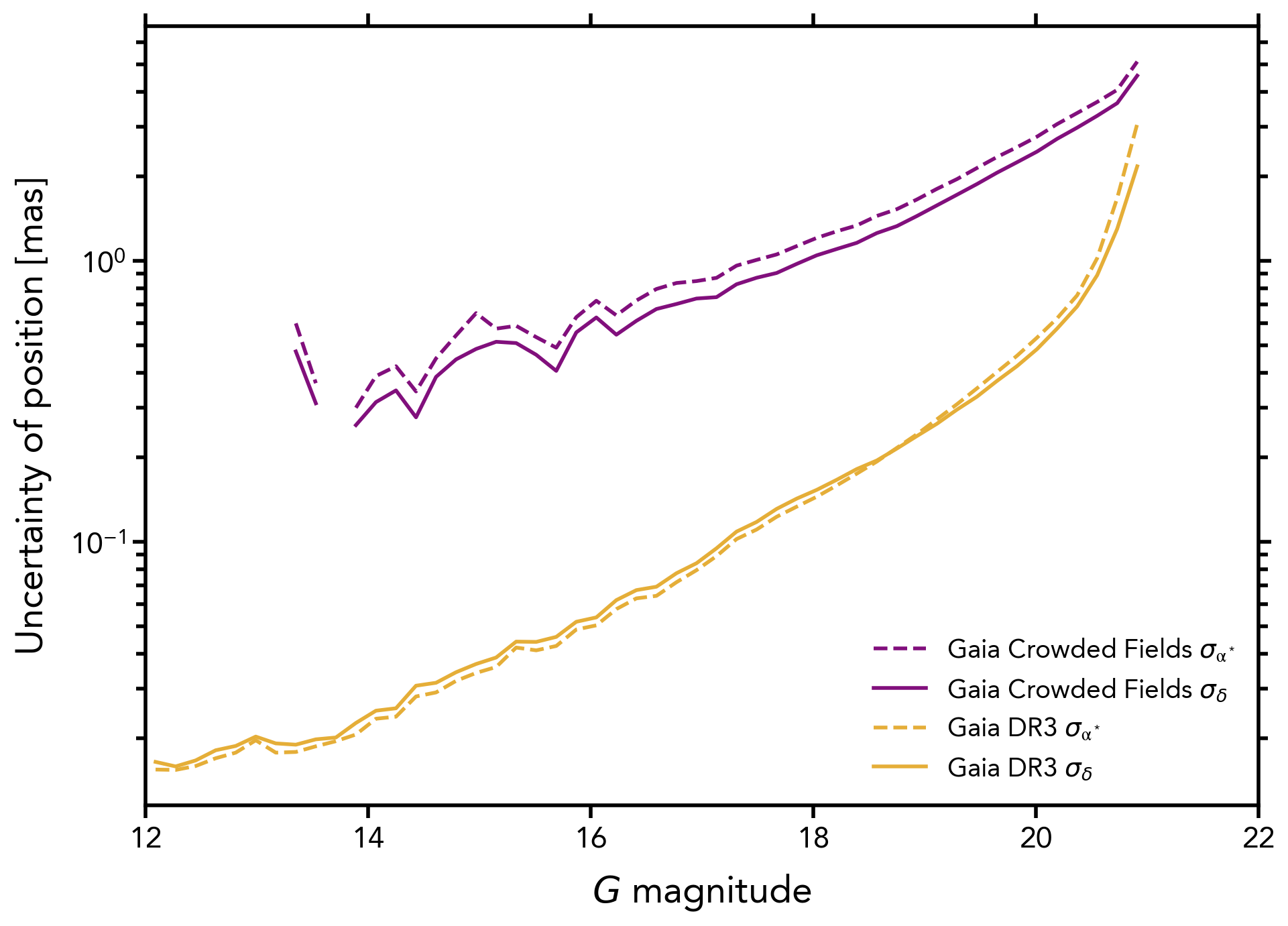}
                \caption{Median positional uncertainties. Uncertainties in right ascension $\sigma_{\alpha^*}$ (dashed lines) and declination $\sigma_{\delta}$ (solid lines) for SIF CF (violet) and {\it Gaia} DR3 (yellow) in $\omega$ Cen.} \label{fig:pos_err}
        \end{center}
\end{figure}

A direct comparison of SIF CF and {\it Gaia} DR3 astrometry is not straightforward. The {\it Gaia} DR3 data have very different input as these sources are created from unbinned and more AF observations per transit; they are generally brighter, situated in less dense environments, and their calibration including the PSF model is better defined. The SIF CF sources instead are based on binned SM data with just one observation per transit; as all sources with {\it Gaia} DR3 counterpart were removed, these are fainter sources situated in denser environments (see Fig. \ref{fig:nn}); and as SM data was originally not meant to be used for scientific exploitation, the calibrations are less accurate.  
However, a few important things can be noticed nevertheless. 

In Fig. \ref{fig:pos_err} we show the median uncertainties of source positions $\sigma_{\alpha^*}$ (dashed line) and $\sigma_{\delta}$ (solid line) as a function of G magnitude in $\omega$ Cen for newly created SIF CF sources (violet) and {\it Gaia} DR3 sources (yellow). Overall, the SIF CF data have a median positional uncertainty of 3.03 mas in right ascension (RA) and 2.69 mas declination (Dec). The reason for the difference between uncertainties in RA and Dec is most likely connected to the asymmetry in scan direction distribution. 

The median positional uncertainties (and also uncertainties in parallax and proper motion, see Fig. \ref{fig:propermotion_err} and Fig. \ref{fig:parallax_err}) are substantially larger for SIF CF data (violet lines) than for {\it Gaia} DR3 (yellow lines). This is not at all unexpected due to the above mentioned limitations.

\subsection{Proper motion}\label{sec:propermotion}

\begin{figure}[t!]
     \begin{center}
     \includegraphics[angle=0, width=1.0\linewidth]{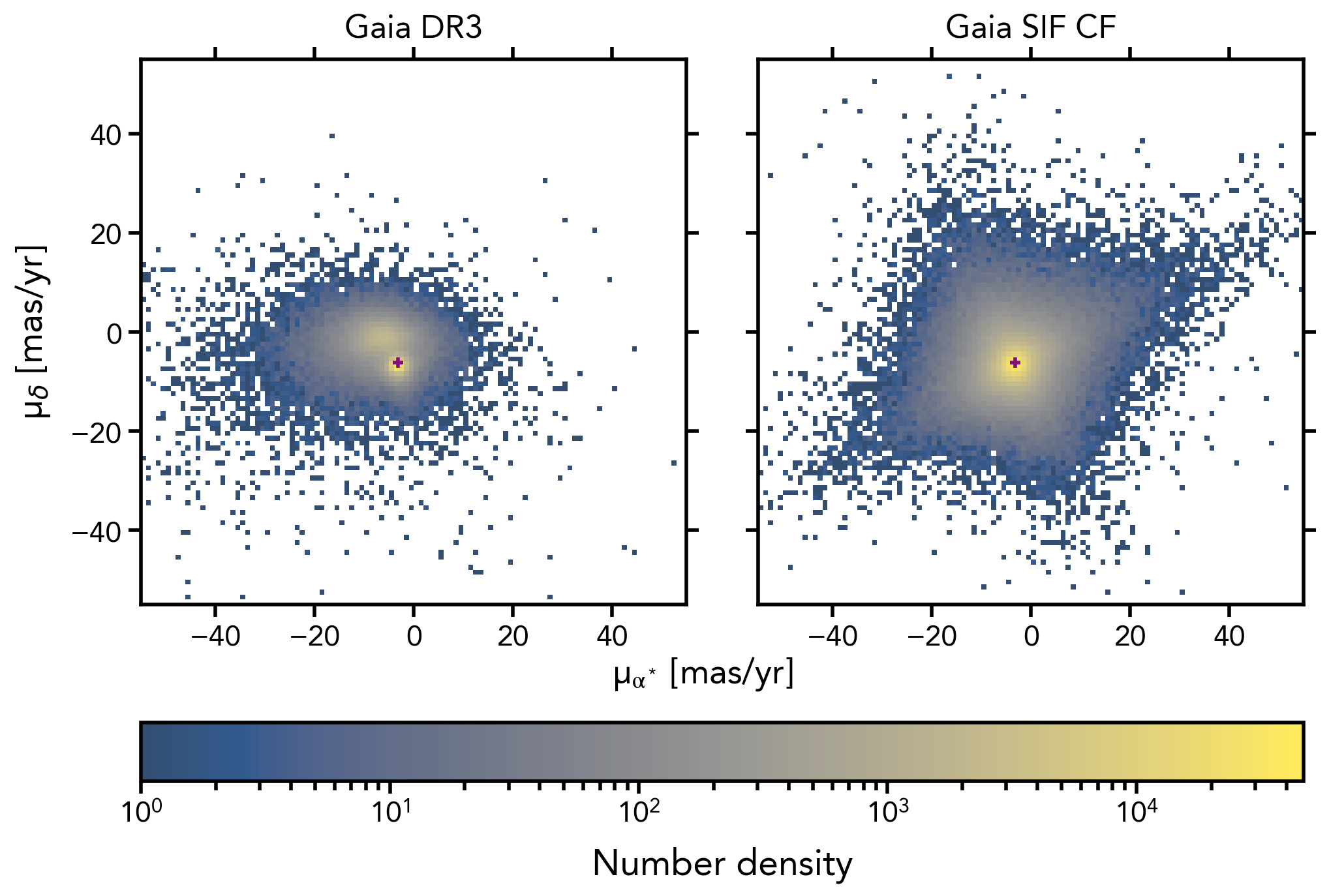}      
     \caption{Distribution of proper motions in $\omega$ Cen coloured by source density. Mean proper motion of SIF CF sources (violet cross). Proper motions for {\it Gaia} DR3 (left) and SIF CF sources (right). } \label{fig:agis_pm}
    \end{center}
\end{figure}
\begin{figure}[t!]
     \begin{center}
     \includegraphics[angle=0, width=1.0\linewidth]{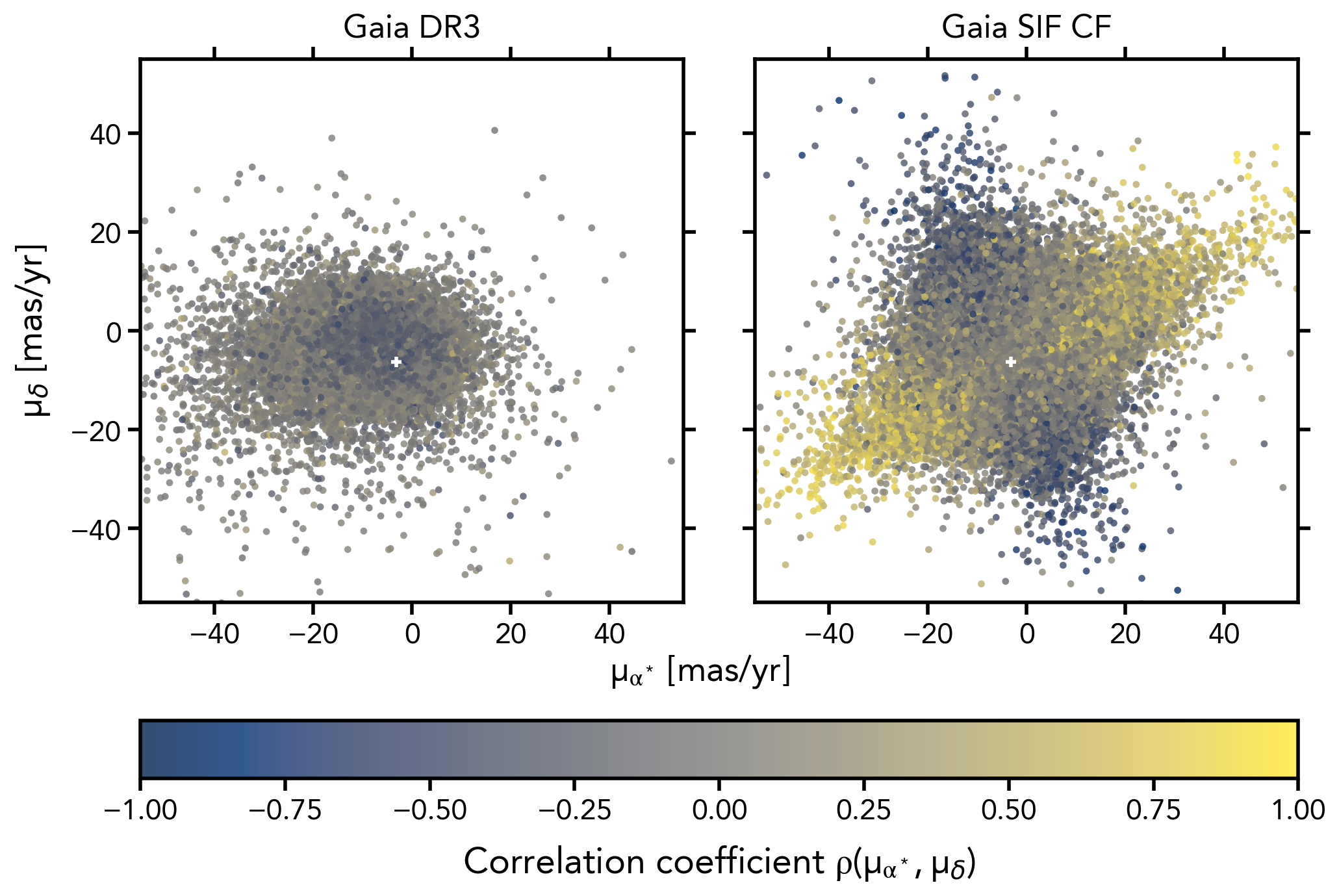} 
     \caption{Distribution of proper motions in $\omega$ Cen coloured by correlation between proper motion in RA and proper motion in Dec. Mean proper motion of SIF CF sources (white cross). Proper motion for {\it Gaia} DR3 sources (left) and SIF CF sources (right). } \label{fig:agis_pm_corr}
    \end{center}
\end{figure}

\Cref{fig:agis_pm} presents the distribution of proper motions for {\it Gaia} DR3 sources (left panel) and SIF CF sources (right panel) in the $\omega$ Cen region. For DR3 sources, we can clearly distinguish two sub-populations -- a dense blob of $\omega$ Cen cluster stars and a more extended distribution of fore- and background stars. In the distribution of the SIF CF sources we see, as expected, primarily cluster members. This is the case, because most SIF CF sources lie in the dense cluster core, where the fraction of fore- and background stars is rather small. The mean proper motion of the SIF CF sources is $\mu_{\alpha*} = -3.21$ mas/yr in RA and $\mu_\delta = -6.24$ mas/yr in Dec. It is marked as violet cross in Fig. \ref{fig:agis_pm}.

We also notice that the distribution of proper motions for SIF CF sources is much more extended than for DR3 sources. This is again due to the SIF CF sources being systematically fainter and detected in a much denser environment, which results in larger proper motion uncertainties.

Additionally, the proper motions of the SIF CF sources reveal an X-shaped feature not visible in the DR3 distribution. The X-shape of this feature is most likely caused by the inhomogeneous distribution of the scanning angles, which leads to a strong correlation between proper motions in RA and Dec. This is visualised in Fig. \ref{fig:agis_pm_corr}, where the colour scale reflects the correlation coefficient between proper motion in right ascension and proper motion in declination. This is a dimensionless quantity in the range $[-1,+1]$. The sources located in the X of the distribution show strong correlations between proper motions in RA and Dec.

\begin{figure}[t!]
        \begin{center}
        \includegraphics[angle=0, width=1.0\linewidth]{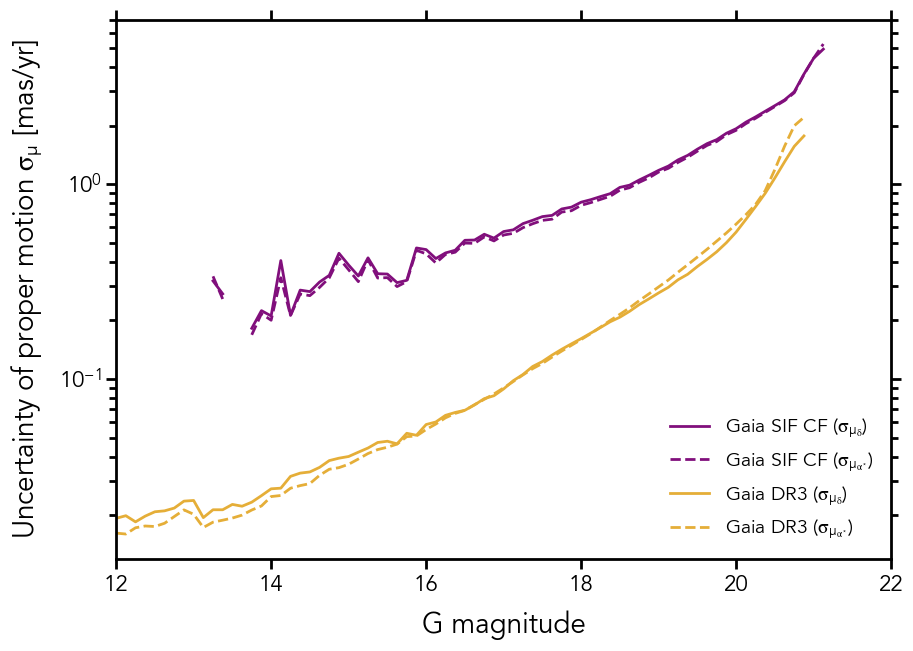}
                \caption{Median proper motion uncertainties as functions of G magnitude. We show $\sigma_{\mu_{\alpha^*}}$ (dashed lines) and $\sigma_{\mu_{\delta}}$ (solid lines) for SIF CF (violet) and {\it Gaia} DR3 (yellow) sources in $\omega$ Cen.} \label{fig:propermotion_err}
        \end{center}
\end{figure}

The median uncertainties of proper motions $\mu_{\alpha^*}$ (dashed lines) and $\mu_{\delta}$ (solid lines) in $\omega$ Cen for {\it Gaia} DR3 (yellow) and SIF CF (violet) sources are given in Fig. \ref{fig:propermotion_err}. As expected (see Sect. \ref{sec:position}), the median proper motion uncertainties are substantially larger for new SIF CF sources (violet) than for {\it Gaia} DR3 sources (yellow). The median proper motion uncertainty for SIF CF sources in $\omega$ Cen amounts to 2.02 mas/year in RA and 2.06 mas/year in Dec. This is less than the systematic proper motion of $\omega$ Cen, but greater than the internal proper motion dispersion for cluster stars, which is just over 0.5 mas/yr in both RA and Dec. An accuracy comparable to the internal proper motion dispersion is reached only for the brighter part of the sample.

\begin{figure}[htb]
        \begin{center}
        \includegraphics[angle=0, width=1.0\linewidth]{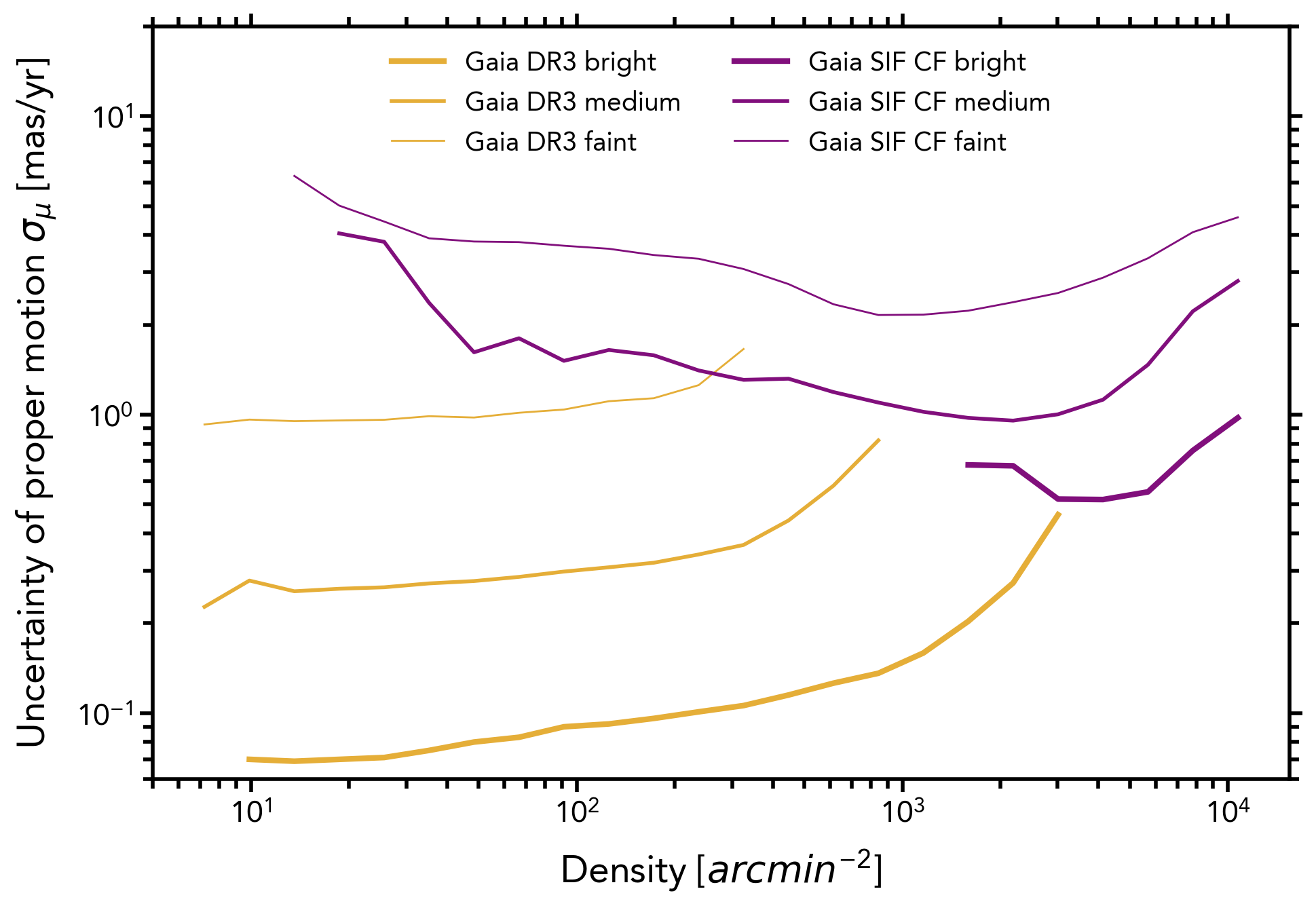} 
                \caption{Median proper motion uncertainties $\sigma_{\mu_{\delta}}$ as functions of source density. Proper motion uncertainties for bright ($G < 18^m$), medium ($18^m \le G < 20^m$), and faint ($G \ge 20^m$) sources in SIF CF (violet) and {\it Gaia} DR3 (yellow).} \label{fig:propermotion_err_density}
        \end{center}
\end{figure}

\Cref{fig:propermotion_err_density} shows the median uncertainties of proper motions $\mu_{\delta}$ as functions of the source density in three magnitude ranges.
The uncertainties for the SIF CF sources (violet) are larger than for the {\it Gaia} DR3 sources (yellow) within the same magnitude range and same density. However, the difference decreases rapidly with increasing density. At the highest densities for which {\it Gaia} DR3 sources of a given magnitude range are still present in the catalogue, the uncertainty difference between SIF CF and {\it Gaia} DR3 sources almost vanishes.

\subsection{Parallax}\label{sec:parallax}

\begin{figure}[t!]
        \begin{center}
        \includegraphics[angle=0, width=1.0\linewidth]{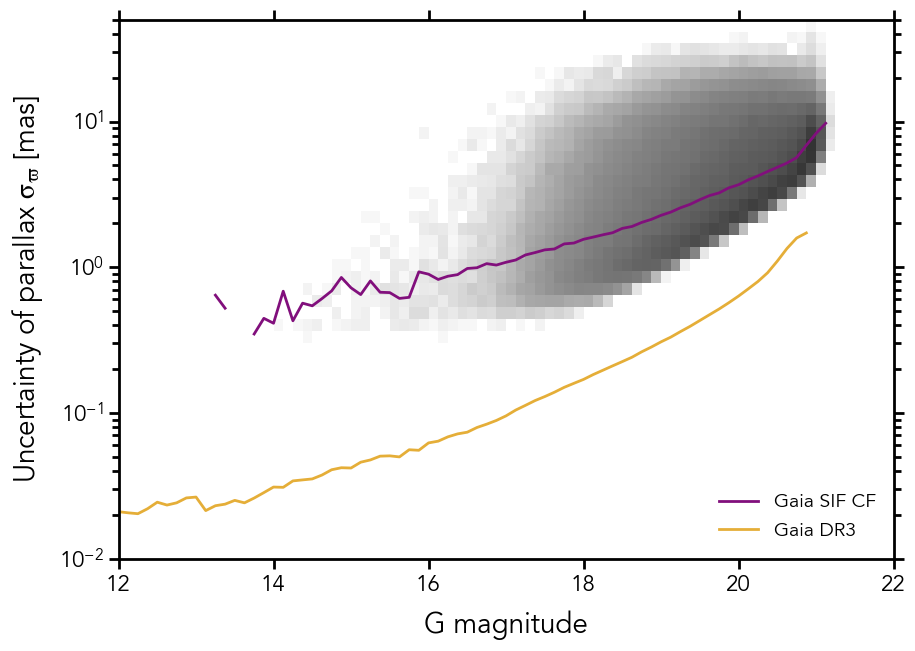}
                \caption{Median parallax uncertainties $\sigma_\varpi$ versus G magnitude in $\omega$ Cen. Uncertainties for SIF CF (violet) and {\it Gaia} DR3 (yellow). Data density in grey. } \label{fig:parallax_err}
        \end{center}
\end{figure}

The distribution of uncertainties for parallax ($\varpi$) is provided in Fig. \ref{fig:parallax_err}.  The uncertainties in the $\omega$ Cen region are plotted with respect to the G magnitude and grey shades mark the density of data points. As for positions and proper motions (compare Sect. \ref{sec:position}), the median uncertainty of the parallax for SIF CF sources, shown as a violet line, is substantially larger than the median uncertainty of the parallax for {\it Gaia} DR3 (yellow). Overall, the median parallax uncertainty for SIF CF sources in $\omega$ Cen amounts to 3.95 mas.

\subsection{Reference data} \label{sec:bellini} 

Thanks to its largely stare-mode observing approach, HST data tend to be much deeper than {\it Gaia} DR3 or SIF CF data. They therefore provide a good opportunity to probe the power of the SIF CF pipeline. 

The higher angular resolution of HST images allows us to test SIF CF deblending abilities. 
We use a dedicated processing of multiple HST images for the $\omega$ Cen cluster core, carried out by \citet{Bellini2017}. In order to better match the {\it Gaia} data, we applied the following corrections to this HST sample:
\begin{itemize}
        \item As G--band proxy, we use the HST F606W filter magnitude with an offset: $G_{Bellini} = F606W - 0^m.07$. 
        \item An astrometric correction of $+12.3$ mas in right ascension and $-11.26$ mas in declination was applied. 
    \item For some HST sources, the photometric uncertainty was not provided. In such cases, we estimated it by means of an empirical function: $\sigma_G = e^{(F606W - 26.59) \cdot 0.675} + 0.0105$ 
\end{itemize}

\begin{figure}[htbp]
        \begin{center}
                \includegraphics[angle=0, width=0.80\linewidth]{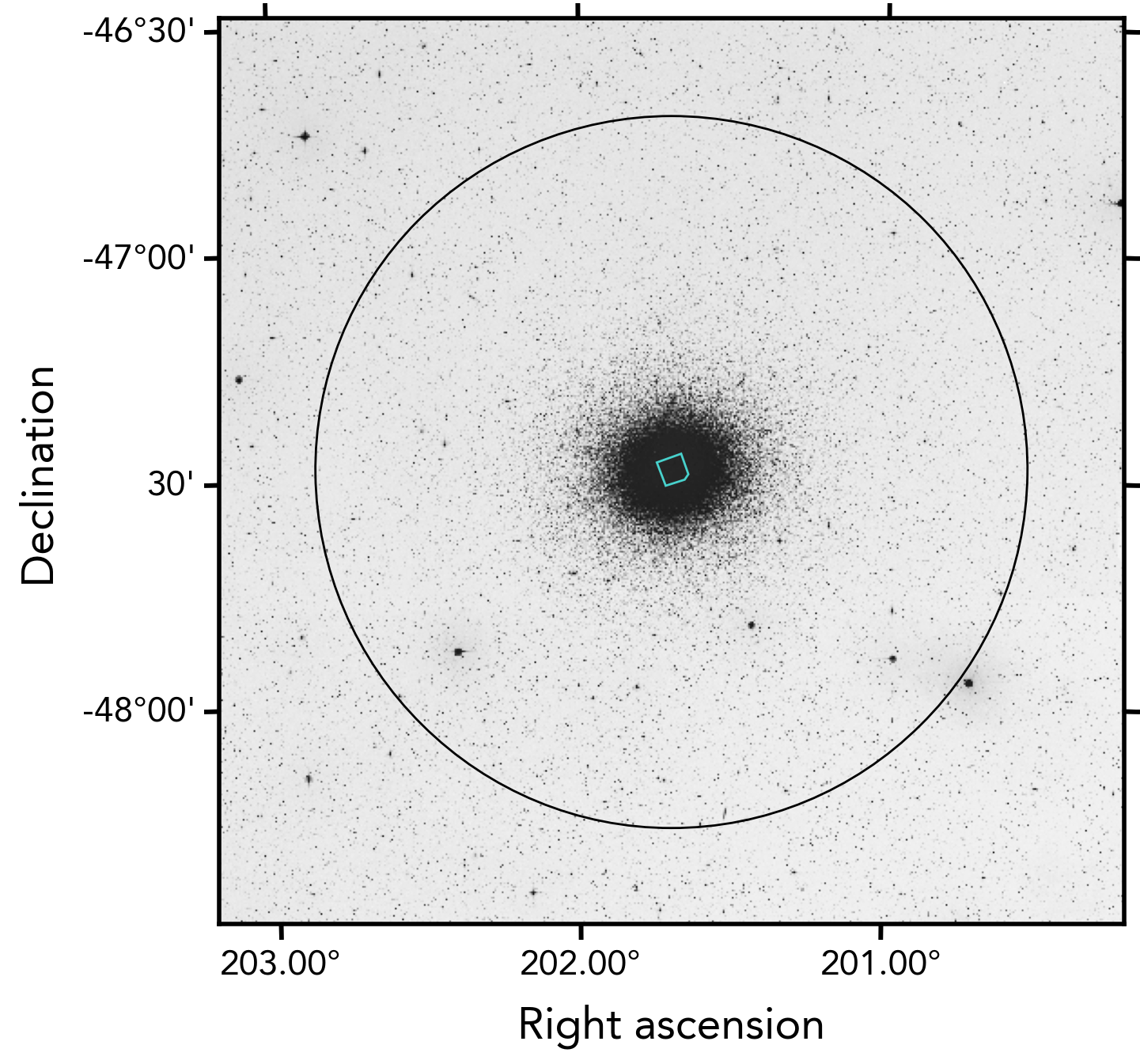}
                \caption{Digitized Sky Survey view of $\omega$ Cen. Extent of the SIF CF region (black circle) and HST reference data (teal square).}\label{fig:overview}
        \end{center}
\end{figure}

As reference for our validation we selected a region in the core of $\omega$ Cen that is slightly smaller than the area covered by \citet{Bellini2017}. The less reliable outer parts were cut, and the area was limited to give best coverage with SIF CF data. In Fig. \ref{fig:overview}, it is shown as teal polygon. For this region, deep HST data are available with which we cannot examine the astrometry, but we can verify the reliability of faint SIF CF source detections in the most crowded parts of $\omega$ Cen.

\subsection{Small-scale completeness}

\label{sec:smallscale}In order to assess the small-scale completeness of the SIF CF catalogue, we calculate the density of source neighbours as a function of their separation from the source. In a survey with infinite resolution, this would be a constant function. In real surveys the density decreases for smaller separations and goes to zero as the separation goes to zero -- it is increasingly more difficult to resolve sources with smaller separations.

\begin{figure}[thbp]
        \begin{center}
                \includegraphics[angle=0, width=\linewidth]{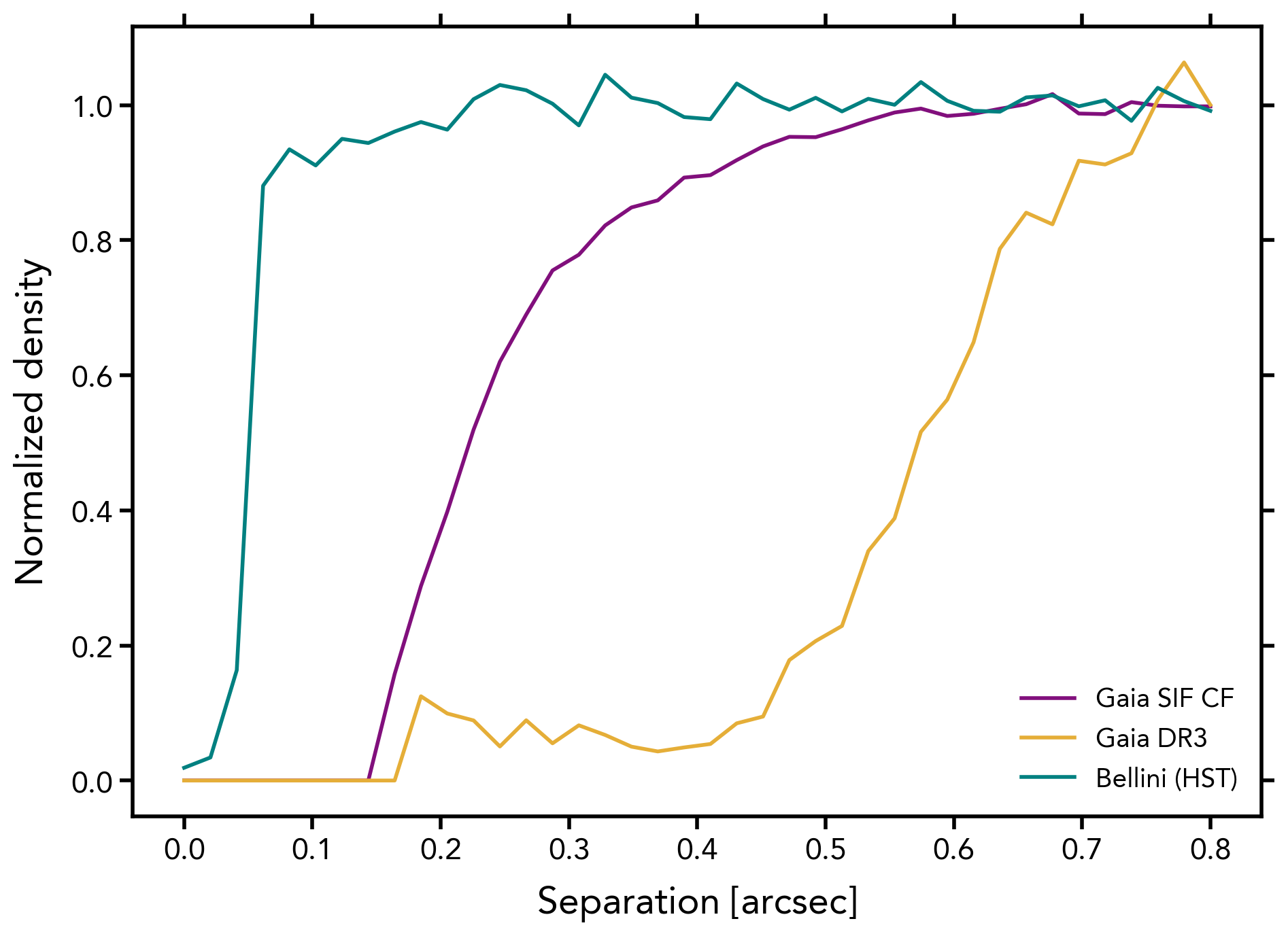}
                \caption{Normalised source density as a function of source separation. Source density for SIF CF (violet), Bellini HST (teal) and {\it Gaia} DR3 data (yellow).}\label{fig:nn}
        \end{center}
\end{figure}

In Fig. \ref{fig:nn} the normalised source density is plotted as a function of the source separation for {\it Gaia} DR3 data (yellow), {\it Gaia} SIF CF data (violet), and the Bellini HST sample (teal, see Sect. \ref{sec:bellini}). The HST images have very good angular resolution, thus it is not surprising that 50\,\% density is reached at a separation of $0\farcs06$ already. {\it Gaia} DR3, on the other hand, reaches 50\,\% density at $0\farcs57$ separation, which roughly corresponds to 10 pixels in AL or 3 pixels in AC. This is consistent with the small-scale completeness reported for {\it Gaia} in \citet{2021A&A...649A...5F}. Unlike the nominal {\it Gaia} pipeline, the SIF CF pipeline can be applied to full images and profits from iterative source detection and parametrisation. The SIF CF dataset therefore reaches 50\,\% density for a separation of $0\farcs22$ -- this is a factor of 2.6 better than {\it Gaia} DR3. It corresponds to 1.7 SM samples in AL and 0.6 samples in AC.

 A more detailed study has shown that the SIF CF pipeline can separate sources down to $0\farcs2$ distance in case their brightness differs by less than three magnitudes, which is the case for the biggest part of the data. If their brightness differs by more than three magnitudes though, the SIF CF resolving power decreases. The reason for this is that with faint sources of say $G=21^m$ and a magnitude difference of more than three, the brighter source already has $G=18^m$ or brighter, which means that these are increasingly bright sources already, and around such sources it becomes more and more complicated to detect $G=21^m$ sources.

In order to describe the resolution limit for SIF CF sources and its dependence on the magnitude, we defined the following approximate relation:
\begin{eqnarray}        \mathrm{Sep} < \left\{
        \begin{aligned}
                &0.2\,\arcsec ,& \textrm{if } \Delta m \leq 3^m \\
                &0.1\,\arcsec \cdot (\Delta m - 1),& \textrm{if } \Delta m > 3^m
        \end{aligned}
        \right. \label{eq:separation}
.\end{eqnarray} 
This limit should be viewed as a soft one though. A considerable number of sources are detected below that limit, and also a number of sources are missed above.

\subsection{Source density and depth}\label{sec:depth}

\begin{figure}[htbp]
        \begin{center}
                \includegraphics[angle=0, width=\linewidth]{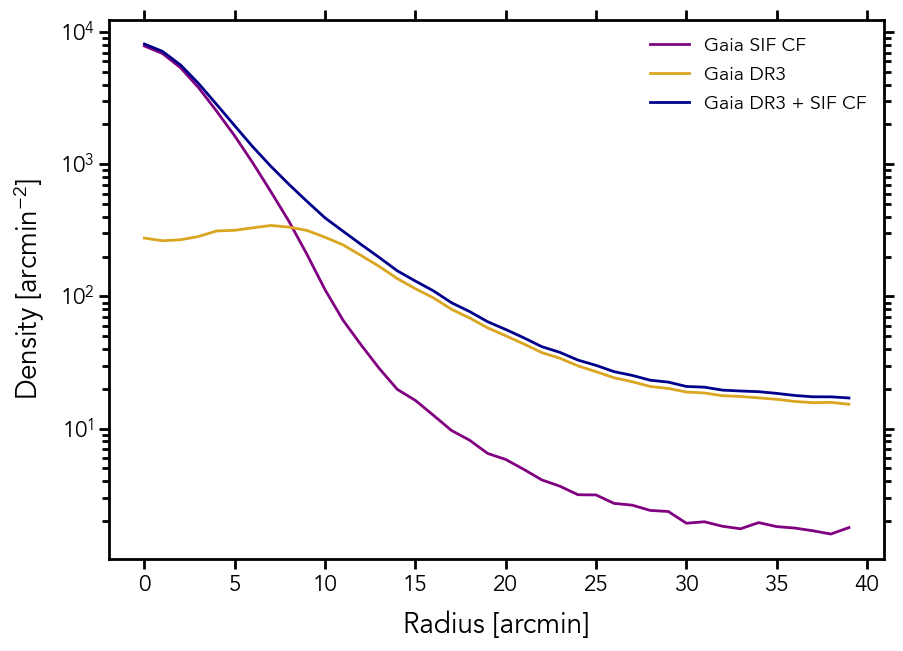}
                \caption{Source density as function of distance from cluster centre  in $\omega$ Cen. Source density for {\it Gaia} DR3 sources (yellow), SIF CF sources (violet), and the combined catalogue (dark blue)} \label{fig:density}
        \end{center}
\end{figure}

The source density as a function of distance from the $\omega$ Cen cluster centre in stars per square arcmin is shown in Fig. \ref{fig:density}. {\it Gaia} DR3 source density is given in yellow, SIF CF source density in violet, and the combined catalogue in dark blue. 
At a radius of about 9\arcmin\ the same number of sources is found in SIF CF and {\it Gaia} DR3 data. Within this radius, SIF CF sources dominate the combined sample. In the central parts of the $\omega$ Cen cluster SIF CF produces about ten times more sources than {\it Gaia} DR3.

\begin{figure}[bthp]
    \begin{center}
            \includegraphics[angle=0, width=\linewidth]{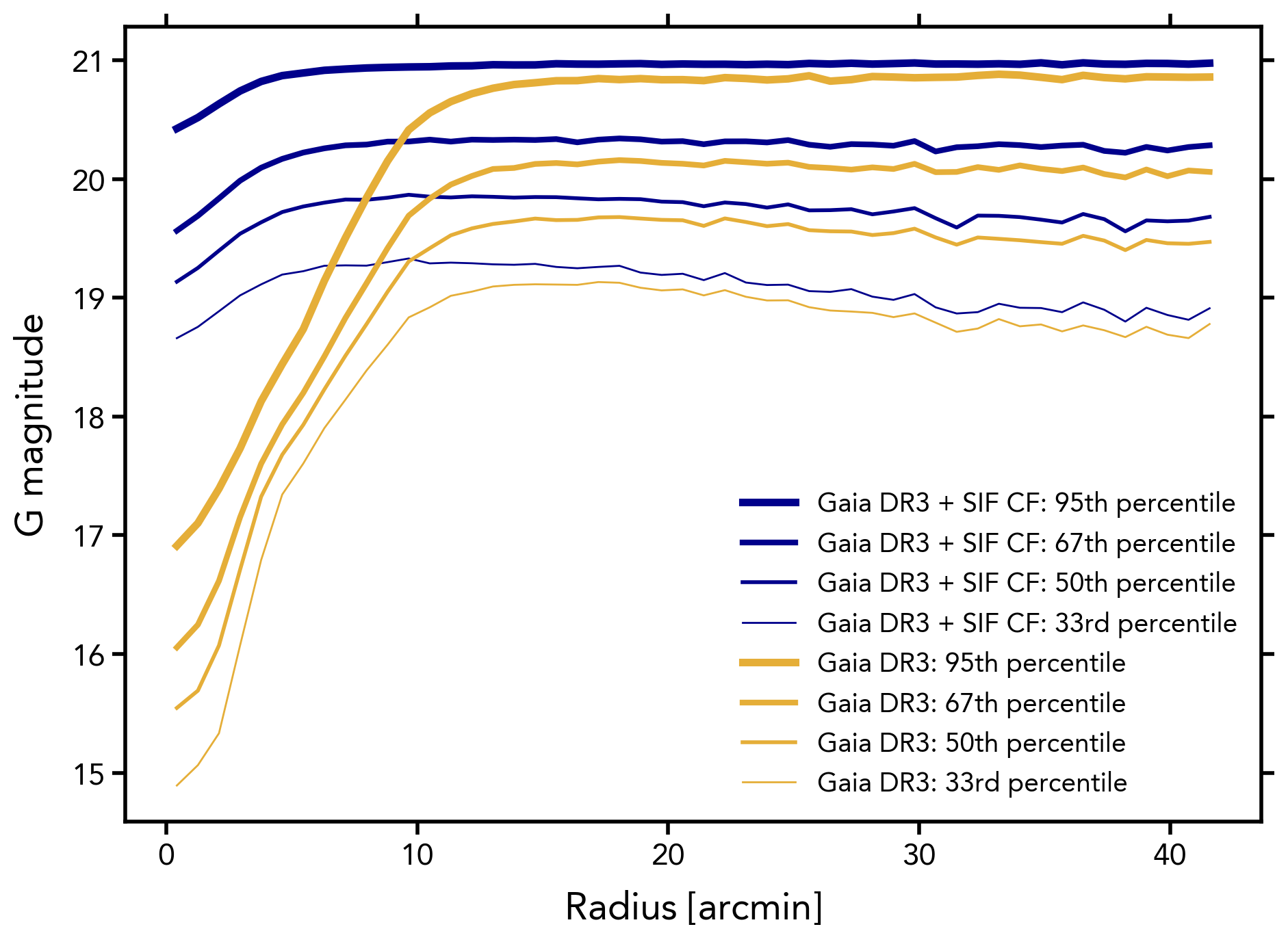}
              \caption{Quantiles of the magnitude distribution versus distance to cluster centre. 33\,\% (thin line), 50\,\%, 67\,\%, and 95\,\% (thick line) for {\it Gaia} DR3 sources (yellow) and the combined catalogue (dark blue).} \label{fig:depth}
   \end{center}
\end{figure}

This is further demonstrated in Fig. \ref{fig:depth}, where quantiles of the magnitude distribution are shown as a function of the distance from the cluster centre. The quantiles are given for {\it Gaia} DR3 alone (yellow) and for the combined catalogue of {\it Gaia} DR3 supplemented with the SIF CF sources (dark blue). As expected, the quantiles for {\it Gaia} DR3 sources and the combined catalogue are relatively close for radii beyond 9\arcmin, where the contribution of new SIF CF sources is small. However, at smaller radii the combined catalogue is considerably deeper. In the cluster core, the 95th percentile of the G magnitude distribution amounts to 16.9 for the nominal {\it Gaia} catalogue, while for the combined catalogue it reaches 20.4. In the densest region of $\omega$ Cen the source magnitudes of the combined catalogue go more than three magnitudes deeper than those of {\it Gaia} DR3 alone.

\subsection{Reliability and completeness}\label{sec:completeness}

\begin{figure*}[t!]
        \begin{center}
                \includegraphics[angle=0, width=1.0\linewidth]{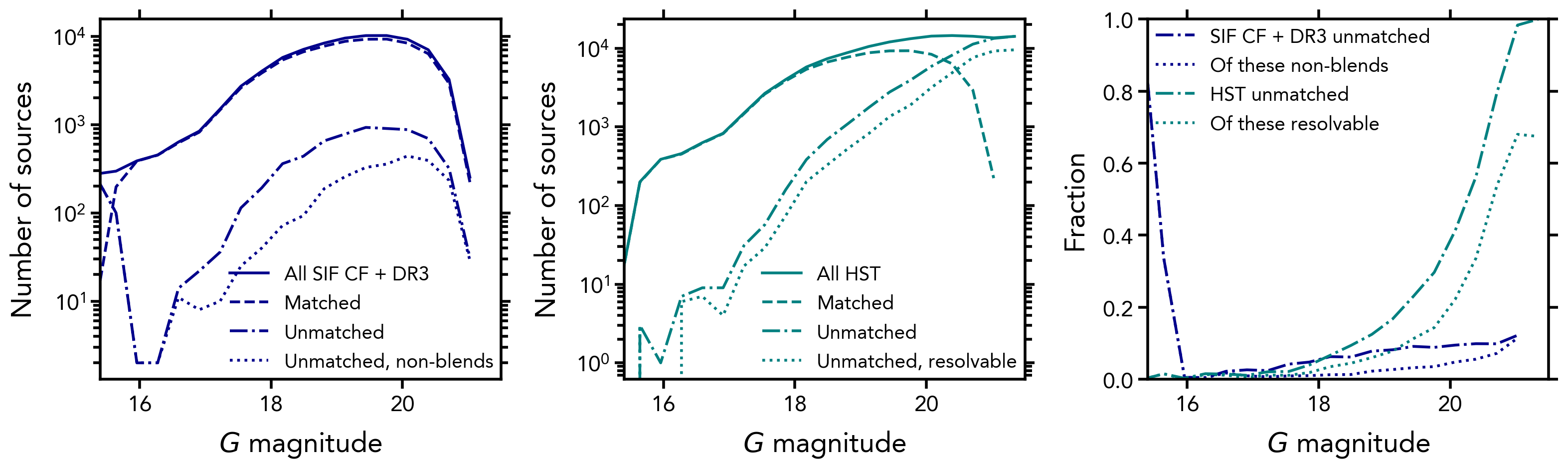}
                \caption{Comparison of SIF CF data extended with {\it Gaia} DR3 and HST data. Left: Distribution of the combined {\it Gaia} SIF CF and {\it Gaia} DR3 sources (solid dark blue) versus {\it Gaia} G magnitude; of these, sources matched (dashed) and not matched (dash-dotted) to HST; of the latter, not blended  sources (dotted).
        Centre: Distribution of all HST sources (solid teal) versus {\it Gaia} G magnitude; HST sources matched (dashed) and not matched (dash-dotted) to SIF CF sources; Unmatched HST sources potentially resolvable by SIF CF (dotted).
        Right: Fraction of combined SIF CF and {\it Gaia} DR3 sources not matched to HST (dash-dotted dark blue); of these, not HST blends (dotted dark blue); fraction of HST sources not matched to SIF CF (dash-dotted teal); fraction of resolvable, unmatched HST sources (dotted teal) versus (estimated) G magnitude}.\label{fig:bellini}
        \end{center}
\end{figure*}

In the core of the $\omega$ Cen cluster, nominal {\it Gaia} DR3 sources become less numerous and less accurate. This is, where SIF CF data are expected to be deeper than nominal data. 

To probe the reliability and completeness of the SIF CF pipeline results, we selected an area in the $\omega$ Cen cluster core for which data are available in \citet{Bellini2017}. The pre-processing applied to this dedicated dataset is described in Sect. \ref{sec:bellini}. We should bear in mind though, that absolute astrometry with HST is either poor or tied to {\it Gaia} data, and should thus be used with care in this validation. The HST photometry can be used to some extent, as the HST F606W magnitudes are with a certain offset close to the {\it Gaia} G--band magnitudes.

To make a proper comparison, we supplemented SIF CF sources with {\it Gaia} DR3 sources, creating a complete {\it Gaia}-and-SIF-CF source list. The contribution of {\it Gaia} DR3 sources is small though -- there are only around 1200 DR3 sources fainter than $16^m$ available in the area under consideration, which is about 1.5\,\% of the SIF CF sources in this area. As the fraction of added {\it Gaia} DR3 sources is really small in this sample, we keep referring to it as SIF CF, even though strictly speaking it is a combined catalogue.

The HST data cover a fraction of the cluster only, but they do go substantially deeper and offer a higher resolution. We can thus assume that the HST data should contain all sources observed in SIF CF. Hence those sources found by SIF CF but not in the HST data are likely spurious or blends, while sources present in HST but missed in SIF CF demonstrate what SIF CF might be missing. 

For Fig. \ref{fig:bellini}, we define 'matches' as pairs for which the separation between HST and SIF CF sources is less than 200 mas, and the following inequality for corresponding source magnitudes $m_{\mathrm{HST}}$ and $m_{\mathrm{SIF}}$ holds: 
\begin{equation}
    |m_{\mathrm{HST}} - m_{\mathrm{SIF}}| < 3 (\sigma_{\mathrm{HST}} + \sigma^*_{\mathrm{SIF}}).\label{ea:match}
\end{equation}
Here, $\sigma_{\mathrm{HST}}$ is the magnitude uncertainty for the HST source, and $\sigma^*_{\mathrm{SIF}}$ is derived from the SIF CF source magnitude uncertainty $\sigma_{\mathrm{SIF}}$ as:
\begin{equation}
    \sigma^*_{\mathrm{SIF}} = \sqrt{ \sigma^2_{\mathrm{SIF}} + \left(0.1 + 10^{4.407 - m_{\mathrm{SIF}} / 3} \right)^2}.
\end{equation}
This complex equation is an empirical relation introduced to account for systematic scatter between SIF CF and HST magnitudes at the bright end due to differences in {\it Gaia} G and HST F606W bands and reduced reliability for HST magnitudes at the brighter end due to saturation.

With this match definition, we find HST matches for 90.6\,\% of the 83\,245 SIF CF sources present in the selected area, suggesting that the vast majority of the SIF CF sources are reliable. This can be seen in Fig. \ref{fig:bellini} left panel, comparing those SIF CF sources matched to HST (dashed dark blue) to all SIF CF sources (solid dark blue line). 

In the 9.4\,\% unmatched SIF CF sources (dash-dotted dark blue), there are 1.8\,\% of sources brighter than 16 mag. Actually most of the SIF CF sources brighter than $16^m$ do not have an HST counterpart. The reason for this is simply that the Bellini data do not completely cover brighter sources. This can be seen in the abrupt cut-off of HST sources (solid teal) in Fig. \ref{fig:bellini} middle panel. It is reflected in the right panel of Fig. \ref{fig:bellini} as well, in the dash-dotted dark blue line that shows the fraction of SIF CF sources not matched to HST sources versus estimated G magnitude. 

For SIF CF sources fainter than $16^m$, the number of sources without HST counterpart (dash-dotted dark blue), and thus the number of possibly spurious sources, amounts to 6\,368. Fig. \ref{fig:bellini} right panel shows, that this is 7.6\,\%  (dash-dotted dark blue) of all SIF CF sources (solid dark blue, left panel) only. 
Some of these are likely blends: We call a SIF CF source blended, if the criteria from Eq. (\ref{ea:match}) holds for the combined flux of all HST sources within $0\farcs2$ from it. This is the case for 2\,400 of the unmatched SIF CF sources. So 38\,\% of the unmatched SIF CF sources fainter than $16^m$ seem to be blends of several HST sources.

Looking at HST sources that could not be matched to a SIF CF source (dash-dotted teal), we find 63\,706 HST sources brighter than mag 21.5 that SIF CF missed. These are 46\,\% of all HST sources in that area. In Fig. \ref{fig:bellini} middle panel, we see that the number of unmatched HST sources increases gradually from 16th magnitude onward. The right panel gives the fraction of unmatched HST sources (dash-dotted teal) as magnitude distribution. It reveals the magnitudes for which SIF CF is missing sources. At mag 20.4, half of all sources present in HST are missed by SIF CF. Looking at the depth of SIF CF data close to the centre of the cluster in Fig. \ref{fig:depth}, this is immediately explained.

However, due to the fact that SIF CF images have a lower resolution than HST, there is a considerable number of unmatched HST sources that SIF CF is simply not able to detect. To estimate this number, we consider as unresolvable all HST sources for which no SIF match exists and a brighter HST source exists within 200 mas. These 24\,019 unresolvable sources constitute 38\,\% of the unmatched HST sources. Fig. \ref{fig:bellini} middle panel shows the resolvable HST sources (dotted). The right panel provides the fraction of unmatched HST sources (teal dash-dotted) and resolvable HST sources (teal dotted) from all HST sources versus estimated G magnitude.

We note that the fact that the fraction of unmatched SIF CF sources that appear to be blends and the fraction of unmatched HST sources that are unresolvable both being approximately 38\% is merely a coincidence.

We stress here that a 200 mas separation corresponds to 50\,\% completeness of SIF catalogue, so there are many more unresolved pairs with larger separations. Our estimate of 38\,\% of the unmatched HST sources actually not being resolvable by SIF CF is therefore a conservative lower limit.

\subsection{Known issues} \label{sec:issues}

\begin{figure*}[t!]
    \centering
    \includegraphics[width=0.49\linewidth]{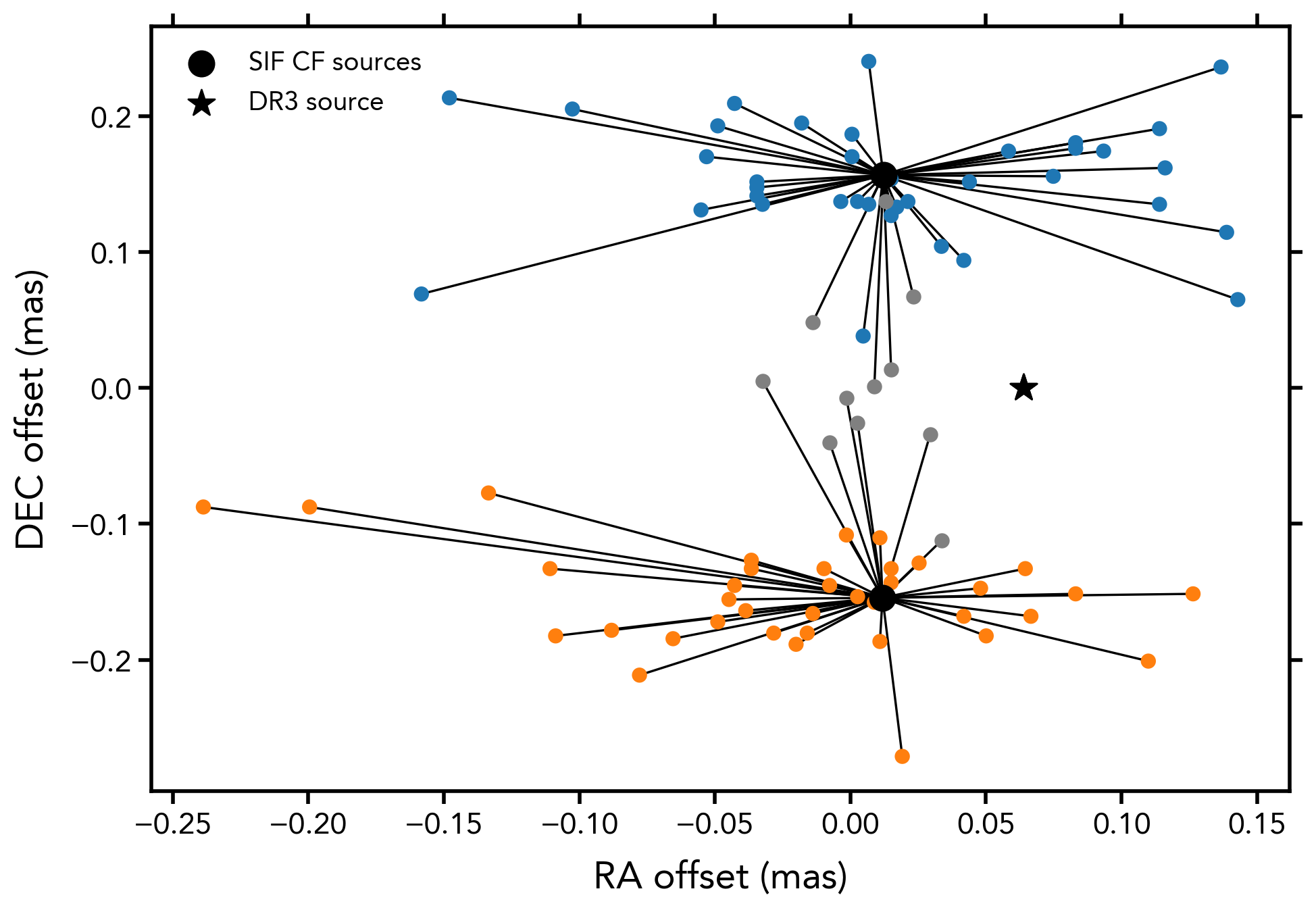}
    \includegraphics[width=0.49\linewidth]{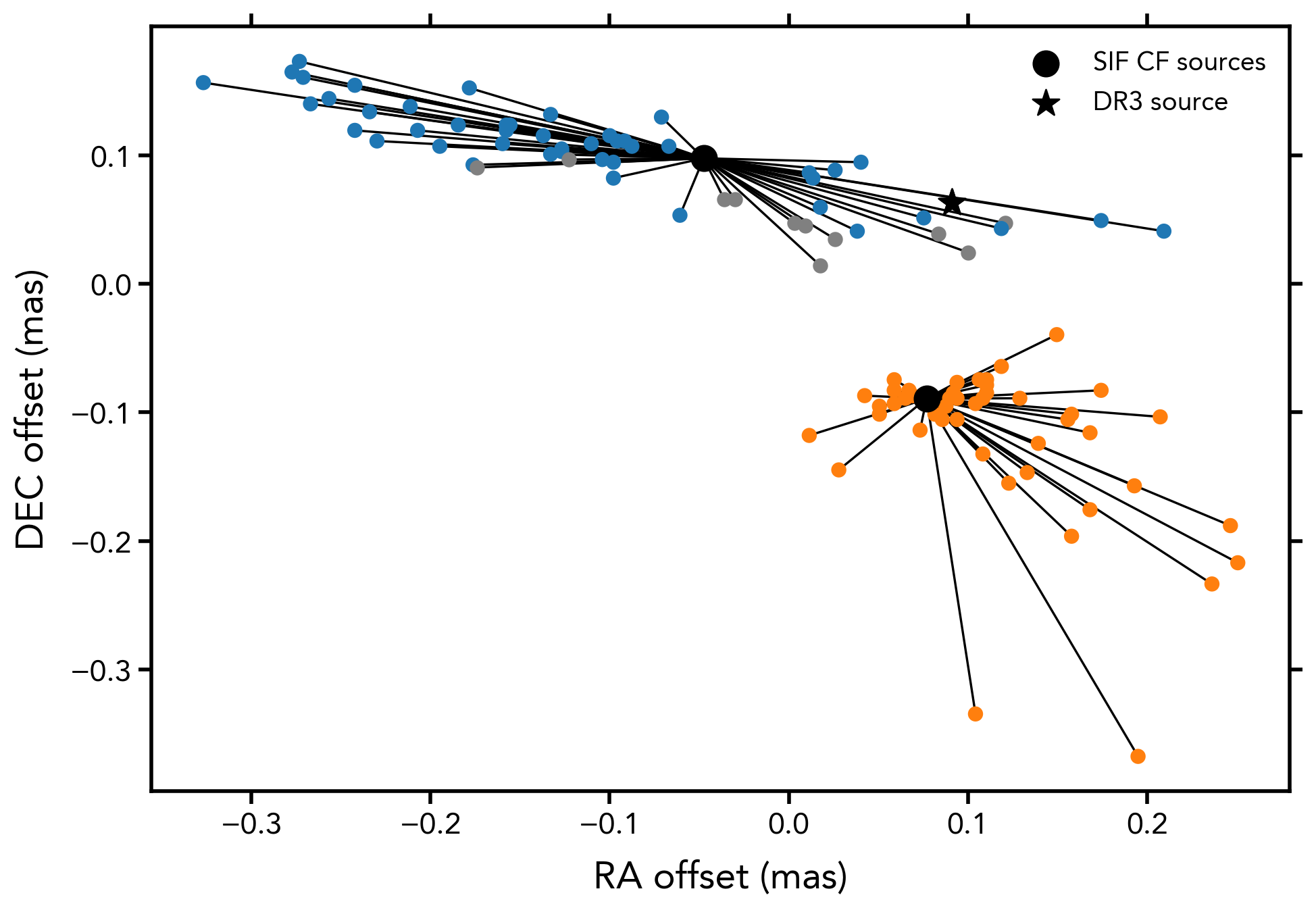}
    \includegraphics[width=0.49\linewidth]{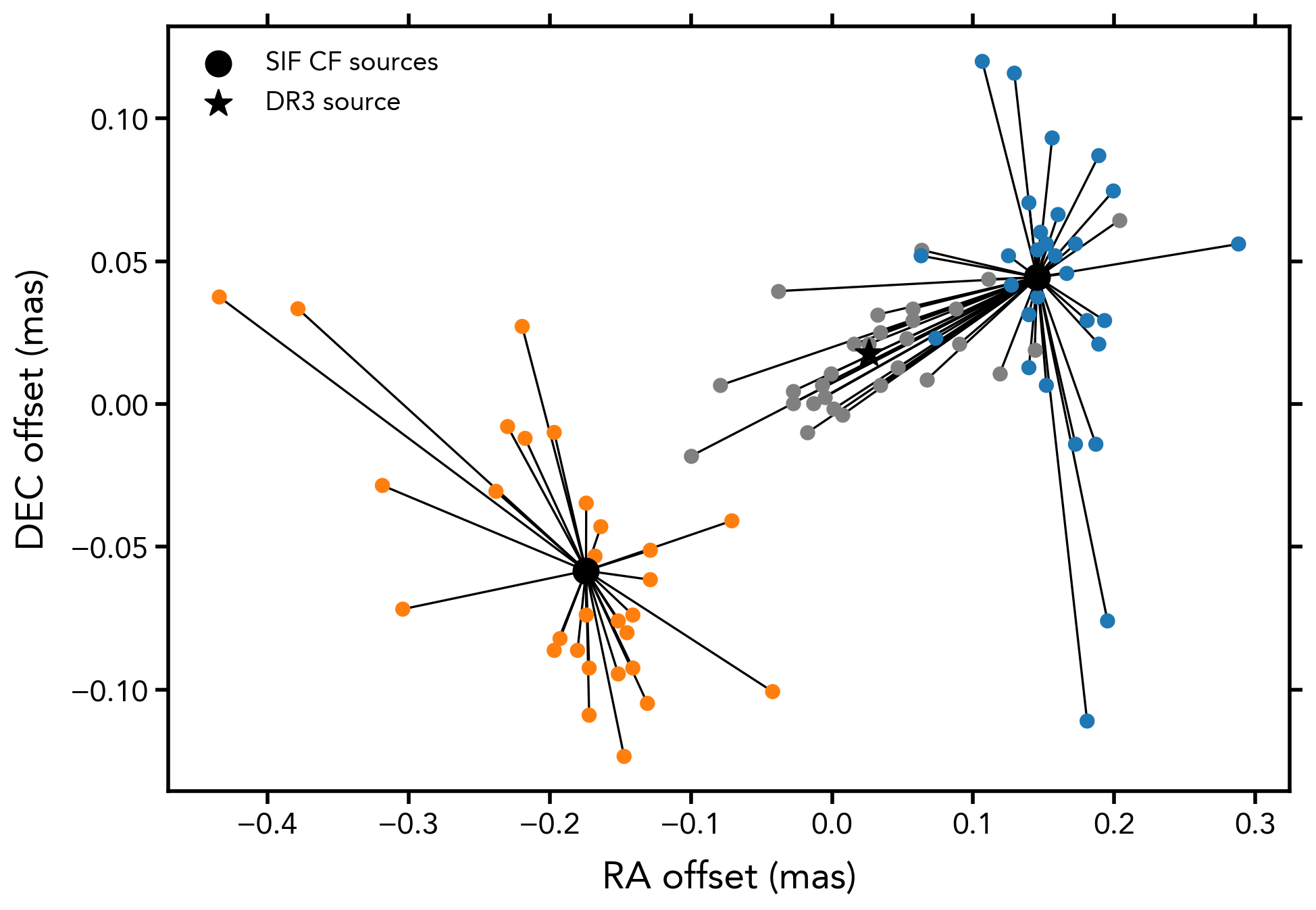}
    \includegraphics[width=0.49\linewidth]{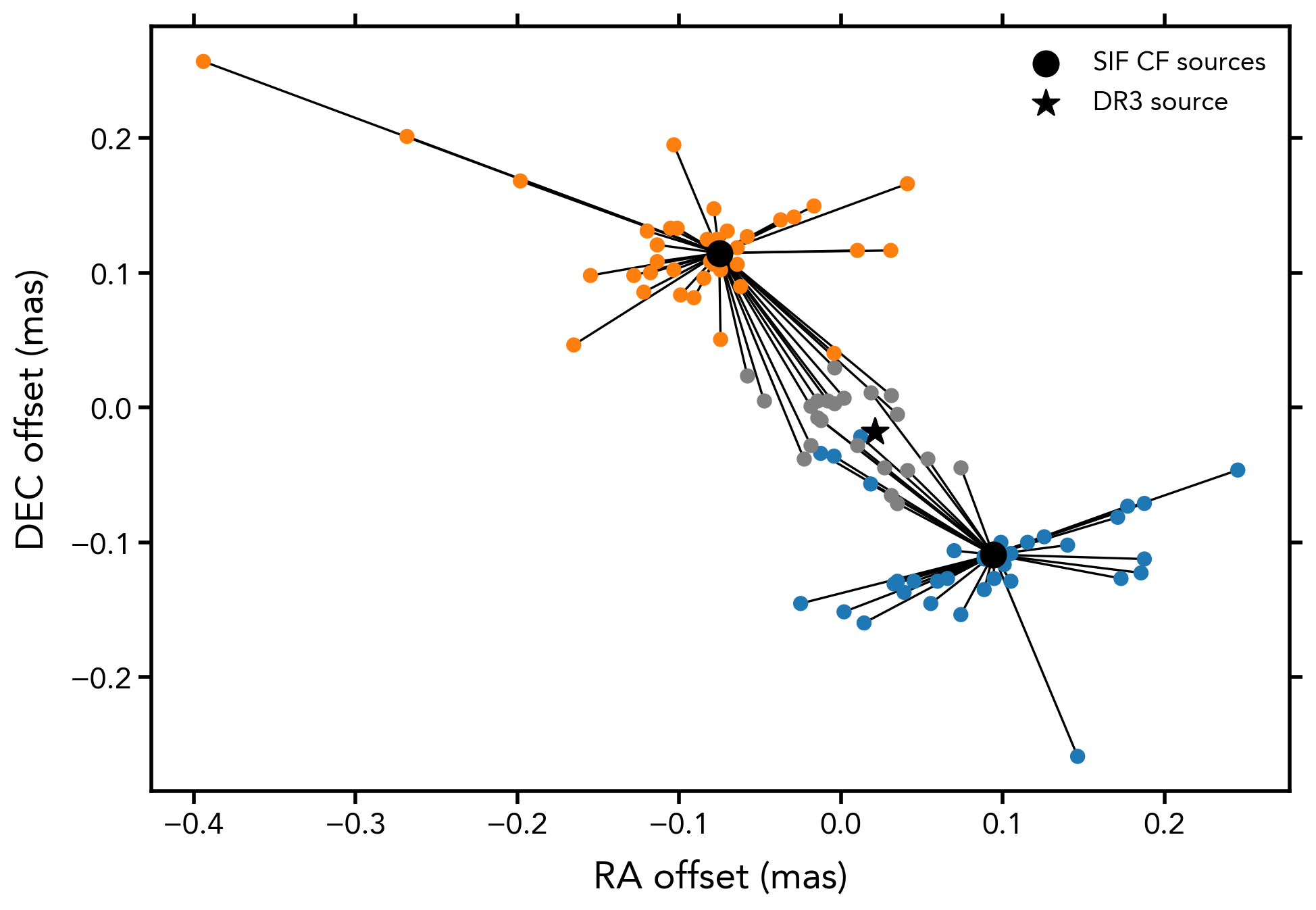}               
        \caption{Four examples of blends in {\it Gaia} DR3 that are de-blended in SIF CF FPR. SIF CF detections (small dots), colour-coded by sourceId for resolved (blue, orange) and blended, unresolved (grey) detections connected to their corresponding SIF CF and rDR3 source (black dots), together with the blended {\it Gaia} DR3 source (black star).} 
    \label{fig:2sources}
\end{figure*}

Even though newly created SIF CF sources are removed from the sample when they match an existing {\it Gaia} DR3 source (see Sect. \ref{sec:xm}), a few sources from the SIF CF FPR catalogue do match a nominal {\it Gaia} DR3 source. When using a positional offset of less than 160 mas as the matching criterion, this is the case for 379 of the 526\,587 published sources. A list of these matches is provided with the paper\footnote{The list of remaining duplicates between the SIF CF and {\it Gaia} DR3 catalogue is available in electronic form at the CDS via anonymous ftp to cdsarc.cds.unistra.fr (130.79.128.5) or via https://cdsarc.cds.unistra.fr/cgi-bin/qcat?J/A+A/} . 

At first sight, this might look like an error, but that is not the case. We first clarify why these sources got created, even though they are located so close to a matching DR3 source, and then explain why they were not filtered, even though sources within a 160 mas distance from brighter sources were removed in the filtering process. 

The reason for the creation of these SIF CF sources in the majority of cases is that SIF CF observed two sources, where in DR3 there was only one: The SIF CF XM module found two clusters of detections and assigned two sources to them (see Fig. \ref{fig:2sources}).
One of these sources, called rDR3 for recreated DR3, is matched to the existing {\it Gaia} DR3 source and thus removed from the output list. The other source, called SIF here, is created as a new SIF CF source and published in this FPR.

In most cases rDR3 is close in magnitude to DR3, while SIF is fainter as expected.  In a handful of cases it is the opposite though, and the kept and disregarded sources should ideally have been swapped.

The position of the unpublished rDR3 source differs slightly from that of the published DR3 source.
During the filtering process, new SIF CF sources are removed, in case their distance to a brighter neighbour is less than $0\farcs16$ (see Sect. \ref{sec:filters}). Here the SIF sources are compared to other new SIF CF sources and rDR3 sources. As the position of a removed rDR3 source and a known DR3 source is not the same, new SIF CF sources can end up within a $0\farcs16$ radius of a DR3 source. This happened in 347 cases.

In the remaining 32 cases the DR3 sources were of too low quality and therefore not included in the SIF CF XM input list. This was revealed in the validation step only and is reported here for clarity.
 
In conclusion, all matches are well explained. While 32 of them are due to low quality {\it Gaia} DR3 input, the remaining 347 matches are actually resolved blends. Furthermore, these examples demonstrate that in some cases SIF CF indeed outperforms the nominal processing in deblending sources.

\section{Archive table}\label{sec:output}

The new SIF CF sources are provided via the {\it Gaia} data archive in an archive table named\linktotablefpr{crowded\_field\_source}. This new table resembles a shortened version of the standard\linktotable{gaia\_source}, featuring the astrometric and photometric data for new sources found by {\it Gaia}'s dedicated SIF CF pipeline in the $\omega$ Cen region. Those SIF CF sources that were matched to an existing {\it Gaia} DR3 source were excluded from publication. 

Apart from the standard fields known from the\linktotable{gaia\_source} table,\linktotablefpr{crowded\_field\_source} contains two additional new fields: 
\begin{itemize}
\item\linktoparamfpr{crowded\_field\_source}{region\_name} identifies the SIF CF region. In this FPR, all published data stem from the same SIF CF region, so all sources will have the same value 'Omega Centauri'.
\item\linktoparamfpr{crowded\_field\_source}{n\_scans} gives the number of scans for the source location on the sky, which is the number of times that the location of this source was contained in an SIF CF image. 
\end{itemize}

Apart from\linktoparamfpr{crowded_field_source}{n\_scans}, the number of potential observations, the table also contains the nominal field\linktoparamfpr{crowded_field_source}{matched\_transits}, which provides the number of actual detections assigned to this source. The division of these two produces the observation fraction, that was used as filter criterion (see Sect. \ref{sec:filters} and Fig. \ref{fig:nobs}). 

For information about all other fields, we refer to the online documentation of the SIF CF {\it Gaia} archive table\linktotablefpr{crowded\_field\_source} as well as the description of the nominal\linktotable{gaia\_source} table, see Section 20.1.1 in \citet{2022gdr3.reptE..20H}. 

The SIF CF FPR data can be downloaded from all known {\it Gaia} archive mirror sites. 
An example query showing how to obtain both {\it Gaia} DR3 and SIF CF sources in $\omega$ Cen can be found in Appendix \ref{sec:query}. Along with this paper we also publish the python code to reproduce most of the plots presented here.\footnote{The python code for most of the plots in this paper is available at \url{https://gitlab.aip.de/dpac_gaia/fpr_plots/}}.

\section{Conclusion}\label{sec:conclusion}

In the $\omega$ Cen region, {\it Gaia}'s new SIF CF pipeline produced 66\,660\,921 source detections, which were matched to 848\,988 sources (Sect. \ref{sec:numbers}). Of these, 321\,698 sources were already present in {\it Gaia} DR3 and 703 sources were filtered for other reasons (see Sect. \ref{sec:filters}). The remaining 526\,587 sources are published in this Focused Product Release. The SIF CF astrometry has been calculated using the nominal AGIS software. The photometry is provided in the photometric system of {\it Gaia} DR3. 

The comparison to the HST \citet{Bellini2017} data shows that in the dense cluster centre most of the SIF CF sources, at least 90.6\,\%, can be matched to an HST source and can thus be regarded as real (Sect. \ref{sec:completeness}, Fig. \ref{fig:bellini}). From the 9.4\,\% SIF CF sources that do not match an HST source, 1.8\,\% are brighter than 16th mag and are simply not included in the HST sample. From the remaining 7.6\,\% unmatched SIF CF sources, 38\,\% correspond to blends that are resolved by HST. 

Regarding the HST sources, we cannot find a SIF CF match for about 46\,\% of them (Sect. \ref{sec:completeness}, Fig. \ref{fig:bellini}). Most of these unmatched HST sources are faint sources. A large fraction (38\,\%) of the unmatched HST sources lie in the vicinity of a brighter source. It is therefore unlikely that these sources could be detected in the SIF CF data.

For obvious reasons, the SIF CF uncertainties for position, parallax, and proper motion are significantly higher than {\it Gaia} DR3 uncertainties (Sects. \ref{sec:position}, \ref{sec:propermotion}, and \ref{sec:parallax}). We report a median positional uncertainty of 3.03 mas and 2.69 mas in RA/Dec, a median parallax uncertainty of 3.95 mas, and a median proper motion uncertainty of 2.02 mas/year in RA and 2.06 mas/year in Dec. Additionally, a strong correlation is reported between proper motion in RA and proper motion in Dec (Sect. \ref{sec:propermotion}).

For stars separated by $\sim$ 200 mas, the published SIF CF FPR data are 50\,\% complete (Fig. \ref{fig:nn}). The HST \citet{Bellini2017} data reveal how many sources are still missing in {\it Gaia} SIF CF: at mag 20.4, half of the HST sources are missed in SIF CF  (Fig. \ref{fig:bellini}, right panel).

Most new SIF CF sources are added in the very dense region of the cluster core. Starting at a radius of 12\arcmin\ from the centre of $\omega$ Cen, SIF CF sources contribute to the combined {\it Gaia} DR3 and SIF CF catalogue. Within 9\arcmin\ from the cluster centre, SIF CF sources dominate the combined catalogue. In the cluster core itself, SIF CF contributes ten times as many sources to the combined catalogue as {\it Gaia} DR3 (Fig. \ref{fig:density}, Sect. \ref{sec:depth}). Here, with respect to {\it Gaia} DR3 alone SIF CF sources increase the completeness of the combined catalogue by more than three magnitudes (Fig. \ref{fig:depth}, Sect. \ref{sec:depth}).

\begin{acknowledgements} 
This work made use of data from the European Space Agency (ESA) mission {\it Gaia} (\url{https://www.cosmos.esa.int/gaia}), processed by the {\it Gaia} Data Processing and Analysis Consortium (DPAC, \url{https://www.cosmos.esa.int/web/gaia/dpac/consortium}). Funding for the DPAC has been provided by national institutions, in particular the institutions participating in the {\it Gaia} Multilateral Agreement. All of the authors are current or past members of the ESA {\it Gaia} mission team and of the {\it Gaia} DPAC, and their work has been supported by 
the German Aerospace Agency (Deutsches Zentrum fur Luft- und Raumfahrt e.V., DLR) grant number 50QG1904 and 50QG1404;
the Spanish Ministry of Economy (MINECO/FEDER, UE) through grants ESP2016-80079-C2-1-R, RTI2018-095076-B-C21 and the Institute of Cosmos Sciences University of Barcelona (ICCUB, Unidad de Excelencia `Mar\'{\i}a de Maeztu’) through grant MDM-2014-0369 and CEX2019-000918-M; 
the Tenure Track Pilot Programme of the Croatian Science Foundation and the Ecole Polytechnique Fédérale de Lausanne and the Project TTP-2018-07-1171 Mining the variable sky, with the funds of the Croatian-Swiss Research Programme;
the European Research Council (ERC) project INTERCLOUDS, grant agreement no. 68211.

The authors thankfully acknowledge the computer resources from MareNostrum, and the technical expertise and assistance provided by the Red Espa\~nola de Supercomputaci\'on at the Barcelona Supercomputing Center, Centro Nacional de Supercomputaci\'on.
\end{acknowledgements}

%

\bibliographystyle{aa} 
\bibliography{references}

\begin{thebibliography}{60}
\expandafter\ifx\csname natexlab\endcsname\relax\def\natexlab#1{#1}\fi

\bibitem[{{Ahn} {et~al.}(2012){Ahn}, {Alexandroff}, {Allende Prieto},
  {Anderson}, {Anderton}, {Andrews}, {Aubourg}, {Bailey}, {Balbinot}, {Barnes},
  \& et~al.}]{SDSS9}
{Ahn}, C.~P., {Alexandroff}, R., {Allende Prieto}, C., {et~al.} 2012, \apjs,
  203, 21

\bibitem[{{Albareti} {et~al.}(2017){Albareti}, {Allende Prieto}, {Almeida},
  {Anders}, {Anderson}, {Andrews}, {Arag{\'o}n-Salamanca},
  {Argudo-Fern{\'a}ndez}, {Armengaud}, {Aubourg}, {Avila-Reese}, {Badenes},
  {Bailey}, {Barbuy}, {Barger}, {Barrera-Ballesteros}, {Bartosz}, {Basu},
  {Bates}, {Battaglia}, {Baumgarten}, {Baur}, {Bautista}, {Beers}, {Belfiore},
  {Bershady}, {Bertran de Lis}, {Bird}, {Bizyaev}, {Blanc}, {Blanton},
  {Blomqvist}, {Bolton}, {Borissova}, {Bovy}, {Brand t}, {Brinkmann},
  {Brownstein}, {Bundy}, {Burtin}, {Busca}, {Orlando Camacho Chavez}, {Cano
  D{\'\i}az}, {Cappellari}, {Carrera}, {Chen}, {Cherinka}, {Cheung},
  {Chiappini}, {Chojnowski}, {Chuang}, {Chung}, {Cirolini}, {Clerc}, {Cohen},
  {Comerford}, {Comparat}, {Correa do Nascimento}, {Cousinou}, {Covey},
  {Crane}, {Croft}, {Cunha}, {Darling}, {Davidson}, {Dawson}, {Da Costa}, {Da
  Silva Ilha}, {Deconto Machado}, {Delubac}, {De Lee}, {De la Macorra}, {De la
  Torre}, {Diamond-Stanic}, {Donor}, {Downes}, {Drory}, {Du}, {Du Mas des
  Bourboux}, {Dwelly}, {Ebelke}, {Eigenbrot}, {Eisenstein}, {Elsworth},
  {Emsellem}, {Eracleous}, {Escoffier}, {Evans}, {Falc{\'o}n-Barroso}, {Fan},
  {Favole}, {Fernandez-Alvar}, {Fernand ez-Trincado}, {Feuillet}, {Fleming},
  {Font-Ribera}, {Freischlad}, {Frinchaboy}, {Fu}, {Gao}, {Garcia},
  {Garcia-Dias}, {Garcia-Hern{\'a}ndez}, {Garcia P{\'e}rez}, {Gaulme}, {Ge},
  {Geisler}, {Gillespie}, {Gil Marin}, {Girardi}, {Goddard}, {Gomez Maqueo
  Chew}, {Gonzalez-Perez}, {Grabowski}, {Green}, {Grier}, {Grier}, {Guo},
  {Guy}, {Hagen}, {Hall}, {Harding}, {Harley}, {Hasselquist}, {Hawley},
  {Hayes}, {Hearty}, {Hekker}, {Hernandez Toledo}, {Ho}, {Hogg},
  {Holley-Bockelmann}, {Holtzman}, {Holzer}, {Hu}, {Huber}, {Hutchinson},
  {Hwang}, {Ibarra-Medel}, {Ivans}, {Ivory}, {Jaehnig}, {Jensen}, {Johnson},
  {Jones}, {Jullo}, {Kallinger}, {Kinemuchi}, {Kirkby}, {Klaene}, {Kneib},
  {Kollmeier}, {Lacerna}, {Lane}, {Lang}, {Laurent}, {Law}, {Leauthaud}, {Le
  Goff}, {Li}, {Li}, {Li}, {Li}, {Liang}, {Liang}, {Lima}, {Lin}, {Lin}, {Lin},
  {Liu}, {Long}, {Lucatello}, {MacDonald}, {MacLeod}, {Mackereth}, {Mahadevan},
  {Maia}, {Maiolino}, {Majewski}, {Malanushenko}, {Malanushenko}, {Mallmann},
  {Manchado}, {Maraston}, {Marques-Chaves}, {Martinez Valpuesta}, {Masters},
  {Mathur}, {McGreer}, {Merloni}, {Merrifield}, {M{\'e}sz{\'a}ros}, {Meza},
  {Miglio}, {Minchev}, {Molaverdikhani}, {Montero-Dorta}, {Mosser}, {Muna},
  {Myers}, {Nair}, {Nandra}, {Ness}, {Newman}, {Nichol}, {Nidever},
  {Nitschelm}, {O'Connell}, {Oravetz}, {Oravetz}, {Pace}, {Padilla},
  {Palanque-Delabrouille}, {Pan}, {Parejko}, {Paris}, {Park}, {Peacock},
  {Peirani}, {Pellejero-Ibanez}, {Penny}, {Percival}, {Percival},
  {Perez-Fournon}, {Petitjean}, {Pieri}, {Pinsonneault}, {Pisani}, {Prada},
  {Prakash}, {Price-Jones}, {Raddick}, {Rahman}, {Raichoor}, {Barboza Rembold},
  {Reyna}, {Rich}, {Richstein}, {Ridl}, {Riffel}, {Riffel}, {Rix}, {Robin},
  {Rockosi}, {Rodr{\'\i}guez-Torres}, {Rodrigues}, {Roe}, {Roman Lopes},
  {Rom{\'a}n-Z{\'u}{\~n}iga}, {Ross}, {Rossi}, {Ruan}, {Ruggeri}, {Runnoe},
  {Salazar-Albornoz}, {Salvato}, {Sanchez}, {Sanchez}, {Sanchez-Gallego},
  {Santiago}, {Schiavon}, {Schimoia}, {Schlafly}, {Schlegel}, {Schneider},
  {Sch{\"o}nrich}, {Schultheis}, {Schwope}, {Seo}, {Serenelli}, {Sesar},
  {Shao}, {Shetrone}, {Shull}, {Silva Aguirre}, {Skrutskie}, {Slosar}, {Smith},
  {Smith}, {Sobeck}, {Somers}, {Souto}, {Stark}, {Stassun}, {Steinmetz},
  {Stello}, {Storchi Bergmann}, {Strauss}, {Streblyanska}, {Stringfellow},
  {Suarez}, {Sun}, {Taghizadeh-Popp}, {Tang}, {Tao}, {Tayar}, {Tembe},
  {Thomas}, {Tinker}, {Tojeiro}, {Tremonti}, {Troup}, {Trump}, {Unda-Sanzana},
  {Valenzuela}, {Van den Bosch}, {Vargas-Maga{\~n}a}, {Vazquez}, {Villanova},
  {Vivek}, {Vogt}, {Wake}, {Walterbos}, {Wang}, {Wang}, {Weaver}, {Weijmans},
  {Weinberg}, {Westfall}, {Whelan}, {Wilcots}, {Wild}, {Williams}, {Wilson},
  {Wood-Vasey}, {Wylezalek}, {Xiao}, {Yan}, {Yang}, {Ybarra}, {Yeche}, {Yuan},
  {Zakamska}, {Zamora}, {Zasowski}, {Zhang}, {Zhao}, {Zhao}, {Zheng}, {Zheng},
  {Zhou}, {Zhu}, {Zinn}, \& {Zou}}]{2017ApJS..233...25A}
{Albareti}, F.~D., {Allende Prieto}, C., {Almeida}, A., {et~al.} 2017, \apjs,
  233, 25

\bibitem[{{Astropy Collaboration} {et~al.}(2018){Astropy Collaboration},
  {Price-Whelan}, {Sip{\H o}cz}, {G{\"u}nther}, {Lim}, {Crawford}, {Conseil},
  {Shupe}, {Craig}, {Dencheva}, {Ginsburg}, {VanderPlas}, {Bradley},
  {P{\'e}rez-Su{\'a}rez}, {de Val-Borro}, {Aldcroft}, {Cruz}, {Robitaille},
  {Tollerud}, {Ardelean}, {Babej}, {Bach}, {Bachetti}, {Bakanov}, {Bamford},
  {Barentsen}, {Barmby}, {Baumbach}, {Berry}, {Biscani}, {Boquien}, {Bostroem},
  {Bouma}, {Brammer}, {Bray}, {Breytenbach}, {Buddelmeijer}, {Burke},
  {Calderone}, {Cano Rodr{\'{\i}}guez}, {Cara}, {Cardoso}, {Cheedella},
  {Copin}, {Corrales}, {Crichton}, {D'Avella}, {Deil}, {Depagne}, {Dietrich},
  {Donath}, {Droettboom}, {Earl}, {Erben}, {Fabbro}, {Ferreira}, {Finethy},
  {Fox}, {Garrison}, {Gibbons}, {Goldstein}, {Gommers}, {Greco}, {Greenfield},
  {Groener}, {Grollier}, {Hagen}, {Hirst}, {Homeier}, {Horton}, {Hosseinzadeh},
  {Hu}, {Hunkeler}, {Ivezi{\'c}}, {Jain}, {Jenness}, {Kanarek}, {Kendrew},
  {Kern}, {Kerzendorf}, {Khvalko}, {King}, {Kirkby}, {Kulkarni}, {Kumar},
  {Lee}, {Lenz}, {Littlefair}, {Ma}, {Macleod}, {Mastropietro}, {McCully},
  {Montagnac}, {Morris}, {Mueller}, {Mumford}, {Muna}, {Murphy}, {Nelson},
  {Nguyen}, {Ninan}, {N{\"o}the}, {Ogaz}, {Oh}, {Parejko}, {Parley}, {Pascual},
  {Patil}, {Patil}, {Plunkett}, {Prochaska}, {Rastogi}, {Reddy Janga},
  {Sabater}, {Sakurikar}, {Seifert}, {Sherbert}, {Sherwood-Taylor}, {Shih},
  {Sick}, {Silbiger}, {Singanamalla}, {Singer}, {Sladen}, {Sooley},
  {Sornarajah}, {Streicher}, {Teuben}, {Thomas}, {Tremblay}, {Turner},
  {Terr{\'o}n}, {van Kerkwijk}, {de la Vega}, {Watkins}, {Weaver}, {Whitmore},
  {Woillez}, {Zabalza}, \& {Astropy Contributors}}]{2018AJ....156..123A}
{Astropy Collaboration}, {Price-Whelan}, A., {Sip{\H o}cz}, B.~M., {et~al.}
  2018, \aj, 156, 123

\bibitem[{{Bellini} {et~al.}(2017){Bellini}, {Anderson}, {Bedin}, {King}, {van
  der Marel}, {Piotto}, \& {Cool}}]{Bellini2017}
{Bellini}, A., {Anderson}, J., {Bedin}, L.~R., {et~al.} 2017, \apj, 842, 6

\bibitem[{{Bertin} \& {Arnouts}(1996)}]{1996A&AS..117..393B}
{Bertin}, E. \& {Arnouts}, S. 1996, \aaps, 117, 393

\bibitem[{{Boch} \& {Fernique}(2014)}]{2014ASPC..485..277B}
{Boch}, T. \& {Fernique}, P. 2014, in Astronomical Society of the Pacific
  Conference Series, Vol. 485, Astronomical Data Analysis Software and Systems
  XXIII, ed. N.~{Manset} \& P.~{Forshay}, 277

\bibitem[{{Bonnarel} {et~al.}(2000){Bonnarel}, {Fernique}, {Bienaym{\'e}},
  {Egret}, {Genova}, {Louys}, {Ochsenbein}, {Wenger}, \&
  {Bartlett}}]{2000A&AS..143...33B}
{Bonnarel}, F., {Fernique}, P., {Bienaym{\'e}}, O., {et~al.} 2000, \aaps, 143,
  33

\bibitem[{{Breddels} \& {Veljanoski}(2018)}]{2018A&A...618A..13B}
{Breddels}, M.~A. \& {Veljanoski}, J. 2018, \aap, 618, A13

\bibitem[{{Casta{\~n}eda} {et~al.}(2015){Casta{\~n}eda}, {Fabricius}, {Torra},
  {Clotet}, {Gonz{\'a}lez}, {Garralda}, \& {Portell}}]{2015hsa8.conf..792C}
{Casta{\~n}eda}, J., {Fabricius}, C., {Torra}, J., {et~al.} 2015, in Highlights
  of Spanish Astrophysics VIII, 792--797

\bibitem[{{Casta{\~n}eda} {et~al.}(2022){Casta{\~n}eda}, {Hobbs}, {Fabricius},
  {Davidson}, {Rowell}, {Lindegren}, {Hambly}, {Bastian}, {Portell}, {Torra},
  {Clotet}, {Smart}, {Mora}, {Biermann}, {L{\"o}ffler}, {Brown}, {Busonero}, \&
  {Riva}}]{2022gdr3.reptE...3C}
{Casta{\~n}eda}, J., {Hobbs}, D., {Fabricius}, C., {et~al.} 2022, {Gaia DR3
  documentation Chapter 3: Pre-processing}, Gaia DR3 documentation, European
  Space Agency; Gaia Data Processing and Analysis Consortium. Online at <A
  href=``https:// gea.esac.esa.int/archive/documentation/GDR3/index.html''>
  https://gea.esac.esa.int/archive/documentation/GDR3/index.html </A>, id. 3

\bibitem[{{Chambers} {et~al.}(2016){Chambers}, {Magnier}, {Metcalfe},
  {Flewelling}, {Huber}, {Waters}, {Denneau}, {Draper}, {Farrow}, {Finkbeiner},
  {Holmberg}, {Koppenhoefer}, {Price}, {Saglia}, {Schlafly}, {Smartt},
  {Sweeney}, {Wainscoat}, {Burgett}, {Grav}, {Heasley}, {Hodapp}, {Jedicke},
  {Kaiser}, {Kudritzki}, {Luppino}, {Lupton}, {Monet}, {Morgan}, {Onaka},
  {Stubbs}, {Tonry}, {Banados}, {Bell}, {Bender}, {Bernard}, {Botticella},
  {Casertano}, {Chastel}, {Chen}, {Chen}, {Cole}, {Deacon}, {Frenk},
  {Fitzsimmons}, {Gezari}, {Goessl}, {Goggia}, {Goldman}, {Grebel}, {Hambly},
  {Hasinger}, {Heavens}, {Heckman}, {Henderson}, {Henning}, {Holman}, {Hopp},
  {Ip}, {Isani}, {Keyes}, {Koekemoer}, {Kotak}, {Long}, {Lucey}, {Liu},
  {Martin}, {McLean}, {Morganson}, {Murphy}, {Nieto-Santisteban}, {Norberg},
  {Peacock}, {Pier}, {Postman}, {Primak}, {Rae}, {Rest}, {Riess}, {Riffeser},
  {Rix}, {Roser}, {Schilbach}, {Schultz}, {Scolnic}, {Szalay}, {Seitz},
  {Shiao}, {Small}, {Smith}, {Soderblom}, {Taylor}, {Thakar}, {Thiel},
  {Thilker}, {Urata}, {Valenti}, {Walter}, {Watters}, {Werner}, {White},
  {Wood-Vasey}, \& {Wyse}}]{panstarrs1}
{Chambers}, K.~C., {Magnier}, E.~A., {Metcalfe}, N., {et~al.} 2016, ArXiv
  e-prints [\eprint[arXiv]{1612.05560}]

\bibitem[{{Cordoni} {et~al.}(2020){Cordoni}, {Milone}, {Marino}, {Da Costa},
  {Dondoglio}, {Jerjen}, {Lagioia}, {Mastrobuono-Battisti}, {Norris}, {Tailo},
  \& {Yong}}]{2020ApJ...898..147C}
{Cordoni}, G., {Milone}, A.~P., {Marino}, A.~F., {et~al.} 2020, \apj, 898, 147

\bibitem[{{de Bruijne} {et~al.}(2022){de Bruijne}, {Babusiaux}, {Brown},
  {Casta{\~n}eda}, {Cheek}, {Crowley}, {De Angeli}, {Drimmel}, {Fabricius},
  {Mart{\'\i}n-Fleitas}, {Gracia-Abril}, {Guerra}, {Hutton}, {Messineo},
  {Mora}, {Nicolas}, {Nienartowicz}, {Panem}, \&
  {Portell}}]{2022gdr3.reptE...1D}
{de Bruijne}, J., {Babusiaux}, C., {Brown}, A., {et~al.} 2022, {Gaia DR3
  documentation Chapter 1: Introduction}, Gaia DR3 documentation, European
  Space Agency; Gaia Data Processing and Analysis Consortium. Online at <A
  href=``https:// gea.esac.esa.int/archive/documentation/GDR3/index.html''>
  https://gea.esac.esa.int/archive/documentation/GDR3/index.html </A>, id. 1

\bibitem[{{Evans} {et~al.}(2022){Evans}, {Strigari}, \&
  {Zivick}}]{2022MNRAS.511.4251E}
{Evans}, A.~J., {Strigari}, L.~E., \& {Zivick}, P. 2022, \mnras, 511, 4251

\bibitem[{{Fabricius} {et~al.}(2002){Fabricius}, {H{\o}g}, {Makarov}, {Mason},
  {Wycoff}, \& {Urban}}]{2002A&A...384..180F}
{Fabricius}, C., {H{\o}g}, E., {Makarov}, V.~V., {et~al.} 2002, \aap, 384, 180

\bibitem[{{Fabricius} {et~al.}(2021){Fabricius}, {Luri}, {Arenou}, {Babusiaux},
  {Helmi}, {Muraveva}, {Reyl{\'e}}, {Spoto}, {Vallenari}, {Antoja}, {Balbinot},
  {Barache}, {Bauchet}, {Bragaglia}, {Busonero}, {Cantat-Gaudin}, {Carrasco},
  {Diakit{\'e}}, {Fabrizio}, {Figueras}, {Garcia-Gutierrez}, {Garofalo},
  {Jordi}, {Kervella}, {Khanna}, {Leclerc}, {Licata}, {Lambert}, {Marrese},
  {Masip}, {Ramos}, {Robichon}, {Robin}, {Romero-G{\'o}mez}, {Rubele}, \&
  {Weiler}}]{2021A&A...649A...5F}
{Fabricius}, C., {Luri}, X., {Arenou}, F., {et~al.} 2021, \aap, 649, A5

\bibitem[{{Flewelling} {et~al.}(2020){Flewelling}, {Magnier}, {Chambers},
  {Heasley}, {Holmberg}, {Huber}, {Sweeney}, {Waters}, {Calamida}, {Casertano},
  {Chen}, {Farrow}, {Hasinger}, {Henderson}, {Long}, {Metcalfe}, {Narayan},
  {Nieto-Santisteban}, {Norberg}, {Rest}, {Saglia}, {Szalay}, {Thakar},
  {Tonry}, {Valenti}, {Werner}, {White}, {Denneau}, {Draper}, {Hodapp},
  {Jedicke}, {Kaiser}, {Kudritzki}, {Price}, {Wainscoat}, {Chastel}, {McLean},
  {Postman}, \& {Shiao}}]{panstarrs1f}
{Flewelling}, H.~A., {Magnier}, E.~A., {Chambers}, K.~C., {et~al.} 2020, \apjs,
  251, 7

\bibitem[{{Gaia Collaboration} {et~al.}(2021){Gaia Collaboration}, {Brown},
  {Vallenari}, {Prusti}, {de Bruijne}, {Babusiaux}, {Biermann}, {Creevey},
  {Evans}, {Eyer}, {Hutton}, {Jansen}, {Jordi}, {Klioner}, {Lammers},
  {Lindegren}, {Luri}, {Mignard}, {Panem}, {Pourbaix}, {Randich}, {Sartoretti},
  {Soubiran}, {Walton}, {Arenou}, {Bailer-Jones}, {Bastian}, {Cropper},
  {Drimmel}, {Katz}, {Lattanzi}, {van Leeuwen}, {Bakker}, {Cacciari},
  {Casta{\~n}eda}, {De Angeli}, {Ducourant}, {Fabricius}, {Fouesneau},
  {Fr{\'e}mat}, {Guerra}, {Guerrier}, {Guiraud}, {Jean-Antoine Piccolo},
  {Masana}, {Messineo}, {Mowlavi}, {Nicolas}, {Nienartowicz}, {Pailler},
  {Panuzzo}, {Riclet}, {Roux}, {Seabroke}, {Sordo}, {Tanga}, {Th{\'e}venin},
  {Gracia-Abril}, {Portell}, {Teyssier}, {Altmann}, {Andrae}, {Bellas-Velidis},
  {Benson}, {Berthier}, {Blomme}, {Brugaletta}, {Burgess}, {Busso}, {Carry},
  {Cellino}, {Cheek}, {Clementini}, {Damerdji}, {Davidson}, {Delchambre},
  {Dell'Oro}, {Fern{\'a}ndez-Hern{\'a}ndez}, {Galluccio}, {Garc{\'\i}a-Lario},
  {Garcia-Reinaldos}, {Gonz{\'a}lez-N{\'u}{\~n}ez}, {Gosset}, {Haigron},
  {Halbwachs}, {Hambly}, {Harrison}, {Hatzidimitriou}, {Heiter},
  {Hern{\'a}ndez}, {Hestroffer}, {Hodgkin}, {Holl}, {Jan{\ss}en}, {Jevardat de
  Fombelle}, {Jordan}, {Krone-Martins}, {Lanzafame}, {L{\"o}ffler}, {Lorca},
  {Manteiga}, {Marchal}, {Marrese}, {Moitinho}, {Mora}, {Muinonen}, {Osborne},
  {Pancino}, {Pauwels}, {Petit}, {Recio-Blanco}, {Richards}, {Riello},
  {Rimoldini}, {Robin}, {Roegiers}, {Rybizki}, {Sarro}, {Siopis}, {Smith},
  {Sozzetti}, {Ulla}, {Utrilla}, {van Leeuwen}, {van Reeven}, {Abbas}, {Abreu
  Aramburu}, {Accart}, {Aerts}, {Aguado}, {Ajaj}, {Altavilla}, {{\'A}lvarez},
  {{\'A}lvarez Cid-Fuentes}, {Alves}, {Anderson}, {Anglada Varela}, {Antoja},
  {Audard}, {Baines}, {Baker}, {Balaguer-N{\'u}{\~n}ez}, {Balbinot}, {Balog},
  {Barache}, {Barbato}, {Barros}, {Barstow}, {Bartolom{\'e}}, {Bassilana},
  {Bauchet}, {Baudesson-Stella}, {Becciani}, {Bellazzini}, {Bernet}, {Bertone},
  {Bianchi}, {Blanco-Cuaresma}, {Boch}, {Bombrun}, {Bossini}, {Bouquillon},
  {Bragaglia}, {Bramante}, {Breedt}, {Bressan}, {Brouillet}, {Bucciarelli},
  {Burlacu}, {Busonero}, {Butkevich}, {Buzzi}, {Caffau}, {Cancelliere},
  {C{\'a}novas}, {Cantat-Gaudin}, {Carballo}, {Carlucci}, {Carnerero},
  {Carrasco}, {Casamiquela}, {Castellani}, {Castro-Ginard}, {Castro Sampol},
  {Chaoul}, {Charlot}, {Chemin}, {Chiavassa}, {Cioni}, {Comoretto}, {Cooper},
  {Cornez}, {Cowell}, {Crifo}, {Crosta}, {Crowley}, {Dafonte}, {Dapergolas},
  {David}, {David}, {de Laverny}, {De Luise}, {De March}, {De Ridder}, {de
  Souza}, {de Teodoro}, {de Torres}, {del Peloso}, {del Pozo}, {Delbo},
  {Delgado}, {Delgado}, {Delisle}, {Di Matteo}, {Diakite}, {Diener},
  {Distefano}, {Dolding}, {Eappachen}, {Edvardsson}, {Enke}, {Esquej}, {Fabre},
  {Fabrizio}, {Faigler}, {Fedorets}, {Fernique}, {Fienga}, {Figueras},
  {Fouron}, {Fragkoudi}, {Fraile}, {Franke}, {Gai}, {Garabato},
  {Garcia-Gutierrez}, {Garc{\'\i}a-Torres}, {Garofalo}, {Gavras}, {Gerlach},
  {Geyer}, {Giacobbe}, {Gilmore}, {Girona}, {Giuffrida}, {Gomel}, {Gomez},
  {Gonzalez-Santamaria}, {Gonz{\'a}lez-Vidal}, {Granvik},
  {Guti{\'e}rrez-S{\'a}nchez}, {Guy}, {Hauser}, {Haywood}, {Helmi}, {Hidalgo},
  {Hilger}, {H{\l}adczuk}, {Hobbs}, {Holland}, {Huckle}, {Jasniewicz},
  {Jonker}, {Juaristi Campillo}, {Julbe}, {Karbevska}, {Kervella}, {Khanna},
  {Kochoska}, {Kontizas}, {Kordopatis}, {Korn}, {Kostrzewa-Rutkowska},
  {Kruszy{\'n}ska}, {Lambert}, {Lanza}, {Lasne}, {Le Campion}, {Le Fustec},
  {Lebreton}, {Lebzelter}, {Leccia}, {Leclerc}, {Lecoeur-Taibi}, {Liao},
  {Licata}, {Lindstr{\o}m}, {Lister}, {Livanou}, {Lobel}, {Madrero Pardo},
  {Managau}, {Mann}, {Marchant}, {Marconi}, {Marcos Santos}, {Marinoni},
  {Marocco}, {Marshall}, {Martin Polo}, {Mart{\'\i}n-Fleitas}, {Masip},
  {Massari}, {Mastrobuono-Battisti}, {Mazeh}, {McMillan}, {Messina},
  {Michalik}, {Millar}, {Mints}, {Molina}, {Molinaro}, {Moln{\'a}r},
  {Montegriffo}, {Mor}, {Morbidelli}, {Morel}, {Morris}, {Mulone}, {Munoz},
  {Muraveva}, {Murphy}, {Musella}, {Noval}, {Ord{\'e}novic}, {Orr{\`u}},
  {Osinde}, {Pagani}, {Pagano}, {Palaversa}, {Palicio}, {Panahi}, {Pawlak},
  {Pe{\~n}alosa Esteller}, {Penttil{\"a}}, {Piersimoni}, {Pineau}, {Plachy},
  {Plum}, {Poggio}, {Poretti}, {Poujoulet}, {Pr{\v{s}}a}, {Pulone}, {Racero},
  {Ragaini}, {Rainer}, {Raiteri}, {Rambaux}, {Ramos}, {Ramos-Lerate}, {Re
  Fiorentin}, {Regibo}, {Reyl{\'e}}, {Ripepi}, {Riva}, {Rixon}, {Robichon},
  {Robin}, {Roelens}, {Rohrbasser}, {Romero-G{\'o}mez}, {Rowell}, {Royer},
  {Rybicki}, {Sadowski}, {Sagrist{\`a} Sell{\'e}s}, {Sahlmann}, {Salgado},
  {Salguero}, {Samaras}, {Sanchez Gimenez}, {Sanna}, {Santove{\~n}a},
  {Sarasso}, {Schultheis}, {Sciacca}, {Segol}, {Segovia}, {S{\'e}gransan},
  {Semeux}, {Shahaf}, {Siddiqui}, {Siebert}, {Siltala}, {Slezak}, {Smart},
  {Solano}, {Solitro}, {Souami}, {Souchay}, {Spagna}, {Spoto}, {Steele},
  {Steidelm{\"u}ller}, {Stephenson}, {S{\"u}veges}, {Szabados}, {Szegedi-Elek},
  {Taris}, {Tauran}, {Taylor}, {Teixeira}, {Thuillot}, {Tonello}, {Torra},
  {Torra}, {Turon}, {Unger}, {Vaillant}, {van Dillen}, {Vanel}, {Vecchiato},
  {Viala}, {Vicente}, {Voutsinas}, {Weiler}, {Wevers}, {Wyrzykowski}, {Yoldas},
  {Yvard}, {Zhao}, {Zorec}, {Zucker}, {Zurbach}, \&
  {Zwitter}}]{2021A&A...650C...3G}
{Gaia Collaboration}, {Brown}, A.~G.~A., {Vallenari}, A., {et~al.} 2021, \aap,
  650, C3

\bibitem[{{Gaia Collaboration} {et~al.}(2016){Gaia Collaboration}, {Prusti},
  {de Bruijne}, {Brown}, {Vallenari}, {Babusiaux}, {Bailer-Jones}, {Bastian},
  {Biermann}, {Evans}, {Eyer}, {Jansen}, {Jordi}, {Klioner}, {Lammers},
  {Lindegren}, {Luri}, {Mignard}, {Milligan}, {Panem}, {Poinsignon},
  {Pourbaix}, {Randich}, {Sarri}, {Sartoretti}, {Siddiqui}, {Soubiran},
  {Valette}, {van Leeuwen}, {Walton}, {Aerts}, {Arenou}, {Cropper}, {Drimmel},
  {H{\o}g}, {Katz}, {Lattanzi}, {O'Mullane}, {Grebel}, {Holland}, {Huc},
  {Passot}, {Bramante}, {Cacciari}, {Casta{\~n}eda}, {Chaoul}, {Cheek}, {De
  Angeli}, {Fabricius}, {Guerra}, {Hern{\'a}ndez}, {Jean-Antoine-Piccolo},
  {Masana}, {Messineo}, {Mowlavi}, {Nienartowicz}, {Ord{\'o}{\~n}ez-Blanco},
  {Panuzzo}, {Portell}, {Richards}, {Riello}, {Seabroke}, {Tanga},
  {Th{\'e}venin}, {Torra}, {Els}, {Gracia-Abril}, {Comoretto},
  {Garcia-Reinaldos}, {Lock}, {Mercier}, {Altmann}, {Andrae}, {Astraatmadja},
  {Bellas-Velidis}, {Benson}, {Berthier}, {Blomme}, {Busso}, {Carry},
  {Cellino}, {Clementini}, {Cowell}, {Creevey}, {Cuypers}, {Davidson}, {De
  Ridder}, {de Torres}, {Delchambre}, {Dell'Oro}, {Ducourant}, {Fr{\'e}mat},
  {Garc{\'\i}a-Torres}, {Gosset}, {Halbwachs}, {Hambly}, {Harrison}, {Hauser},
  {Hestroffer}, {Hodgkin}, {Huckle}, {Hutton}, {Jasniewicz}, {Jordan},
  {Kontizas}, {Korn}, {Lanzafame}, {Manteiga}, {Moitinho}, {Muinonen},
  {Osinde}, {Pancino}, {Pauwels}, {Petit}, {Recio-Blanco}, {Robin}, {Sarro},
  {Siopis}, {Smith}, {Smith}, {Sozzetti}, {Thuillot}, {van Reeven}, {Viala},
  {Abbas}, {Abreu Aramburu}, {Accart}, {Aguado}, {Allan}, {Allasia},
  {Altavilla}, {{\'A}lvarez}, {Alves}, {Anderson}, {Andrei}, {Anglada Varela},
  {Antiche}, {Antoja}, {Ant{\'o}n}, {Arcay}, {Atzei}, {Ayache}, {Bach},
  {Baker}, {Balaguer-N{\'u}{\~n}ez}, {Barache}, {Barata}, {Barbier}, {Barblan},
  {Baroni}, {Barrado y Navascu{\'e}s}, {Barros}, {Barstow}, {Becciani},
  {Bellazzini}, {Bellei}, {Bello Garc{\'\i}a}, {Belokurov}, {Bendjoya},
  {Berihuete}, {Bianchi}, {Bienaym{\'e}}, {Billebaud}, {Blagorodnova},
  {Blanco-Cuaresma}, {Boch}, {Bombrun}, {Borrachero}, {Bouquillon}, {Bourda},
  {Bouy}, {Bragaglia}, {Breddels}, {Brouillet}, {Br{\"u}semeister},
  {Bucciarelli}, {Budnik}, {Burgess}, {Burgon}, {Burlacu}, {Busonero}, {Buzzi},
  {Caffau}, {Cambras}, {Campbell}, {Cancelliere}, {Cantat-Gaudin}, {Carlucci},
  {Carrasco}, {Castellani}, {Charlot}, {Charnas}, {Charvet}, {Chassat},
  {Chiavassa}, {Clotet}, {Cocozza}, {Collins}, {Collins}, {Costigan}, {Crifo},
  {Cross}, {Crosta}, {Crowley}, {Dafonte}, {Damerdji}, {Dapergolas}, {David},
  {David}, {De Cat}, {de Felice}, {de Laverny}, {De Luise}, {De March}, {de
  Martino}, {de Souza}, {Debosscher}, {del Pozo}, {Delbo}, {Delgado},
  {Delgado}, {di Marco}, {Di Matteo}, {Diakite}, {Distefano}, {Dolding}, {Dos
  Anjos}, {Drazinos}, {Dur{\'a}n}, {Dzigan}, {Ecale}, {Edvardsson}, {Enke},
  {Erdmann}, {Escolar}, {Espina}, {Evans}, {Eynard Bontemps}, {Fabre},
  {Fabrizio}, {Faigler}, {Falc{\~a}o}, {Farr{\`a}s Casas}, {Faye}, {Federici},
  {Fedorets}, {Fern{\'a}ndez-Hern{\'a}ndez}, {Fernique}, {Fienga}, {Figueras},
  {Filippi}, {Findeisen}, {Fonti}, {Fouesneau}, {Fraile}, {Fraser}, {Fuchs},
  {Furnell}, {Gai}, {Galleti}, {Galluccio}, {Garabato}, {Garc{\'\i}a-Sedano},
  {Gar{\'e}}, {Garofalo}, {Garralda}, {Gavras}, {Gerssen}, {Geyer}, {Gilmore},
  {Girona}, {Giuffrida}, {Gomes}, {Gonz{\'a}lez-Marcos},
  {Gonz{\'a}lez-N{\'u}{\~n}ez}, {Gonz{\'a}lez-Vidal}, {Granvik}, {Guerrier},
  {Guillout}, {Guiraud}, {G{\'u}rpide}, {Guti{\'e}rrez-S{\'a}nchez}, {Guy},
  {Haigron}, {Hatzidimitriou}, {Haywood}, {Heiter}, {Helmi}, {Hobbs},
  {Hofmann}, {Holl}, {Holland}, {Hunt}, {Hypki}, {Icardi}, {Irwin}, {Jevardat
  de Fombelle}, {Jofr{\'e}}, {Jonker}, {Jorissen}, {Julbe}, {Karampelas},
  {Kochoska}, {Kohley}, {Kolenberg}, {Kontizas}, {Koposov}, {Kordopatis},
  {Koubsky}, {Kowalczyk}, {Krone-Martins}, {Kudryashova}, {Kull}, {Bachchan},
  {Lacoste-Seris}, {Lanza}, {Lavigne}, {Le Poncin-Lafitte}, {Lebreton},
  {Lebzelter}, {Leccia}, {Leclerc}, {Lecoeur-Taibi}, {Lemaitre}, {Lenhardt},
  {Leroux}, {Liao}, {Licata}, {Lindstr{\o}m}, {Lister}, {Livanou}, {Lobel},
  {L{\"o}ffler}, {L{\'o}pez}, {Lopez-Lozano}, {Lorenz}, {Loureiro},
  {MacDonald}, {Magalh{\~a}es Fernandes}, {Managau}, {Mann}, {Mantelet},
  {Marchal}, {Marchant}, {Marconi}, {Marie}, {Marinoni}, {Marrese},
  {Marschalk{\'o}}, {Marshall}, {Mart{\'\i}n-Fleitas}, {Martino}, {Mary},
  {Matijevi{\v{c}}}, {Mazeh}, {McMillan}, {Messina}, {Mestre}, {Michalik},
  {Millar}, {Miranda}, {Molina}, {Molinaro}, {Molinaro}, {Moln{\'a}r},
  {Moniez}, {Montegriffo}, {Monteiro}, {Mor}, {Mora}, {Morbidelli}, {Morel},
  {Morgenthaler}, {Morley}, {Morris}, {Mulone}, {Muraveva}, {Musella},
  {Narbonne}, {Nelemans}, {Nicastro}, {Noval}, {Ord{\'e}novic},
  {Ordieres-Mer{\'e}}, {Osborne}, {Pagani}, {Pagano}, {Pailler}, {Palacin},
  {Palaversa}, {Parsons}, {Paulsen}, {Pecoraro}, {Pedrosa}, {Pentik{\"a}inen},
  {Pereira}, {Pichon}, {Piersimoni}, {Pineau}, {Plachy}, {Plum}, {Poujoulet},
  {Pr{\v{s}}a}, {Pulone}, {Ragaini}, {Rago}, {Rambaux}, {Ramos-Lerate},
  {Ranalli}, {Rauw}, {Read}, {Regibo}, {Renk}, {Reyl{\'e}}, {Ribeiro},
  {Rimoldini}, {Ripepi}, {Riva}, {Rixon}, {Roelens}, {Romero-G{\'o}mez},
  {Rowell}, {Royer}, {Rudolph}, {Ruiz-Dern}, {Sadowski}, {Sagrist{\`a}
  Sell{\'e}s}, {Sahlmann}, {Salgado}, {Salguero}, {Sarasso}, {Savietto},
  {Schnorhk}, {Schultheis}, {Sciacca}, {Segol}, {Segovia}, {Segransan},
  {Serpell}, {Shih}, {Smareglia}, {Smart}, {Smith}, {Solano}, {Solitro},
  {Sordo}, {Soria Nieto}, {Souchay}, {Spagna}, {Spoto}, {Stampa}, {Steele},
  {Steidelm{\"u}ller}, {Stephenson}, {Stoev}, {Suess}, {S{\"u}veges}, {Surdej},
  {Szabados}, {Szegedi-Elek}, {Tapiador}, {Taris}, {Tauran}, {Taylor},
  {Teixeira}, {Terrett}, {Tingley}, {Trager}, {Turon}, {Ulla}, {Utrilla},
  {Valentini}, {van Elteren}, {Van Hemelryck}, {van Leeuwen}, {Varadi},
  {Vecchiato}, {Veljanoski}, {Via}, {Vicente}, {Vogt}, {Voss}, {Votruba},
  {Voutsinas}, {Walmsley}, {Weiler}, {Weingrill}, {Werner}, {Wevers},
  {Whitehead}, {Wyrzykowski}, {Yoldas}, {{\v{Z}}erjal}, {Zucker}, {Zurbach},
  {Zwitter}, {Alecu}, {Allen}, {Allende Prieto}, {Amorim},
  {Anglada-Escud{\'e}}, {Arsenijevic}, {Azaz}, {Balm}, {Beck}, {Bernstein},
  {Bigot}, {Bijaoui}, {Blasco}, {Bonfigli}, {Bono}, {Boudreault}, {Bressan},
  {Brown}, {Brunet}, {Bunclark}, {Buonanno}, {Butkevich}, {Carret}, {Carrion},
  {Chemin}, {Ch{\'e}reau}, {Corcione}, {Darmigny}, {de Boer}, {de Teodoro}, {de
  Zeeuw}, {Delle Luche}, {Domingues}, {Dubath}, {Fodor}, {Fr{\'e}zouls},
  {Fries}, {Fustes}, {Fyfe}, {Gallardo}, {Gallegos}, {Gardiol}, {Gebran},
  {Gomboc}, {G{\'o}mez}, {Grux}, {Gueguen}, {Heyrovsky}, {Hoar}, {Iannicola},
  {Isasi Parache}, {Janotto}, {Joliet}, {Jonckheere}, {Keil}, {Kim},
  {Klagyivik}, {Klar}, {Knude}, {Kochukhov}, {Kolka}, {Kos}, {Kutka}, {Lainey},
  {LeBouquin}, {Liu}, {Loreggia}, {Makarov}, {Marseille}, {Martayan},
  {Martinez-Rubi}, {Massart}, {Meynadier}, {Mignot}, {Munari}, {Nguyen},
  {Nordlander}, {Ocvirk}, {O'Flaherty}, {Olias Sanz}, {Ortiz}, {Osorio},
  {Oszkiewicz}, {Ouzounis}, {Palmer}, {Park}, {Pasquato}, {Peltzer}, {Peralta},
  {P{\'e}turaud}, {Pieniluoma}, {Pigozzi}, {Poels}, {Prat}, {Prod'homme},
  {Raison}, {Rebordao}, {Risquez}, {Rocca-Volmerange}, {Rosen}, {Ruiz-Fuertes},
  {Russo}, {Sembay}, {Serraller Vizcaino}, {Short}, {Siebert}, {Silva},
  {Sinachopoulos}, {Slezak}, {Soffel}, {Sosnowska}, {Strai{\v{z}}ys}, {ter
  Linden}, {Terrell}, {Theil}, {Tiede}, {Troisi}, {Tsalmantza}, {Tur},
  {Vaccari}, {Vachier}, {Valles}, {Van Hamme}, {Veltz}, {Virtanen}, {Wallut},
  {Wichmann}, {Wilkinson}, {Ziaeepour}, \& {Zschocke}}]{2016A&A...595A...1G}
{Gaia Collaboration}, {Prusti}, T., {de Bruijne}, J.~H.~J., {et~al.} 2016,
  \aap, 595, A1

\bibitem[{{Gaia Collaboration} {et~al.}(2023){Gaia Collaboration}, {Vallenari},
  {Brown}, {Prusti}, {de Bruijne}, {Arenou}, {Babusiaux}, {Biermann},
  {Creevey}, {Ducourant}, {Evans}, {Eyer}, {Guerra}, {Hutton}, {Jordi},
  {Klioner}, {Lammers}, {Lindegren}, {Luri}, {Mignard}, {Panem}, {Pourbaix},
  {Randich}, {Sartoretti}, {Soubiran}, {Tanga}, {Walton}, {Bailer-Jones},
  {Bastian}, {Drimmel}, {Jansen}, {Katz}, {Lattanzi}, {van Leeuwen}, {Bakker},
  {Cacciari}, {Casta{\~n}eda}, {De Angeli}, {Fabricius}, {Fouesneau},
  {Fr{\'e}mat}, {Galluccio}, {Guerrier}, {Heiter}, {Masana}, {Messineo},
  {Mowlavi}, {Nicolas}, {Nienartowicz}, {Pailler}, {Panuzzo}, {Riclet}, {Roux},
  {Seabroke}, {Sordo}, {Th{\'e}venin}, {Gracia-Abril}, {Portell}, {Teyssier},
  {Altmann}, {Andrae}, {Audard}, {Bellas-Velidis}, {Benson}, {Berthier},
  {Blomme}, {Burgess}, {Busonero}, {Busso}, {C{\'a}novas}, {Carry}, {Cellino},
  {Cheek}, {Clementini}, {Damerdji}, {Davidson}, {de Teodoro}, {Nu{\~n}ez
  Campos}, {Delchambre}, {Dell'Oro}, {Esquej}, {Fern{\'a}ndez-Hern{\'a}ndez},
  {Fraile}, {Garabato}, {Garc{\'\i}a-Lario}, {Gosset}, {Haigron}, {Halbwachs},
  {Hambly}, {Harrison}, {Hern{\'a}ndez}, {Hestroffer}, {Hodgkin}, {Holl},
  {Jan{\ss}en}, {Jevardat de Fombelle}, {Jordan}, {Krone-Martins}, {Lanzafame},
  {L{\"o}ffler}, {Marchal}, {Marrese}, {Moitinho}, {Muinonen}, {Osborne},
  {Pancino}, {Pauwels}, {Recio-Blanco}, {Reyl{\'e}}, {Riello}, {Rimoldini},
  {Roegiers}, {Rybizki}, {Sarro}, {Siopis}, {Smith}, {Sozzetti}, {Utrilla},
  {van Leeuwen}, {Abbas}, {{\'A}brah{\'a}m}, {Abreu Aramburu}, {Aerts},
  {Aguado}, {Ajaj}, {Aldea-Montero}, {Altavilla}, {{\'A}lvarez}, {Alves},
  {Anders}, {Anderson}, {Anglada Varela}, {Antoja}, {Baines}, {Baker},
  {Balaguer-N{\'u}{\~n}ez}, {Balbinot}, {Balog}, {Barache}, {Barbato},
  {Barros}, {Barstow}, {Bartolom{\'e}}, {Bassilana}, {Bauchet}, {Becciani},
  {Bellazzini}, {Berihuete}, {Bernet}, {Bertone}, {Bianchi}, {Binnenfeld},
  {Blanco-Cuaresma}, {Blazere}, {Boch}, {Bombrun}, {Bossini}, {Bouquillon},
  {Bragaglia}, {Bramante}, {Breedt}, {Bressan}, {Brouillet}, {Brugaletta},
  {Bucciarelli}, {Burlacu}, {Butkevich}, {Buzzi}, {Caffau}, {Cancelliere},
  {Cantat-Gaudin}, {Carballo}, {Carlucci}, {Carnerero}, {Carrasco},
  {Casamiquela}, {Castellani}, {Castro-Ginard}, {Chaoul}, {Charlot}, {Chemin},
  {Chiaramida}, {Chiavassa}, {Chornay}, {Comoretto}, {Contursi}, {Cooper},
  {Cornez}, {Cowell}, {Crifo}, {Cropper}, {Crosta}, {Crowley}, {Dafonte},
  {Dapergolas}, {David}, {David}, {de Laverny}, {De Luise}, {De March}, {De
  Ridder}, {de Souza}, {de Torres}, {del Peloso}, {del Pozo}, {Delbo},
  {Delgado}, {Delisle}, {Demouchy}, {Dharmawardena}, {Di Matteo}, {Diakite},
  {Diener}, {Distefano}, {Dolding}, {Edvardsson}, {Enke}, {Fabre}, {Fabrizio},
  {Faigler}, {Fedorets}, {Fernique}, {Fienga}, {Figueras}, {Fournier},
  {Fouron}, {Fragkoudi}, {Gai}, {Garcia-Gutierrez}, {Garcia-Reinaldos},
  {Garc{\'\i}a-Torres}, {Garofalo}, {Gavel}, {Gavras}, {Gerlach}, {Geyer},
  {Giacobbe}, {Gilmore}, {Girona}, {Giuffrida}, {Gomel}, {Gomez},
  {Gonz{\'a}lez-N{\'u}{\~n}ez}, {Gonz{\'a}lez-Santamar{\'\i}a},
  {Gonz{\'a}lez-Vidal}, {Granvik}, {Guillout}, {Guiraud},
  {Guti{\'e}rrez-S{\'a}nchez}, {Guy}, {Hatzidimitriou}, {Hauser}, {Haywood},
  {Helmer}, {Helmi}, {Sarmiento}, {Hidalgo}, {Hilger}, {H{\l}adczuk}, {Hobbs},
  {Holland}, {Huckle}, {Jardine}, {Jasniewicz}, {Jean-Antoine Piccolo},
  {Jim{\'e}nez-Arranz}, {Jorissen}, {Juaristi Campillo}, {Julbe}, {Karbevska},
  {Kervella}, {Khanna}, {Kontizas}, {Kordopatis}, {Korn}, {K{\'o}sp{\'a}l},
  {Kostrzewa-Rutkowska}, {Kruszy{\'n}ska}, {Kun}, {Laizeau}, {Lambert},
  {Lanza}, {Lasne}, {Le Campion}, {Lebreton}, {Lebzelter}, {Leccia}, {Leclerc},
  {Lecoeur-Taibi}, {Liao}, {Licata}, {Lindstr{\o}m}, {Lister}, {Livanou},
  {Lobel}, {Lorca}, {Loup}, {Madrero Pardo}, {Magdaleno Romeo}, {Managau},
  {Mann}, {Manteiga}, {Marchant}, {Marconi}, {Marcos}, {Marcos Santos},
  {Mar{\'\i}n Pina}, {Marinoni}, {Marocco}, {Marshall}, {Martin Polo},
  {Mart{\'\i}n-Fleitas}, {Marton}, {Mary}, {Masip}, {Massari},
  {Mastrobuono-Battisti}, {Mazeh}, {McMillan}, {Messina}, {Michalik}, {Millar},
  {Mints}, {Molina}, {Molinaro}, {Moln{\'a}r}, {Monari}, {Mongui{\'o}},
  {Montegriffo}, {Montero}, {Mor}, {Mora}, {Morbidelli}, {Morel}, {Morris},
  {Muraveva}, {Murphy}, {Musella}, {Nagy}, {Noval}, {Oca{\~n}a}, {Ogden},
  {Ordenovic}, {Osinde}, {Pagani}, {Pagano}, {Palaversa}, {Palicio},
  {Pallas-Quintela}, {Panahi}, {Payne-Wardenaar}, {Pe{\~n}alosa Esteller},
  {Penttil{\"a}}, {Pichon}, {Piersimoni}, {Pineau}, {Plachy}, {Plum}, {Poggio},
  {Pr{\v{s}}a}, {Pulone}, {Racero}, {Ragaini}, {Rainer}, {Raiteri}, {Rambaux},
  {Ramos}, {Ramos-Lerate}, {Re Fiorentin}, {Regibo}, {Richards}, {Rios Diaz},
  {Ripepi}, {Riva}, {Rix}, {Rixon}, {Robichon}, {Robin}, {Robin}, {Roelens},
  {Rogues}, {Rohrbasser}, {Romero-G{\'o}mez}, {Rowell}, {Royer}, {Ruz Mieres},
  {Rybicki}, {Sadowski}, {S{\'a}ez N{\'u}{\~n}ez}, {Sagrist{\`a} Sell{\'e}s},
  {Sahlmann}, {Salguero}, {Samaras}, {Sanchez Gimenez}, {Sanna},
  {Santove{\~n}a}, {Sarasso}, {Schultheis}, {Sciacca}, {Segol}, {Segovia},
  {S{\'e}gransan}, {Semeux}, {Shahaf}, {Siddiqui}, {Siebert}, {Siltala},
  {Silvelo}, {Slezak}, {Slezak}, {Smart}, {Snaith}, {Solano}, {Solitro},
  {Souami}, {Souchay}, {Spagna}, {Spina}, {Spoto}, {Steele},
  {Steidelm{\"u}ller}, {Stephenson}, {S{\"u}veges}, {Surdej}, {Szabados},
  {Szegedi-Elek}, {Taris}, {Taylor}, {Teixeira}, {Tolomei}, {Tonello}, {Torra},
  {Torra}, {Torralba Elipe}, {Trabucchi}, {Tsounis}, {Turon}, {Ulla}, {Unger},
  {Vaillant}, {van Dillen}, {van Reeven}, {Vanel}, {Vecchiato}, {Viala},
  {Vicente}, {Voutsinas}, {Weiler}, {Wevers}, {Wyrzykowski}, {Yoldas}, {Yvard},
  {Zhao}, {Zorec}, {Zucker}, \& {Zwitter}}]{2023A&A...674A...1G}
{Gaia Collaboration}, {Vallenari}, A., {Brown}, A.~G.~A., {et~al.} 2023, \aap,
  674, A1

\bibitem[{{Gilmore} {et~al.}(2022){Gilmore}, {Randich}, {Worley}, {Hourihane},
  {Gonneau}, {Sacco}, \& {et al.}}]{GES_final_release_paper_1}
{Gilmore}, G., {Randich}, S., {Worley}, C.~C., {et~al.} 2022, \aap\ in press

\bibitem[{{G{\'o}rski} {et~al.}(2005){G{\'o}rski}, {Hivon}, {Banday},
  {Wandelt}, {Hansen}, {Reinecke}, \& {Bartelmann}}]{2005ApJ...622..759G}
{G{\'o}rski}, K.~M., {Hivon}, E., {Banday}, A.~J., {et~al.} 2005, \apj, 622,
  759

\bibitem[{{Hambly} {et~al.}(2022){Hambly}, {Andrae}, {De Angeli}, {Antonio},
  {Arenou}, {Audard}, {Babusiaux}, {Bailer-Jones}, {Bakker}, {Bastian},
  {Bauchet}, {Bellas-Velidis}, {Blomme}, {Bombrun}, {Brouillet}, {Brugaletta},
  {de Bruijne}, {Busonero}, {Busso}, {Carballo}, {Carnerero}, {Clementini},
  {Creevey}, {Damerdji}, {Delchambre}, {Distefano}, {Drimmel}, {Ducourant},
  {Duran}, {Fabricius}, {Eyer}, {Faigler}, {Findeisen}, {Jevardat de Fombelle},
  {Fouesneau}, {Fr{\'e}mat}, {Galluccio}, {Garabato}, {Gavras}, {Giuffrida},
  {Gomel}, {Gonz{\'a}lez}, {Gonz{\'a}lez-N{\'u}{\~n}ez}, {Gosset},
  {Gracia-Abril}, {Halbwachs}, {Harrison}, {Heiter}, {Hernandez}, {Hestroffer},
  {Hobbs}, {Hodgkin}, {Holl}, {Hutton}, {Katz}, {Klioner}, {Leccia},
  {Lebreton}, {Lecoeur-Ta{\"\i}bi}, {van Leeuwen}, {Lindegren}, {Lobel},
  {Luri}, {Mantelet}, {Marrese}, {Marinoni}, {Marshall}, {Masana}, {Mazeh},
  {Michalik}, {Molinaro}, {Mora}, {Mowlavi}, {Nienartowicz}, {Ordenovic},
  {Panahi}, {Pancino}, {Pauwels}, {Pichon}, {Portell}, {Pourbaix}, {Raiteri},
  {Recio-Blanco}, {De Ridder}, {Riello}, {Rimoldini}, {Ripepi}, {Rixon},
  {Robin}, {Rybizki}, {Sartoretti}, {Sarro Baro}, {Seabroke}, {Segovia
  Serrato}, {Siopis}, {Smart}, {Soubiran}, {Sozzetti}, {Spoto}, {Tanga},
  {Teyssier}, {Utrilla}, {Masip Vela}, {Wyrzykowski}, \&
  {Zucker}}]{2022gdr3.reptE..20H}
{Hambly}, N., {Andrae}, R., {De Angeli}, F., {et~al.} 2022, {Gaia DR3
  documentation Chapter 20: Datamodel description}, Gaia DR3 documentation,
  European Space Agency; Gaia Data Processing and Analysis Consortium. Online
  at <A href=``https://
  gea.esac.esa.int/archive/documentation/GDR3/index.html''>
  https://gea.esac.esa.int/archive/documentation/GDR3/index.html </A>, id. 20

\bibitem[{{Henden} {et~al.}(2016){Henden}, {Templeton}, {Terrell}, {Smith},
  {Levine}, \& {Welch}}]{apass9}
{Henden}, A.~A., {Templeton}, M., {Terrell}, D., {et~al.} 2016, VizieR Online
  Data Catalogue, 2336

\bibitem[{{Hobbs} {et~al.}(2022){Hobbs}, {Lindegren}, {Bastian}, {Klioner},
  {Butkevich}, {Stephenson}, {Hernandez}, {Lammers}, {Bombrun}, {Mignard},
  {Altmann}, {Davidson}, {de Bruijne}, {Fern{\'a}ndez-Hern{\'a}ndez},
  {Siddiqui}, \& {Utrilla}}]{2022gdr3.reptE...4H}
{Hobbs}, D., {Lindegren}, L., {Bastian}, U., {et~al.} 2022, {Gaia DR3
  documentation Chapter 4: Astrometric data}, Gaia DR3 documentation, European
  Space Agency; Gaia Data Processing and Analysis Consortium. Online at <A
  href=``https:// gea.esac.esa.int/archive/documentation/GDR3/index.html''>
  https://gea.esac.esa.int/archive/documentation/GDR3/index.html </A>, id. 4

\bibitem[{{H{\o}g} {et~al.}(2000){H{\o}g}, {Fabricius}, {Makarov}, {Urban},
  {Corbin}, {Wycoff}, {Bastian}, {Schwekendiek}, \&
  {Wicenec}}]{2000A&A...355L..27H}
{H{\o}g}, E., {Fabricius}, C., {Makarov}, V.~V., {et~al.} 2000, \aap, 355, L27

\bibitem[{{Huber} {et~al.}(2016){Huber}, {Bryson}, {Haas}, {Barclay},
  {Barentsen}, {Howell}, {Sharma}, {Stello}, \&
  {Thompson}}]{epic-2016ApJS..224....2H}
{Huber}, D., {Bryson}, S.~T., {Haas}, M.~R., {et~al.} 2016, \apjs, 224, 2

\bibitem[{Hunter(2007)}]{Hunter:2007}
Hunter, J.~D. 2007, Computing In Science \& Engineering, 9, 90

\bibitem[{{Ibata} {et~al.}(2019{\natexlab{a}}){Ibata}, {Bellazzini}, {Malhan},
  {Martin}, \& {Bianchini}}]{2019NatAs...3..667I}
{Ibata}, R.~A., {Bellazzini}, M., {Malhan}, K., {Martin}, N., \& {Bianchini},
  P. 2019{\natexlab{a}}, Nature Astronomy, 3, 667

\bibitem[{{Ibata} {et~al.}(2019{\natexlab{b}}){Ibata}, {Malhan}, \&
  {Martin}}]{2019ApJ...872..152I}
{Ibata}, R.~A., {Malhan}, K., \& {Martin}, N.~F. 2019{\natexlab{b}}, \apj, 872,
  152

\bibitem[{{Lasker} {et~al.}(2008){Lasker}, {Lattanzi}, {McLean}, {Bucciarelli},
  {Drimmel}, {Garcia}, {Greene}, {Guglielmetti}, {Hanley}, {Hawkins},
  {Laidler}, {Loomis}, {Meakes}, {Mignani}, {Morbidelli}, {Morrison},
  {Pannunzio}, {Rosenberg}, {Sarasso}, {Smart}, {Spagna}, {Sturch},
  {Volpicelli}, {White}, {Wolfe}, \& {Zacchei}}]{2008AJ....136..735L}
{Lasker}, B.~M., {Lattanzi}, M.~G., {McLean}, B.~J., {et~al.} 2008, \aj, 136,
  735

\bibitem[{{Lindegren} {et~al.}(2021){Lindegren}, {Klioner}, {Hern{\'a}ndez},
  {Bombrun}, {Ramos-Lerate}, {Steidelm{\"u}ller}, {Bastian}, {Biermann}, {de
  Torres}, {Gerlach}, {Geyer}, {Hilger}, {Hobbs}, {Lammers}, {McMillan},
  {Stephenson}, {Casta{\~n}eda}, {Davidson}, {Fabricius}, {Gracia-Abril},
  {Portell}, {Rowell}, {Teyssier}, {Torra}, {Bartolom{\'e}}, {Clotet},
  {Garralda}, {Gonz{\'a}lez-Vidal}, {Torra}, {Abbas}, {Altmann}, {Anglada
  Varela}, {Balaguer-N{\'u}{\~n}ez}, {Balog}, {Barache}, {Becciani}, {Bernet},
  {Bertone}, {Bianchi}, {Bouquillon}, {Brown}, {Bucciarelli}, {Busonero},
  {Butkevich}, {Buzzi}, {Cancelliere}, {Carlucci}, {Charlot}, {Cioni},
  {Crosta}, {Crowley}, {del Peloso}, {del Pozo}, {Drimmel}, {Esquej}, {Fienga},
  {Fraile}, {Gai}, {Garcia-Reinaldos}, {Guerra}, {Hambly}, {Hauser},
  {Jan{\ss}en}, {Jordan}, {Kostrzewa-Rutkowska}, {Lattanzi}, {Liao}, {Licata},
  {Lister}, {L{\"o}ffler}, {Marchant}, {Masip}, {Mignard}, {Mints}, {Molina},
  {Mora}, {Morbidelli}, {Murphy}, {Pagani}, {Panuzzo}, {Pe{\~n}alosa Esteller},
  {Poggio}, {Re Fiorentin}, {Riva}, {Sagrist{\`a} Sell{\'e}s}, {Sanchez
  Gimenez}, {Sarasso}, {Sciacca}, {Siddiqui}, {Smart}, {Souami}, {Spagna},
  {Steele}, {Taris}, {Utrilla}, {van Reeven}, \& {Vecchiato}}]{lindegren2021}
{Lindegren}, L., {Klioner}, S.~A., {Hern{\'a}ndez}, J., {et~al.} 2021, \aap,
  649, A2

\bibitem[{{Luo} {et~al.}(2015){Luo}, {Zhao}, {Zhao}, {Deng}, {Liu}, {Jing},
  {Wang}, {Zhang}, {Shi}, {Cui}, {Chu}, {Li}, {Bai}, {Wu}, {Cai}, {Cao}, {Cao},
  {Carlin}, {Chen}, {Chen}, {Chen}, {Chen}, {Chen}, {Chen}, {Chen},
  {Christlieb}, {Chu}, {Cui}, {Dong}, {Du}, {Fan}, {Feng}, {Fu}, {Gao}, {Gong},
  {Gu}, {Guo}, {Han}, {He}, {Hou}, {Hou}, {Hou}, {Hu}, {Hu}, {Hu}, {Huo},
  {Jia}, {Jiang}, {Jiang}, {Jiang}, {Jin}, {Kong}, {Kong}, {Lei}, {Li}, {Li},
  {Li}, {Li}, {Li}, {Li}, {Li}, {Li}, {Li}, {Li}, {Li}, {Li}, {Liang}, {Lin},
  {Liu}, {Liu}, {Liu}, {Liu}, {Lu}, {Luo}, {Mao}, {Newberg}, {Ni}, {Qi}, {Qi},
  {Shen}, {Shi}, {Song}, {Song}, {Su}, {Su}, {Tang}, {Tao}, {Tian}, {Wang},
  {Wang}, {Wang}, {Wang}, {Wang}, {Wang}, {Wang}, {Wang}, {Wang}, {Wang},
  {Wang}, {Wang}, {Wang}, {Wang}, {Wang}, {Wang}, {Wang}, {Wang}, {Wang},
  {Wang}, {Wei}, {Wei}, {Wu}, {Wu}, {Wu}, {Wu}, {Xing}, {Xu}, {Xu}, {Xu},
  {Yan}, {Yang}, {Yang}, {Yang}, {Yang}, {Yao}, {Yu}, {Yuan}, {Yuan}, {Yuan},
  {Yuan}, {Zhai}, {Zhang}, {Zhang}, {Zhang}, {Zhang}, {Zhang}, {Zhang},
  {Zhang}, {Zhang}, {Zhao}, {Zhou}, {Zhou}, {Zhu}, {Zhu}, {Zou}, \&
  {Zuo}}]{LamostDR1}
{Luo}, A.~L., {Zhao}, Y.-H., {Zhao}, G., {et~al.} 2015, Research in Astronomy
  and Astrophysics, 15, 1095

\bibitem[{{Magnier} {et~al.}(2020{\natexlab{a}}){Magnier}, {Chambers},
  {Flewelling}, {Hoblitt}, {Huber}, {Price}, {Sweeney}, {Waters}, {Denneau},
  {Draper}, {Hodapp}, {Jedicke}, {Kaiser}, {Kudritzki}, {Metcalfe}, {Stubbs},
  \& {Wainscoat}}]{panstarrs1b}
{Magnier}, E.~A., {Chambers}, K.~C., {Flewelling}, H.~A., {et~al.}
  2020{\natexlab{a}}, \apjs, 251, 3

\bibitem[{{Magnier} {et~al.}(2020{\natexlab{b}}){Magnier}, {Schlafly},
  {Finkbeiner}, {Tonry}, {Goldman}, {R{\"o}ser}, {Schilbach}, {Casertano},
  {Chambers}, {Flewelling}, {Huber}, {Price}, {Sweeney}, {Waters}, {Denneau},
  {Draper}, {Hodapp}, {Jedicke}, {Kaiser}, {Kudritzki}, {Metcalfe}, {Stubbs},
  \& {Wainscoat}}]{panstarrs1e}
{Magnier}, E.~A., {Schlafly}, E.~F., {Finkbeiner}, D.~P., {et~al.}
  2020{\natexlab{b}}, \apjs, 251, 6

\bibitem[{{Magnier} {et~al.}(2020{\natexlab{c}}){Magnier}, {Sweeney},
  {Chambers}, {Flewelling}, {Huber}, {Price}, {Waters}, {Denneau}, {Draper},
  {Farrow}, {Jedicke}, {Hodapp}, {Kaiser}, {Kudritzki}, {Metcalfe}, {Stubbs},
  \& {Wainscoat}}]{panstarrs1d}
{Magnier}, E.~A., {Sweeney}, W.~E., {Chambers}, K.~C., {et~al.}
  2020{\natexlab{c}}, \apjs, 251, 5

\bibitem[{{Ochsenbein} {et~al.}(2000){Ochsenbein}, {Bauer}, \&
  {Marcout}}]{2000A&AS..143...23O}
{Ochsenbein}, F., {Bauer}, P., \& {Marcout}, J. 2000, \aaps, 143, 23

\bibitem[{{Onken} {et~al.}(2019){Onken}, {Wolf}, {Bessell}, {Chang}, {Da
  Costa}, {Luvaul}, {Mackey}, {Schmidt}, \& {Shao}}]{2019PASA...36...33O}
{Onken}, C.~A., {Wolf}, C., {Bessell}, M.~S., {et~al.} 2019, \pasa, 36, e033

\bibitem[{P\'erez \& Granger(2007)}]{PER-GRA:2007}
P\'erez, F. \& Granger, B.~E. 2007, Computing in Science and Engineering, 9, 21

\bibitem[{{Prabhu} {et~al.}(2022){Prabhu}, {Subramaniam}, {Sahu}, {Chung},
  {Leigh}, {Dalessandro}, {Chatterjee}, {Rao}, {Shara}, {C{\^o}t{\'e}},
  {Choudhury}, {Pandey}, {Valcarce}, {Singh}, {Postma}, {Rani},
  {Bandyopadhyay}, {Geller}, {Hutchings}, {Puzia}, {Simunovic}, {Sohn},
  {Thirupathi}, \& {Yadav}}]{2022ApJ...939L..20P}
{Prabhu}, D.~S., {Subramaniam}, A., {Sahu}, S., {et~al.} 2022, \apjl, 939, L20

\bibitem[{{R Core Team}(2013)}]{RManual}
{R Core Team}. 2013, {R: A Language and Environment for Statistical Computing},
  R Foundation for Statistical Computing, Vienna, Austria

\bibitem[{{Randich} {et~al.}(2022){Randich}, {Gilmore}, {Magrini}, {Sacco},
  {Jackson}, {Jeffries}, \& {et al.}}]{GES_final_release_paper_2}
{Randich}, S., {Gilmore}, G., {Magrini}, L., {et~al.} 2022, \aap\ in press

\bibitem[{{Riello} {et~al.}(2021){Riello}, {De Angeli}, {Evans}, {Montegriffo},
  {Carrasco}, {Busso}, {Palaversa}, {Burgess}, {Diener}, {Davidson}, {Rowell},
  {Fabricius}, {Jordi}, {Bellazzini}, {Pancino}, {Harrison}, {Cacciari}, {van
  Leeuwen}, {Hambly}, {Hodgkin}, {Osborne}, {Altavilla}, {Barstow}, {Brown},
  {Castellani}, {Cowell}, {De Luise}, {Gilmore}, {Giuffrida}, {Hidalgo},
  {Holland}, {Marinoni}, {Pagani}, {Piersimoni}, {Pulone}, {Ragaini}, {Rainer},
  {Richards}, {Sanna}, {Walton}, {Weiler}, \& {Yoldas}}]{2021A&A...649A...3R}
{Riello}, M., {De Angeli}, F., {Evans}, D.~W., {et~al.} 2021, \aap, 649, A3

\bibitem[{{Roeser} {et~al.}(2010){Roeser}, {Demleitner}, \&
  {Schilbach}}]{2010AJ....139.2440R}
{Roeser}, S., {Demleitner}, M., \& {Schilbach}, E. 2010, \aj, 139, 2440

\bibitem[{{Rowell} {et~al.}(2021){Rowell}, {Davidson}, {Lindegren}, {van
  Leeuwen}, {Casta{\~n}eda}, {Fabricius}, {Bastian}, {Hambly}, {Hern{\'a}ndez},
  {Bombrun}, {Evans}, {De Angeli}, {Riello}, {Busonero}, {Crowley}, {Mora},
  {Lammers}, {Gracia}, {Portell}, {Biermann}, \& {Brown}}]{2021A&A...649A..11R}
{Rowell}, N., {Davidson}, M., {Lindegren}, L., {et~al.} 2021, \aap, 649, A11

\bibitem[{{Sanna} {et~al.}(2020){Sanna}, {Pancino}, {Zocchi}, {Ferraro}, \&
  {Stetson}}]{2020A&A...637A..46S}
{Sanna}, N., {Pancino}, E., {Zocchi}, A., {Ferraro}, F.~R., \& {Stetson}, P.~B.
  2020, \aap, 637, A46

\bibitem[{{Scalco} {et~al.}(2021){Scalco}, {Bellini}, {Bedin}, {Anderson},
  {Rosati}, {Libralato}, {Salaris}, {Vesperini}, {Nardiello}, {Apai},
  {Burgasser}, \& {Gerasimov}}]{2021MNRAS.505.3549S}
{Scalco}, M., {Bellini}, A., {Bedin}, L.~R., {et~al.} 2021, \mnras, 505, 3549

\bibitem[{{Skrutskie} {et~al.}(2006){Skrutskie}, {Cutri}, {Stiening},
  {Weinberg}, {Schneider}, {Carpenter}, {Beichman}, {Capps}, {Chester},
  {Elias}, {Huchra}, {Liebert}, {Lonsdale}, {Monet}, {Price}, {Seitzer},
  {Jarrett}, {Kirkpatrick}, {Gizis}, {Howard}, {Evans}, {Fowler}, {Fullmer},
  {Hurt}, {Light}, {Kopan}, {Marsh}, {McCallon}, {Tam}, {Van Dyk}, \&
  {Wheelock}}]{2006AJ....131.1163S}
{Skrutskie}, M.~F., {Cutri}, R.~M., {Stiening}, R., {et~al.} 2006, \aj, 131,
  1163

\bibitem[{{Soltis} {et~al.}(2021){Soltis}, {Casertano}, \&
  {Riess}}]{2021ApJ...908L...5S}
{Soltis}, J., {Casertano}, S., \& {Riess}, A.~G. 2021, \apjl, 908, L5

\bibitem[{{Steinmetz} {et~al.}(2020{\natexlab{a}}){Steinmetz}, {Guiglion},
  {McMillan}, {Matijevi{\v{c}}}, {Enke}, {Kordopatis}, {Zwitter}, {Valentini},
  {Chiappini}, {Casagrande}, {Wojno}, {Anguiano}, {Bienaym{\'e}}, {Bijaoui},
  {Binney}, {Burton}, {Cass}, {de Laverny}, {Fiegert}, {Freeman}, {Fulbright},
  {Gibson}, {Gilmore}, {Grebel}, {Helmi}, {Kunder}, {Munari}, {Navarro},
  {Parker}, {Ruchti}, {Recio-Blanco}, {Reid}, {Seabroke}, {Siviero}, {Siebert},
  {Stupar}, {Watson}, {Williams}, {Wyse}, {Anders}, {Antoja}, {Birko},
  {Bland-Hawthorn}, {Bossini}, {Garc{\'\i}a}, {Carrillo}, {Chaplin},
  {Elsworth}, {Famaey}, {Gerhard}, {Jofre}, {Just}, {Mathur}, {Miglio},
  {Minchev}, {Monari}, {Mosser}, {Ritter}, {Rodrigues}, {Scholz}, {Sharma},
  {Sysoliatina}, \& {RAVE Collaboration}}]{2020AJ....160...83S}
{Steinmetz}, M., {Guiglion}, G., {McMillan}, P.~J., {et~al.}
  2020{\natexlab{a}}, \aj, 160, 83

\bibitem[{{Steinmetz} {et~al.}(2020{\natexlab{b}}){Steinmetz},
  {Matijevi{\v{c}}}, {Enke}, {Zwitter}, {Guiglion}, {McMillan}, {Kordopatis},
  {Valentini}, {Chiappini}, {Casagrande}, {Wojno}, {Anguiano}, {Bienaym{\'e}},
  {Bijaoui}, {Binney}, {Burton}, {Cass}, {de Laverny}, {Fiegert}, {Freeman},
  {Fulbright}, {Gibson}, {Gilmore}, {Grebel}, {Helmi}, {Kunder}, {Munari},
  {Navarro}, {Parker}, {Ruchti}, {Recio-Blanco}, {Reid}, {Seabroke}, {Siviero},
  {Siebert}, {Stupar}, {Watson}, {Williams}, {Wyse}, {Anders}, {Antoja},
  {Birko}, {Bland-Hawthorn}, {Bossini}, {Garc{\'\i}a}, {Carrillo}, {Chaplin},
  {Elsworth}, {Famaey}, {Gerhard}, {Jofre}, {Just}, {Mathur}, {Miglio},
  {Minchev}, {Monari}, {Mosser}, {Ritter}, {Rodrigues}, {Scholz}, {Sharma},
  {Sysoliatina}, \& {RAVE Collaboration}}]{rave6a}
{Steinmetz}, M., {Matijevi{\v{c}}}, G., {Enke}, H., {et~al.}
  2020{\natexlab{b}}, \aj, 160, 82

\bibitem[{{Stetson}(1987)}]{1987PASP...99..191S}
{Stetson}, P.~B. 1987, \pasp, 99, 191

\bibitem[{{Taylor}(2005)}]{2005ASPC..347...29T}
{Taylor}, M.~B. 2005, in Astronomical Society of the Pacific Conference Series,
  Vol. 347, Astronomical Data Analysis Software and Systems XIV, ed.
  P.~{Shopbell}, M.~{Britton}, \& R.~{Ebert}, 29

\bibitem[{{Taylor}(2006)}]{2006ASPC..351..666T}
{Taylor}, M.~B. 2006, in Astronomical Society of the Pacific Conference Series,
  Vol. 351, Astronomical Data Analysis Software and Systems XV, ed.
  C.~{Gabriel}, C.~{Arviset}, D.~{Ponz}, \& S.~{Enrique}, 666

\bibitem[{{Torra} {et~al.}(2021){Torra}, {Casta{\~n}eda}, {Fabricius},
  {Lindegren}, {Clotet}, {Gonz{\'a}lez-Vidal}, {Bartolom{\'e}}, {Bastian},
  {Bernet}, {Biermann}, {Garralda}, {G{\'u}rpide}, {Lammers}, {Portell}, \&
  {Torra}}]{2021A&A...649A..10T}
{Torra}, F., {Casta{\~n}eda}, J., {Fabricius}, C., {et~al.} 2021, \aap, 649,
  A10

\bibitem[{{van Leeuwen}(2007)}]{2007A&A...474..653V}
{van Leeuwen}, F. 2007, \aap, 474, 653

\bibitem[{{Waters} {et~al.}(2020){Waters}, {Magnier}, {Price}, {Chambers},
  {Burgett}, {Draper}, {Flewelling}, {Hodapp}, {Huber}, {Jedicke}, {Kaiser},
  {Kudritzki}, {Lupton}, {Metcalfe}, {Rest}, {Sweeney}, {Tonry}, {Wainscoat},
  \& {Wood-Vasey}}]{panstarrs1c}
{Waters}, C.~Z., {Magnier}, E.~A., {Price}, P.~A., {et~al.} 2020, \apjs, 251, 4

\bibitem[{{Wenger} {et~al.}(2000){Wenger}, {Ochsenbein}, {Egret}, {Dubois},
  {Bonnarel}, {Borde}, {Genova}, {Jasniewicz}, {Lalo{\"e}}, {Lesteven}, \&
  {Monier}}]{2000AAS..143....9W}
{Wenger}, M., {Ochsenbein}, F., {Egret}, D., {et~al.} 2000, \aaps, 143, 9

\bibitem[{{Zacharias} {et~al.}(2015){Zacharias}, {Finch}, {Subasavage},
  {Bredthauer}, {Crockett}, {Divittorio}, {Ferguson}, {Harris}, {Harris},
  {Henden}, {Kilian}, {Munn}, {Rafferty}, {Rhodes}, {Schultheiss}, {Tilleman},
  \& {Wieder}}]{urat1}
{Zacharias}, N., {Finch}, C., {Subasavage}, J., {et~al.} 2015, \aj, 150, 101

\bibitem[{{Zacharias} {et~al.}(2013){Zacharias}, {Finch}, {Girard}, {Henden},
  {Bartlett}, {Monet}, \& {Zacharias}}]{2013AJ....145...44Z}
{Zacharias}, N., {Finch}, C.~T., {Girard}, T.~M., {et~al.} 2013, \aj, 145, 44

\end{thebibliography}

\begin{appendix}

\section{List of acronyms}
\begin{table}[htbp]
    \caption{Acronyms used in this paper}\label{tab:acronyms}
        \centering    
    \begin{tabular}{ p{2cm} p{6cm}}
    \hline\hline 
    \textbf{Acronym} & \textbf{Description}  \\
    \hline
    AC & Across scan direction \\ 
    ADQL & Astronomical Data Query Language \\
    AF & Astrometric Field \\ 
    AGIS & Astrometric Global Iterative Solution \\ 
    AL & Along scan direction \\ 
    BP & Blue Photometer \\ 
    CCD & Charge-Coupled Device \\ 
    CF & Crowded Field \\ 
    DAOPHOT & Code for crowded field stellar photometry\\ 
    DEC & Declination \\ 
    DPAC & Data Processing and Analysis Consortium \\ 
    DR3 & Data Release 3 \\ 
    EDR3 & Data Release 3 Early \\ 
    ESA & European Space Agency \\ 
    ESAC & European Space Astronomy Centre \\ 
    FOV & Field of View \\ 
    FPR & Focused Product Release \\ 
    HST & Hubble Space Telescope \\ 
    IDU & Intermediate Data Update \\ 
    IPD & Image Parameter Determination \\ 
    LMC & Large magellanic cloud \\ 
    NGC & New General Catalogue \\ 
    PSF & Point Spread Function \\ 
    RA & Right Ascension \\ 
    RP & Red Photometer \\ 
    RVS & Radial Velocity Spectrometer \\ 
    SIF & Service Interface Function \\ 
    SM & Sky Mapper \\ 
    SMC & Small magellanic cloud \\ 
    TDI & Time Delayed Integration \\ 
    VBS & Very Bright Star \\ 
    XM & Crossmatch \\ \hline
    \end{tabular}
\end{table}

\section{{\it Gaia} archive query}\label{sec:query}

Example, showing how to retrieve all {\it Gaia} SIF CF and {\it Gaia} DR3 sources within 0.8 degree radius around $\omega$ Cen cluster centre from the {\it Gaia} archive via Astronomical Data Query Language (ADQL). For DR3 data, the field\linktoparamfpr{crowded_field_source}{n\_scans} does not exist, so a value of -1 is used there.

\noindent
\begin{verbatim}
SELECT source_id, ref_epoch, ra, ra_error, dec, 
dec_error, pmra, pmra_error, pmdec, pmdec_error, 
phot_g_mean_mag, l, b, n_scans, 
'fpr' as origin
FROM gaiafpr.crowded_field_source
UNION
SELECT source_id, ref_epoch, ra, ra_error, dec, 
dec_error, pmra, pmra_error, pmdec, pmdec_error, 
phot_g_mean_mag, l, b, -1 as nscans, 
'dr3' as origin
FROM gaiadr3.gaia_source
WHERE
1=CONTAINS(POINT('ICRS', ra, dec), 
CIRCLE('ICRS', 
201.69399972775088, -47.484610741298994, 0.8))
\end{verbatim}

\section{Acknowledgements}
\section*{Acknowledgements\label{sec:acknowl}}
\addcontentsline{toc}{section}{Acknowledgements}

This work presents results from the European Space Agency (ESA) space mission \gaia. \gaia\ data are being processed by the \gaia\ Data Processing and Analysis Consortium (DPAC). Funding for the DPAC is provided by national institutions, in particular the institutions participating in the \gaia\ MultiLateral Agreement (MLA). The \gaia\ mission website is \url{https://www.cosmos.esa.int/gaia}. The \gaia\ archive website is \url{https://archives.esac.esa.int/gaia}.

The \gaia\ mission and data processing have financially been supported by, in alphabetical order by country:
\begin{itemize}
\item the Algerian Centre de Recherche en Astronomie, Astrophysique et G\'{e}ophysique of Bouzareah Observatory;
\item the Austrian Fonds zur F\"{o}rderung der wissenschaftlichen Forschung (FWF) Hertha Firnberg Programme through grants T359, P20046, and P23737;
\item the BELgian federal Science Policy Office (BELSPO) through various PROgramme de D\'{e}veloppement d'Exp\'{e}riences scientifiques (PRODEX) grants of the European Space Agency (ESA), the Research Foundation Flanders (Fonds Wetenschappelijk Onderzoek) through grant VS.091.16N, the Fonds de la Recherche Scientifique (FNRS), and the Research Council of Katholieke Universiteit (KU) Leuven through grant C16/18/005 (Pushing AsteRoseismology to the next level with TESS, GaiA, and the Sloan DIgital Sky SurvEy -- PARADISE);
\item the Brazil-France exchange programmes Funda\c{c}\~{a}o de Amparo \`{a} Pesquisa do Estado de S\~{a}o Paulo (FAPESP) and Coordena\c{c}\~{a}o de Aperfeicoamento de Pessoal de N\'{\i}vel Superior (CAPES) - Comit\'{e} Fran\c{c}ais d'Evaluation de la Coop\'{e}ration Universitaire et Scientifique avec le Br\'{e}sil (COFECUB);
\item the Chilean Agencia Nacional de Investigaci\'{o}n y Desarrollo (ANID) through Fondo Nacional de Desarrollo Cient\'{\i}fico y Tecnol\'{o}gico (FONDECYT) Regular Project 1210992 (L.~Chemin);
\item the National Natural Science Foundation of China (NSFC) through grants 11573054, 11703065, and 12173069, the China Scholarship Council through grant 201806040200, and the Natural Science Foundation of Shanghai through grant 21ZR1474100;  
\item the Tenure Track Pilot Programme of the Croatian Science Foundation and the \'{E}cole Polytechnique F\'{e}d\'{e}rale de Lausanne and the project TTP-2018-07-1171 `Mining the Variable Sky', with the funds of the Croatian-Swiss Research Programme;
\item the Czech-Republic Ministry of Education, Youth, and Sports through grant LG 15010 and INTER-EXCELLENCE grant LTAUSA18093, and the Czech Space Office through ESA PECS contract 98058;
\item the Danish Ministry of Science;
\item the Estonian Ministry of Education and Research through grant IUT40-1;
\item the European Commission's Sixth Framework Programme through the European Leadership in Space Astrometry (\href{https://www.cosmos.esa.int/web/gaia/elsa-rtn-programme}{ELSA}) Marie Curie Research Training Network (MRTN-CT-2006-033481), through Marie Curie project PIOF-GA-2009-255267 (Space AsteroSeismology \& RR Lyrae stars, SAS-RRL), and through a Marie Curie Transfer-of-Knowledge (ToK) fellowship (MTKD-CT-2004-014188); the European Commission's Seventh Framework Programme through grant FP7-606740 (FP7-SPACE-2013-1) for the \gaia\ European Network for Improved data User Services (\href{https://gaia.ub.edu/twiki/do/view/GENIUS/}{GENIUS}) and through grant 264895 for the \gaia\ Research for European Astronomy Training (\href{https://www.cosmos.esa.int/web/gaia/great-programme}{GREAT-ITN}) network;
\item the European Cooperation in Science and Technology (COST) through COST Action CA18104 `Revealing the Milky Way with \gaia\ (MW-\gaia)';
\item the European Research Council (ERC) through grants 320360, 647208, and 834148 and through the European Union's Horizon 2020 research and innovation and excellent science programmes through Marie Sk{\l}odowska-Curie grants 687378 (Small Bodies: Near and Far), 682115 (Using the Magellanic Clouds to Understand the Interaction of Galaxies), 695099 (A sub-percent distance scale from binaries and Cepheids -- CepBin), 716155 (Structured ACCREtion Disks -- SACCRED), 745617 (Our Galaxy at full HD -- Gal-HD), 895174 (The build-up and fate of self-gravitating systems in the Universe), 951549 (Sub-percent calibration of the extragalactic distance scale in the era of big surveys -- UniverScale), 101004214 (Innovative Scientific Data Exploration and Exploitation Applications for Space Sciences -- EXPLORE), 101004719 (OPTICON-RadioNET Pilot), 101055318 (The 3D motion of the Interstellar Medium with ESO and ESA telescopes -- ISM-FLOW), and 101063193 (Evolutionary Mechanisms in the Milky waY; the Gaia Data Release 3 revolution -- EMMY);
\item the European Science Foundation (ESF), in the framework of the \gaia\ Research for European Astronomy Training Research Network Programme (\href{https://www.cosmos.esa.int/web/gaia/great-programme}{GREAT-ESF});
\item the European Space Agency (ESA) in the framework of the \gaia\ project, through the Plan for European Cooperating States (PECS) programme through contracts C98090 and 4000106398/12/NL/KML for Hungary, through contract 4000115263/15/NL/IB for Germany, through PROgramme de D\'{e}veloppement d'Exp\'{e}riences scientifiques (PRODEX) grants  4000132054 for Hungary and through contract 4000132226/20/ES/CM;
\item the Academy of Finland through grants 299543, 307157, 325805, 328654, 336546, and 345115 and the Magnus Ehrnrooth Foundation;
\item the French Centre National d'\'{E}tudes Spatiales (CNES), the Agence Nationale de la Recherche (ANR) through grant ANR-10-IDEX-0001-02 for the `Investissements d'avenir' programme, through grant ANR-15-CE31-0007 for project `Modelling the Milky Way in the \gaia\ era' (MOD4\gaia), through grant ANR-14-CE33-0014-01 for project `The Milky Way disc formation in the \gaia\ era' (ARCHEOGAL), through grant ANR-15-CE31-0012-01 for project `Unlocking the potential of Cepheids as primary distance calibrators' (UnlockCepheids), through grant ANR-19-CE31-0017 for project `Secular evolution of galaxies' (SEGAL), and through grant ANR-18-CE31-0006 for project `Galactic Dark Matter' (GaDaMa), the Centre National de la Recherche Scientifique (CNRS) and its SNO \gaia\ of the Institut des Sciences de l'Univers (INSU), its Programmes Nationaux: Cosmologie et Galaxies (PNCG), Gravitation R\'{e}f\'{e}rences Astronomie M\'{e}trologie (PNGRAM), Plan\'{e}tologie (PNP), Physique et Chimie du Milieu Interstellaire (PCMI), and Physique Stellaire (PNPS), supported by INSU along with the Institut National de Physique  (INP) and the Institut National de Physique nucl\'{e}aire et de Physique des Particules (IN2P3), and co-funded by CNES; the `Action F\'{e}d\'{e}ratrice \gaia' of the Observatoire de Paris, and the R\'{e}gion de Franche-Comt\'{e};
\item the German Aerospace Agency (Deutsches Zentrum f\"{u}r Luft- und Raumfahrt e.V., DLR) through grants 50QG0501, 50QG0601, 50QG0602, 50QG0701, 50QG0901, 50QG1001, 50QG1101, 50\-QG1401, 50QG1402, 50QG1403, 50QG1404, 50QG1904, 50QG2101, 50QG2102, and 50QG2202, and the Centre for Information Services and High Performance Computing (ZIH) at the Technische Universit\"{a}t Dresden for generous allocations of computer time;
\item the Hungarian Academy of Sciences through the J\'anos Bolyai Research Scholarship (G. Marton and Z. Nagy), the Lend\"{u}let Programme grants LP2014-17 and LP2018-7 and the Hungarian National Research, Development, and Innovation Office (NKFIH) through grant KKP-137523 (`SeismoLab');
\item the Science Foundation Ireland (SFI) through a Royal Society - SFI University Research Fellowship (M.~Fraser);
\item the Israel Ministry of Science and Technology through grant 3-18143 and the Israel Science Foundation (ISF) through grant 1404/22;
\item the Agenzia Spaziale Italiana (ASI) through contracts I/037/08/0, I/058/10/0, 2014-025-R.0, 2014-025-R.1.2015, and 2018-24-HH.0 and its addendum 2018-24-HH.1-2022 to the Italian Istituto Nazionale di Astrofisica (INAF), contract 2014-049-R.0/1/2, 2022-14-HH.0 to INAF for the Space Science Data Centre (SSDC, formerly known as the ASI Science Data Center, ASDC), contracts I/008/10/0, 2013/030/I.0, 2013-030-I.0.1-2015, and 2016-17-I.0 to the Aerospace Logistics Technology Engineering Company (ALTEC S.p.A.), INAF, and the Italian Ministry of Education, University, and Research (Ministero dell'Istruzione, dell'Universit\`{a} e della Ricerca) through the Premiale project `MIning The Cosmos Big Data and Innovative Italian Technology for Frontier Astrophysics and Cosmology' (MITiC);
\item the Netherlands Organisation for Scientific Research (NWO) through grant NWO-M-614.061.414, through a VICI grant (A.~Helmi), and through a Spinoza prize (A.~Helmi), and the Netherlands Research School for Astronomy (NOVA);
\item the Polish National Science Centre through HARMONIA grant 2018/30/M/ST9/00311 and DAINA grant 2017/27/L/ST9/03221 and the Ministry of Science and Higher Education (MNiSW) through grant DIR/WK/2018/12;
\item the Portuguese Funda\c{c}\~{a}o para a Ci\^{e}ncia e a Tecnologia (FCT) through national funds, grants 2022.06962.PTDC and 2022.03993.PTDC, and work contract DL 57/2016/CP1364/CT0006, grants UIDB/04434/2020 and 
UIDP/04434/2020 for the Instituto de Astrof\'{\i}sica e Ci\^{e}ncias do Espa\c{c}o (IA), grants UIDB/00408/2020 and UIDP/00408/2020 for LASIGE, and grants UIDB/00099/2020 and UIDP/00099/2020 for the Centro de Astrof\'{\i}sica e Gravita\c{c}\~{a}o (CENTRA);  
\item the Slovenian Research Agency through grant P1-0188;
\item the Spanish Ministry of Economy (MINECO/FEDER, UE), the Spanish Ministry of Science and Innovation (MCIN), the Spanish Ministry of Education, Culture, and Sports, and the Spanish Government through grants BES-2016-078499, BES-2017-083126, BES-C-2017-0085, ESP2016-80079-C2-1-R, FPU16/03827, RTI2018-095076-B-C22, PID2021-122842OB-C22, PDC2021-121059-C22,  and TIN2015-65316-P (`Computaci\'{o}n de Altas Prestaciones VII'), the Juan de la Cierva Incorporaci\'{o}n Programme (FJCI-2015-2671 and IJC2019-04862-I for F.~Anders), the Severo Ochoa Centre of Excellence Programme (SEV2015-0493) and MCIN/AEI/10.13039/501100011033/ EU FEDER and Next Generation EU/PRTR (PRTR-C17.I1); the European Union through European Regional Development Fund `A way of making Europe' through grants PID2021-122842OB-C21, PID2021-125451NA-I00, CNS2022-13523 and RTI2018-095076-B-C21, the Institute of Cosmos Sciences University of Barcelona (ICCUB, Unidad de Excelencia `Mar\'{\i}a de Maeztu') through grant CEX2019-000918-M, the University of Barcelona's official doctoral programme for the development of an R+D+i project through an Ajuts de Personal Investigador en Formaci\'{o} (APIF) grant, the \href{https://svo.cab.inta-csic.es/}{Spanish Virtual Observatory} project funded by MCIN/AEI/10.13039/501100011033/ through grant PID2020-112949GB-I00; the Centro de Investigaci\'{o}n en Tecnolog\'{\i}as de la Informaci\'{o}n y las Comunicaciones (CITIC), funded by the Xunta de Galicia through the collaboration agreement to reinforce CIGUS research centers, research consolidation grant ED431B 2021/36 and scholarships from Xunta de Galicia and the EU - ESF ED481A-2019/155 and ED481A 2021/296; the Red Espa\~{n}ola de Supercomputaci\'{o}n (RES) computer resources at MareNostrum, the Barcelona Supercomputing Centre - Centro Nacional de Supercomputaci\'{o}n (BSC-CNS) through activities AECT-2017-2-0002, AECT-2017-3-0006, AECT-2018-1-0017, AECT-2018-2-0013, AECT-2018-3-0011, AECT-2019-1-0010, AECT-2019-2-0014, AECT-2019-3-0003, AECT-2020-1-0004, and DATA-2020-1-0010, the Departament d'Innovaci\'{o}, Universitats i Empresa de la Generalitat de Catalunya through grant 2014-SGR-1051 for project `Models de Programaci\'{o} i Entorns d'Execuci\'{o} Parallels' (MPEXPAR), and Ramon y Cajal Fellowships RYC2018-025968-I,  RYC2021-031683-I and RYC2021-033762-I, funded by MICIN/AEI/10.13039/501100011033 and by the European Union NextGenerationEU/PRTR and the European Science Foundation (`Investing in your future'); the Port d'Informaci\'{o} Cient\'{i}fica (PIC), through a collaboration between the Centro de Investigaciones Energ\'{e}ticas, Medioambientales y Tecnol\'{o}gicas (CIEMAT) and the Institut de F\'{i}sica d’Altes Energies (IFAE), supported by the call for grants for Scientific and Technical Equipment 2021 of the State Program for Knowledge Generation and Scientific and Technological Strengthening of the R+D+i System, financed by MCIN/AEI/ 10.13039/501100011033 and the EU NextGeneration/PRTR (Hadoop Cluster for the comprehensive management of massive scientific data, reference EQC2021-007479-P);
\item the Swedish National Space Agency (SNSA/Rymdstyrelsen);
\item the Swiss State Secretariat for Education, Research, and Innovation through the Swiss Activit\'{e}s Nationales Compl\'{e}mentaires and the Swiss National Science Foundation through an Eccellenza Professorial Fellowship (award PCEFP2\_194638 for R.~Anderson);
\item the United Kingdom Particle Physics and Astronomy Research Council (PPARC), the United Kingdom Science and Technology Facilities Council (STFC), and the United Kingdom Space Agency (UKSA) through the following grants to the University of Bristol, Brunel University London, the Open University, the University of Cambridge, the University of Edinburgh, the University of Leicester, the Mullard Space Sciences Laboratory of University College London, and the United Kingdom Rutherford Appleton Laboratory (RAL): PP/D006503/1, PP/D006511/1, PP/D006546/1, PP/D006570/1, PP/D006791/1, ST/I000852/1, ST/J005045/1, ST/K00056X/1, ST/K000209/1, ST/K000756/1, ST/K000578/1, ST/L002388/1, ST/L006553/1, ST/L006561/1, ST/N000595/1, ST/N000641/1, ST/N000978/1, ST/N001117/1, ST/S000089/1, ST/S000976/1, ST/S000984/1, ST/S001123/1, ST/S001948/1, ST/S001980/1, ST/S002103/1, ST/V000969/1, ST/W002469/1, ST/W002493/1, ST/W002671/1, ST/W002809/1, EP/V520342/1, ST/X00158X/1, ST/X001601/1, ST/X001636/1, ST/X001687/1, ST/X002667/1, ST/X002683/1 and ST/X002969/1.
\end{itemize}

The \gaia\ project and data processing have made use of:
\begin{itemize}
\item the Set of Identifications, Measurements, and Bibliography for Astronomical Data \citep[SIMBAD,][]{2000AAS..143....9W}, the `Aladin sky atlas' \citep{2000A&AS..143...33B,2014ASPC..485..277B}, and the VizieR catalogue access tool \citep{2000A&AS..143...23O}, all operated at the Centre de Donn\'{e}es astronomiques de Strasbourg (\href{http://cds.u-strasbg.fr/}{CDS});
\item the National Aeronautics and Space Administration (NASA) Astrophysics Data System (\href{http://adsabs.harvard.edu/abstract_service.html}{ADS});
\item the SPace ENVironment Information System (SPENVIS), initiated by the Space Environment and Effects Section (TEC-EES) of ESA and developed by the Belgian Institute for Space Aeronomy (BIRA-IASB) under ESA contract through ESA’s General Support Technologies Programme (GSTP), administered by the BELgian federal Science Policy Office (BELSPO);
\item the software products \href{http://www.starlink.ac.uk/topcat/}{TOPCAT}, \href{http://www.starlink.ac.uk/stil}{STIL}, and \href{http://www.starlink.ac.uk/stilts}{STILTS} \citep{2005ASPC..347...29T,2006ASPC..351..666T};
\item Matplotlib \citep{Hunter:2007};
\item IPython \citep{PER-GRA:2007};  
\item Astropy, a community-developed core Python package for Astronomy \citep{2018AJ....156..123A};
\item R \citep{RManual};
\item the HEALPix package \citep[][\url{http://healpix.sourceforge.net/}]{2005ApJ...622..759G};
\item Vaex \citep{2018A&A...618A..13B};
\item the \hip-2 catalogue \citep{2007A&A...474..653V}. The \hip and \tyc catalogues were constructed under the responsibility of large scientific teams collaborating with ESA. The Consortia Leaders were Lennart Lindegren (Lund, Sweden: NDAC) and Jean Kovalevsky (Grasse, France: FAST), together responsible for the \hip Catalogue; Erik H{\o}g (Copenhagen, Denmark: TDAC) responsible for the \tyc Catalogue; and Catherine Turon (Meudon, France: INCA) responsible for the \hip Input Catalogue (HIC);  
\item the \tyctwo catalogue \citep{2000A&A...355L..27H}, the construction of which was supported by the Velux Foundation of 1981 and the Danish Space Board;
\item The Tycho double star catalogue \citep[TDSC,][]{2002A&A...384..180F}, based on observations made with the ESA \hip astrometry satellite, as supported by the Danish Space Board and the United States Naval Observatory through their double-star programme;
\item data products from the Two Micron All Sky Survey \citep[2MASS,][]{2006AJ....131.1163S}, which is a joint project of the University of Massachusetts and the Infrared Processing and Analysis Center (IPAC) / California Institute of Technology, funded by the National Aeronautics and Space Administration (NASA) and the National Science Foundation (NSF) of the USA;
\item the ninth data release of the AAVSO Photometric All-Sky Survey (\href{https://www.aavso.org/apass}{APASS}, \citealt{apass9}), funded by the Robert Martin Ayers Sciences Fund;
\item the first data release of the Pan-STARRS survey \citep{panstarrs1,panstarrs1b,panstarrs1c,panstarrs1d,panstarrs1e,panstarrs1f}. The Pan-STARRS1 Surveys (PS1) and the PS1 public science archive have been made possible through contributions by the Institute for Astronomy, the University of Hawaii, the Pan-STARRS Project Office, the Max-Planck Society and its participating institutes, the Max Planck Institute for Astronomy, Heidelberg and the Max Planck Institute for Extraterrestrial Physics, Garching, The Johns Hopkins University, Durham University, the University of Edinburgh, the Queen's University Belfast, the Harvard-Smithsonian Center for Astrophysics, the Las Cumbres Observatory Global Telescope Network Incorporated, the National Central University of Taiwan, the Space Telescope Science Institute, the National Aeronautics and Space Administration (NASA) through grant NNX08AR22G issued through the Planetary Science Division of the NASA Science Mission Directorate, the National Science Foundation through grant AST-1238877, the University of Maryland, Eotvos Lorand University (ELTE), the Los Alamos National Laboratory, and the Gordon and Betty Moore Foundation;
\item the second release of the Guide Star Catalogue \citep[GSC2.3,][]{2008AJ....136..735L}. The Guide Star Catalogue II is a joint project of the Space Telescope Science Institute (STScI) and the Osservatorio Astrofisico di Torino (OATo). STScI is operated by the Association of Universities for Research in Astronomy (AURA), for the National Aeronautics and Space Administration (NASA) under contract NAS5-26555. OATo is operated by the Italian National Institute for Astrophysics (INAF). Additional support was provided by the European Southern Observatory (ESO), the Space Telescope European Coordinating Facility (STECF), the International GEMINI project, and the European Space Agency (ESA) Astrophysics Division (nowadays SCI-S);
\item the eXtended, Large (XL) version of the catalogue of Positions and Proper Motions \citep[PPM-XL,][]{2010AJ....139.2440R};
\item data products from the Wide-field Infrared Survey Explorer (WISE), which is a joint project of the University of California, Los Angeles, and the Jet Propulsion Laboratory/California Institute of Technology, and NEOWISE, which is a project of the Jet Propulsion Laboratory/California Institute of Technology. WISE and NEOWISE are funded by the National Aeronautics and Space Administration (NASA);
\item the first data release of the United States Naval Observatory (USNO) Robotic Astrometric Telescope \citep[URAT-1,][]{urat1};
\item the fourth data release of the United States Naval Observatory (USNO) CCD Astrograph Catalogue \citep[UCAC-4,][]{2013AJ....145...44Z};
\item the sixth and final data release of the Radial Velocity Experiment \citep[RAVE DR6,][]{2020AJ....160...83S,rave6a}. Funding for RAVE has been provided by the Leibniz Institute for Astrophysics Potsdam (AIP), the Australian Astronomical Observatory, the Australian National University, the Australian Research Council, the French National Research Agency, the German Research Foundation (SPP 1177 and SFB 881), the European Research Council (ERC-StG 240271 Galactica), the Istituto Nazionale di Astrofisica at Padova, the Johns Hopkins University, the National Science Foundation of the USA (AST-0908326), the W.M.\ Keck foundation, the Macquarie University, the Netherlands Research School for Astronomy, the Natural Sciences and Engineering Research Council of Canada, the Slovenian Research Agency, the Swiss National Science Foundation, the Science \& Technology Facilities Council of the UK, Opticon, Strasbourg Observatory, and the Universities of Basel, Groningen, Heidelberg, and Sydney. The RAVE website is at \url{https://www.rave-survey.org/};
\item the first data release of the Large sky Area Multi-Object Fibre Spectroscopic Telescope \citep[LAMOST DR1,][]{LamostDR1};
\item the K2 Ecliptic Plane Input Catalogue \citep[EPIC,][]{epic-2016ApJS..224....2H};
\item the ninth data release of the Sloan Digitial Sky Survey \citep[SDSS DR9,][]{SDSS9}. Funding for SDSS-III has been provided by the Alfred P. Sloan Foundation, the Participating Institutions, the National Science Foundation, and the United States Department of Energy Office of Science. The SDSS-III website is \url{http://www.sdss3.org/}. SDSS-III is managed by the Astrophysical Research Consortium for the Participating Institutions of the SDSS-III Collaboration including the University of Arizona, the Brazilian Participation Group, Brookhaven National Laboratory, Carnegie Mellon University, University of Florida, the French Participation Group, the German Participation Group, Harvard University, the Instituto de Astrof\'{\i}sica de Canarias, the Michigan State/Notre Dame/JINA Participation Group, Johns Hopkins University, Lawrence Berkeley National Laboratory, Max Planck Institute for Astrophysics, Max Planck Institute for Extraterrestrial Physics, New Mexico State University, New York University, Ohio State University, Pennsylvania State University, University of Portsmouth, Princeton University, the Spanish Participation Group, University of Tokyo, University of Utah, Vanderbilt University, University of Virginia, University of Washington, and Yale University;
\item the thirteenth release of the Sloan Digital Sky Survey \citep[SDSS DR13,][]{2017ApJS..233...25A}. Funding for SDSS-IV has been provided by the Alfred P. Sloan Foundation, the United States Department of Energy Office of Science, and the Participating Institutions. SDSS-IV acknowledges support and resources from the Center for High-Performance Computing at the University of Utah. The SDSS web site is \url{https://www.sdss.org/}. SDSS-IV is managed by the Astrophysical Research Consortium for the Participating Institutions of the SDSS Collaboration including the Brazilian Participation Group, the Carnegie Institution for Science, Carnegie Mellon University, the Chilean Participation Group, the French Participation Group, Harvard-Smithsonian Center for Astrophysics, Instituto de Astrof\'isica de Canarias, The Johns Hopkins University, Kavli Institute for the Physics and Mathematics of the Universe (IPMU) / University of Tokyo, the Korean Participation Group, Lawrence Berkeley National Laboratory, Leibniz Institut f\"ur Astrophysik Potsdam (AIP),  Max-Planck-Institut f\"ur Astronomie (MPIA Heidelberg), Max-Planck-Institut f\"ur Astrophysik (MPA Garching), Max-Planck-Institut f\"ur Extraterrestrische Physik (MPE), National Astronomical Observatories of China, New Mexico State University, New York University, University of Notre Dame, Observat\'ario Nacional / MCTI, The Ohio State University, Pennsylvania State University, Shanghai Astronomical Observatory, United Kingdom Participation Group, Universidad Nacional Aut\'onoma de M\'{e}xico, University of Arizona, University of Colorado Boulder, University of Oxford, University of Portsmouth, University of Utah, University of Virginia, University of Washington, University of Wisconsin, Vanderbilt University, and Yale University;
\item the second release of the SkyMapper catalogue \citep[SkyMapper DR2,][Digital Object Identifier 10.25914/5ce60d31ce759]{2019PASA...36...33O}. The national facility capability for SkyMapper has been funded through grant LE130100104 from the Australian Research Council (ARC) Linkage Infrastructure, Equipment, and Facilities (LIEF) programme, awarded to the University of Sydney, the Australian National University, Swinburne University of Technology, the University of Queensland, the University of Western Australia, the University of Melbourne, Curtin University of Technology, Monash University, and the Australian Astronomical Observatory. SkyMapper is owned and operated by The Australian National University's Research School of Astronomy and Astrophysics. The survey data were processed and provided by the SkyMapper Team at the Australian National University. The SkyMapper node of the All-Sky Virtual Observatory (ASVO) is hosted at the National Computational Infrastructure (NCI). Development and support the SkyMapper node of the ASVO has been funded in part by Astronomy Australia Limited (AAL) and the Australian Government through the Commonwealth's Education Investment Fund (EIF) and National Collaborative Research Infrastructure Strategy (NCRIS), particularly the National eResearch Collaboration Tools and Resources (NeCTAR) and the Australian National Data Service Projects (ANDS);
\item the \gaia-ESO Public Spectroscopic Survey \citep[GES,][]{GES_final_release_paper_1,GES_final_release_paper_2}. The \gaia-ESO Survey is based on data products from observations made with ESO Telescopes at the La Silla Paranal Observatory under programme ID 188.B-3002. Public data releases are available through the \href{https://www.gaia-eso.eu/data-products/public-data-releases}{ESO Science Portal}. The project has received funding from the Leverhulme Trust (project RPG-2012-541), the European Research Council (project ERC-2012-AdG 320360-\gaia-ESO-MW), and the Istituto Nazionale di Astrofisica, INAF (2012: CRA 1.05.01.09.16; 2013: CRA 1.05.06.02.07).
\end{itemize}

The GBOT programme (\href{https://gea.esac.esa.int/archive/documentation/GDR3/Data_processing/chap_cu3ast/sec_cu3ast_prop/ssec_cu3ast_prop_gbot.html}{GBOT}) uses observations collected at (i) the European Organisation for Astronomical Research in the Southern Hemisphere (ESO) with the VLT Survey Telescope (VST), under ESO programmes
092.B-0165,
093.B-0236,
094.B-0181,
095.B-0046,
096.B-0162,
097.B-0304,
098.B-0030,
099.B-0034,
0100.B-0131,
0101.B-0156,
0102.B-0174, 
0103.B-0165,
0104.B-0081,
0106.20ZA.001 (OmegaCam),
0106.20ZA.002 (FORS2),
0108.21YF;
and under INAF programs 110.256C,
112.266Q;
%
%
and (ii) the Liverpool Telescope, which is operated on the island of La Palma by Liverpool John Moores University in the Spanish Observatorio del Roque de los Muchachos of the Instituto de Astrof\'{\i}sica de Canarias with financial support from the United Kingdom Science and Technology Facilities Council, and (iii) telescopes of the Las Cumbres Observatory Global Telescope Network.

In addition to the currently active DPAC (and ESA science) authors of the peer-reviewed papers accompanying the data release, there are large numbers of former DPAC members who made significant contributions to the (preparations of the) data processing. 
In addition to the DPAC consortium, past and present, there are numerous people, mostly in ESA and in industry, who have made or continue to make essential contributions to \gaia, for instance those employed in science and mission operations or in the design, manufacturing, integration, and testing of the spacecraft and its modules, subsystems, and units. Many of those will remain unnamed yet spent countless hours, occasionally during nights, weekends, and public holidays, in cold offices and dark clean rooms. They are acknowledged in the Gaia Documentation.


\end{appendix}

\end{document}